\documentclass[showpacs,prd,twocolumn,eqsecnum,superscriptaddress,tightenlines, showpacs,preprintnumbers,a4paper]{revtex4}

\usepackage{amsfonts}
\usepackage{amsmath}
\usepackage{graphicx}
\usepackage{verbatim}
\usepackage{color}
\usepackage{subfigure}
\usepackage[colorlinks=true]{hyperref}
\usepackage{epstopdf}
\usepackage{bm}
\usepackage{bm}

\def \ft{\widetilde}
\def \Dt{{\Delta t}}
\def \d {\mathrm{d}}

\begin{document}

\title{Multibaseline gravitational wave radiometry}

\author{Dipongkar Talukder}
\email[]{talukder\_d@wsu.edu}
\affiliation{Department of Physics and Astronomy, Washington State University, Pullman, Washington 99164-2814, USA}

\author{Sanjit Mitra}
\email[]{smitra@ligo.caltech.edu}
\affiliation{Jet Propulsion Laboratory, California Institute of Technology, 4800 Oak Grove Drive, Pasadena, California 91109, USA}
\affiliation{LIGO Laboratory, California Institute of Technology, MS 18-34, Pasadena, California 91125, USA}
\affiliation{Observatoire de la C\^ote dÕAzur, BP 4229, 06304 Nice Cedex 4, France}

\author{Sukanta Bose}
\email[]{sukanta@wsu.edu}
\affiliation{Department of Physics and Astronomy, Washington State University, Pullman, Washington 99164-2814, USA}

\pacs{95.55.Ym, 04.30.Db, 98.80.-k, 97.80.-d}

\date{\today}

\begin{abstract}
We present a statistic for the detection of stochastic gravitational wave backgrounds (SGWBs) using radiometry with a network of multiple baselines. We also quantitatively compare the sensitivities of existing baselines and their network to SGWBs. We assess how the measurement accuracy of signal parameters, e.g., the sky position of a localized source, can improve when using a network of baselines, as compared to any of the single participating baselines. The search statistic itself is derived from the likelihood ratio of the cross correlation of the data across all possible baselines in a detector network and is optimal in Gaussian noise. Specifically, it is the likelihood ratio maximized over the strength of the SGWB and is called the maximized-likelihood ratio (MLR). One of the main advantages of using the MLR over past search strategies for inferring the presence or absence of a signal is that the former does not require the deconvolution of the cross correlation statistic. Therefore, it does not suffer from errors inherent to the deconvolution procedure and is especially useful for detecting weak sources. In the limit of a single baseline, it reduces to the detection statistic studied by \citet{Ballmer} and \citet{Mitra}. Unlike past studies, here the MLR statistic enables us to compare {\em quantitatively} the performances of a variety of baselines searching for a SGWB signal in (simulated) data. Although we use simulated noise and SGWB signals for making these comparisons, our method can be straightforwardly applied on real data.
\end{abstract}

\preprint{[{\color{red} LIGO-P1000123}]}

\maketitle

\section{Introduction}
Just like the discoveries of the cosmic microwave background and pulsars in the electromagnetic spectrum, a discovery of unknown sources by Earth-based detectors such as LIGO and Virgo in the gravitational wave (GW) spectrum by serendipity is an interesting prospect. The LIGO Scientific Collaboration and the Virgo Collaboration are addressing it by searching for both transient signals, or ``bursts,'' and long-duration signals in the data from their detectors. Here, we focus on a subset of the latter type that can be modeled as a stochastic background. The search for an isotropic stochastic GW background has caught significant attention due to its cosmological significance. This primordial GW background is a direct probe of cosmological inflation \cite{Abbott:2009ws}. However, the astrophysical background, arising in the nearby Universe~\cite{Coward}, e.g., from an unresolved superposition of GW signals from multiple sources, such as low-mass x-ray binaries or, even, coalescing compact objects, is possibly much stronger than the primordial background and is anisotropic.

A variety of data analysis strategies to search for an anisotropic GW background have been proposed and implemented in the past~\cite{AllenOttewill,cornish,kudohI,kudohII,taruyaIII,Cannon:2007br}. These searches are usually performed in two types of bases in the sky, namely, the pixel and spherical harmonic bases. Use of the radiometer technique for searching a GW background was proposed in Ref. \cite{LazzariniWeiss} and was implemented in the pixel basis on data from LIGO's fourth science run~\cite{S4Radiometer}. An elaborate study of this method, including the maximum-likelihood (ML) estimation of the true anisotropy of GW background by deconvolving the observed sky map, was presented in \citet{Mitra}. Even though the pixel-based search is promising and simpler to understand, it is not the best basis for probing sources with angular spreads greater than the angular resolution of the GW radiometer. The spherical harmonic basis is expected to be better suited for detecting such sources \cite{ThraneEtal}. Past attempts at probing the GW anisotropy in the spherical harmonic basis were essentially studies of the periodic modulation of the observed background in the detector baselines. Recently, a general ML formalism was developed to search for the GW anisotropies in any basis, including the spherical harmonic basis, using a network of ground-based GW interferometers \cite{ThraneEtal}. The pixel-based search is a specific application of this formalism. One of the main goals of this paper is to perform a thorough comparison of the expected performances of individual baselines and the whole network in detecting an astrophysical stochastic gravitational wave background (SGWB) and in estimating its parameters. The pixel basis is used for this study.

Even though a pixel-based search is optimal for a localized source, the resolution of the source is limited by the length of the radiometer baselines, the orientation of the detectors, and their individual sensitivities.
Probing a stochastic GW background with 
energy distributed across the pixelated sky demands a statistically meaningful integration of the energies received in every pixel. In order to accomplish this, we extend the maximized-likelihood ratio (MLR) statistic for a single baseline to incorporate a network of detectors or, equivalently, multiple baselines. The rest of the paper is devoted to studying the performance of individual GW detector baselines and the whole network by comparing different figures of merit for their performance, e.g., sensitivity, accuracy in localizing sources, sky coverage, and faithful extraction from the data of the sky distribution of a stochastic 
background.

The paper is organized as follows: In Sec.~\ref{sec:statistic}, we develop and study the efficiency of an optimal all-sky search statistic for anisotropic SGWBs that obviates the solving of the inverse problem, which may not always be well-posed. In Sec.~\ref{sec:performance}, we compare the performance of a network with that of its individual baselines using a variety of figures of merit. 
In Sec.~\ref{sec:conclusion}, we conclude by summarizing the implications of this work on ongoing SGWB searches and by highlighting future directions in GW radiometry.

\section{Optimal search statistic}
\label{sec:statistic}

\subsection{Statistical properties of the signal and detector noise}
In the transverse traceless gauge, the spatial part of the metric perturbations due to a SGWB can be written as a superposition of plane waves
\begin{equation}\label{eq3.1}
h_{ab}(t,{\bf{r}})=\int_{-\infty}^{\infty} df \int_{S^{2}} d\hat{\Omega}\;e_{ab}^{A}(\hat{\Omega})\; \tilde{h}_{A}(f,\hat{\Omega})e^{i 2 \pi f(t+\hat{\Omega}\cdot {\bf r}/c)}\; ,
\end{equation}
where $a$ and $b$ are spatial indices, $e_{ab}^{A}(\hat{\Omega})$ are the components of the gravitational wave polarization tensors, and $\hat{\Omega}$ is a unit vector pointing in the direction of wave propagation. Note that the Einstein summation convention for repeated indices has been applied to the polarization index $A=\{+,\times \}$. Since the GW strain $h_{ab}(t,{\bf r})$ is real, the complex Fourier amplitudes $\tilde{h}_{A}(f,\hat{\Omega})$ satisfy the reality condition, $\tilde{h}_{A}(-f,\hat{\Omega})=\tilde{h}_{A}^{*}(f,\hat{\Omega})$.\\

The polarization tensors can be defined in terms of the spherical polar coordinates, 
$\theta\in[0,\pi]$ and $\phi\in[0,2\pi]$, on the sky. Let 
as follows
\begin{eqnarray}\label{eq3.2}
\hat{\Omega} &=& \sin\theta\cos\phi\,\hat{x}+\sin\theta\sin\phi\,\hat{y}+\cos\theta\,\hat{z}\;, \nonumber \\
\hat{m} &=& \cos\theta\cos\phi\,\hat{x}+\cos\theta\sin\phi\,\hat{y}-\sin\theta\,\hat{z}\; , \nonumber\\
\hat{n} &=& -\sin\phi\,\hat{x}+\cos\phi\,\hat{y}\; ,
\end{eqnarray}
such that $\{\hat{m},\hat{n},\hat{\Omega}\}$ form a right-handed system of unit vectors. The axes are defined as follows: For a fixed but arbitrarily chosen origin of time $t=0$, $\hat{x}$ is directed toward the intersection of the equator and the longitude $\phi=0$, $\hat{z}$ points at the Celestial North Pole, and $\hat{y}$ is chosen orthogonal to the previous two axes, forming a right-handed triad. Then, the polarization tensor components $e_{ab}^{A}(\hat{\Omega})$ are defined as
\begin{eqnarray}\label{eq3.3}
e_{ab}^{+}(\hat{\Omega}) &=& \hat{m}_{a}\hat{m}_{b}-\hat{n}_{a}\hat{n}_{b}\; , \nonumber \\
e_{ab}^{\times}(\hat{\Omega}) &=& \hat{m}_{a}\hat{n}_{b}+\hat{n}_{a}\hat{m}_{b}\; ,
\end{eqnarray}
in the aforementioned right-handed orthogonal basis.

Understanding the signal excited in an interferometric detector by a SGWB is helped by specifying the detector's location and orientation in the above orthogonal basis. Let the $I^{\mbox{th}}$ GW detector be located at ${\bf r}_{I}(t)$, and let $\hat{X}_{I}(t)$ and $\hat{Y}_{I}(t)$ be the unit vectors pointing along its arms. These three detector location and orientation vectors are all time-dependent due to the Earth's rotation. Then, the components of the $I^{\mbox{th}}$ detector tensor are given by
\begin{equation}\label{eq3.4}
d_{I}^{ab}(t)=\frac{1}{2}\left[\hat{X}_{I}^{a}(t)\,\hat{X}_{I}^{b}(t)-\hat{Y}_{I}^{a}(t)\,\hat{Y}_{I}^{b}(t)\right]\;,
\end{equation}
and 
\begin{equation}\label{eq3.5}
h_{I}(t)=h_{ab}(t,{\bf r}_{I}(t))\,d_{I}^{ab}(t)\, ,
\end{equation}
is the strain 
in it due to the SGWB.

The response of a detector to the polarization component $A$ of a wave incident from direction $\hat{\Omega}$ is given by the antenna-pattern function
\begin{equation}\label{eq3.6}
F_{I}^{A}(\hat{\Omega},t) \equiv d_{I}^{ab}(t)\,e_{ab}^{A}(\hat{\Omega})\; ,
\end{equation}
where we assumed the Einstein summation convention over the repeated indices $a$ and $b$. Contracting (\ref{eq3.1}) with the detector tensor $d_{I}$, the GW strain signal in the $I^{\mbox{th}}$ detector can be expressed as
\begin{equation}\label{eq3.7}
h_{I}(t)=\int_{-\infty}^{\infty}df\int_{S^{2}} d \hat{\Omega}\,\tilde{h}_{A}(f,\hat{\Omega})F_{I}^{A}(\hat{\Omega},t)e^{i 2\pi f(t+\hat{\Omega}\cdot{\bf r}_{I}(t)/c)}\,,
\end{equation}
in terms of the antenna-pattern functions.

The Fourier components of the strain $\tilde{h}_{A}(f,\hat{\Omega})$ describing a stochastic GW background are random variables whose expectation values define the statistical properties of the background. Without loss of generality we assume that these components have zero mean:
\begin{equation}\label{eq3.8}
\langle\tilde{h}_{A}(f,\hat{\Omega})\rangle=0\,,
\end{equation}
where the angular brackets denote statistical average.
In the presence of a signal, the time series of the $I^{\mbox{th}}$ detector's output $x_I(t)$ is a sum of the GW signal $h_{I}(t)$ and the detector noise $n_{I}(t)$:
\begin{equation}\label{eq3.9}
x_{I}(t)=h_{I}(t)+n_{I}(t)\,.
\end{equation}
Statistically, the gravitational wave strain $h_{I}(t)$ are uncorrelated with the detector noise, implying that
\begin{equation}\label{eq3.10}
\langle h_{I}(t)\,n_{J}(t^{\prime})\rangle = 0\, .
\end{equation}
We also assume that the noise is Gaussian with zero mean,  i.e., $\langle n_{I}(t)\rangle =0$, and is uncorrelated in different detectors, namely,
\begin{equation}\label{eq3.11}
\langle  n_{I}(f)\, n_{J}(f^{\prime})\rangle = \frac{1}{2}\,\delta(f-f^{\prime})\,\delta_{IJ}\,\xi_{(I)}(f) \,,
\end{equation}
where $\xi_{(I)}$ is the one-sided noise power spectral density (PSD) of the $I^{\text{th}}$ detector. The last assumption is not unreasonable when the detectors are widely separated across the globe.

\subsection{Cross correlation statistic}
Since the targeted source is stochastic,
we search for its GW signal by looking for correlated patterns in the data of two or more detectors after accounting for time delays and detector responses consistent with a given sky location. This is done by cross-correlating the data $x_I(t)$ from the detectors, taken in pairs, with a sky-position-dependent time-frequency filter $\tilde{Q}^k(t;f)$, labeled by the sky-position index $k$. The cross correlation statistic combined for the observation period $T$ for the data $x_{1,2}(t)$ from two detectors or, equivalently, for a baseline is defined as
\begin{equation}\label{eq3.12}
S^k = 4\,\Delta t\sum_{t=0}^{T}\int_{-\infty}^{\infty} df~\tilde{x}_1^*(t;f)\tilde{x}_2(t;f)\tilde{Q}^k(t;f)\,,
\end{equation}
where $\tilde{x}_I(t;f)$ is the short-term Fourier transform of $x_I(t)$, over time interval $\Delta t$, and is defined as in Ref. \cite{Mitra} as
\begin{equation}
 \ft{x}_I(t;f) \ := \ \int_{t-\Dt/2}^{t+\Dt/2} \d t' \, x_I(t') \, e^{-2\pi i f t'} \,. 
\label{eq:defChunkFT}
\end{equation}
The filter that maximizes the signal-to-noise ratio (SNR) associated with this statistic is a scalar, square-integrable function on the sky \cite{Mitra} and, hence, can be resolved linearly in an appropriate basis, such as a pixel basis or the spherical harmonic basis. In the former case, $k$ is the pixel index.\\

Let an astrophysical GW background be modeled such that the Fourier components of its GW strain $\tilde{h}_A^k(f)$ of polarization $A$ from the $k^{\text{th}}$ sky-position obey
\begin{equation}\label{eq3.13}
\langle \tilde{h}_A^{k*}(f) \tilde{h}_{A'}^{k'}(f') \rangle = \delta_{AA'}~\delta(f-f')
\delta^{kk'} {\mathcal P}^k_{(A)} \,H(f)\,,
\end{equation}
where ${\mathcal P}^k_{(A)}$ is a dimensionless measure of the signal strength, and $H(f)$ is its two-sided power spectral density, with units of ${\rm Hz}^{-1}$ \cite{Mitra}. Here, we assume the signal to be stochastic and uncorrelated in the two polarizations, different frequencies, and different sky locations. In the presence of a signal in the detector data, the cross correlation statistic is
\begin{equation}\label{eq3.14}
S^k = {\mathcal B}_{+k^{\prime}}^k {\mathcal P}_{(+)}^{k^{\prime}} + {\mathcal B}_{\times k^{\prime}}^k {\mathcal P}_{(\times)}^{k^{\prime}} + n^{k} \,,
\end{equation}
where the {\em beam function} ${\mathcal B}_{A~k^{\prime}}^k$ is analogous to the point-spread function that maps the power in the {\em object} (or sky) plane to that in the {\em image} plane. Above $n^k$ is the noise in the $k^{\text{th}}$ sky position, and $S^k$ is termed as the dirty map \cite{Mitra}. We define ${\bm{\mathcal P}}_{(A)}$ as a vector, with ${\mathcal P}_{(A)}^{k}$ as its $k^{\text{th}}$ component, and ${\bm{\mathcal B}}_{A}$ as a matrix, with ${\mathcal B}_{A\,k^{\prime}}^{k}$ as its $(k,k^{\prime})^{\text{th}}$ element.

\subsection{Detection statistic}
\label{subsec:statistic}

To get a single detection statistic, one must combine the measurements of $S^{k}$ for all $k$. When the detector noises are Gaussian and uncorrelated, an assumption borne out to sufficient approximation for our purposes, the $n^k$ are Gaussian with a nontrivial covariance matrix, ${\bf{N}}$, determined by the beam functions. The exact form of ${\bf N}$ is discussed below. \\

If an astrophysical GW background signal, characterized by the pixel-strength vector ${\bm{\mathcal{P}}}$, is present in the data, then the probability density function of the  radiometer output ${\bf{S}}$ is given by
\begin{eqnarray}\label{eq3.15}
p({\bf{S}}|{\bm{\mathcal{P}}}) &=& (2\pi)^{-N_{\text{pix}}/2}\nonumber \\
&\times&\exp[-\frac{1}{2}(({\bf{S}}-{\bm{\mathcal B}}\cdot{\bm{\mathcal{P}}})^{T}\cdot {\bf{N}}^{-1}\cdot ({\bf{S}}-{\bm{\mathcal B}}\cdot {\bm{\mathcal{P}}})\nonumber \\
&+& {\text{Tr}}[\ln {\bf{N}}])]\; ,
\end{eqnarray}
whereas in the absence of a signal it is
\begin{equation}\label{eq3.16}
p({\bf{n}}) = (2\pi)^{-N_{\text{pix}}/2}\exp[-\frac{1}{2}({\bf{n}}^{T}\cdot {\bf{N}}^{-1}\cdot {\bf{n}} + {\text{Tr}}[\ln {\bf{N}}])]\; .
\end{equation}
By the Neyman-Pearson criterion, the optimal detection statistic is the likelihood ratio $p({\bf{S}}|{\bm{\mathcal P}})/p({\bf{n}})$ \cite{Helstrom}.\\

For an unpolarized background from a source distributed across multiple pixels and quantified by the signal-strength vector ${\bm{\mathcal P}}= {\bm{\mathcal P}}_{(+)}=\bm{{\mathcal P}}_{(\times)}$, the log-likelihood ratio maximized over ${\mathcal P }\equiv \| {\bm{\mathcal P }}\|$ is
\begin{eqnarray}\label{eq3.17}
\lambda  &=& \frac{S^k ({\bf N}^{-1})_{k k^{\prime}} ({\bm{\mathcal B} \cdot \hat{\bm{\mathcal P}}})^{k^{\prime}}} {\sqrt{ ({\bm{\mathcal B}} \cdot \hat{\bm{\mathcal P}})^{q}({{\bf N}}^{-1})_{q r} ({\bm{\mathcal B}} \cdot \hat{\bm{\mathcal P}})^{r} } } \; ,\nonumber\\
&=& \frac{S_{k}\hat{\mathcal P}^{k}}{\sqrt{\hat{\mathcal P}^{q}{\mathcal B}_{qr}\hat{\mathcal P}^{r}}}\; ,
\end{eqnarray}
where $\hat{{\bm{\mathcal P}}}$ is the unit vector along ${\bm{\mathcal P}}$. The beam matrix for an unpolarized source is given by 
\begin{eqnarray}\label{eq3.18}
{\mathcal B}_{p q} & = &  {\mathcal B}_{+ pq}+{\mathcal B}_{\times pq}  \, , \label{eq3.19} \\
& = & 8\,\Delta f\, \Delta t \, \sum_{t=0}^{T}\Gamma(\hat{\Omega}_{q},t)\Gamma(\hat{\Omega}_{p},t) \nonumber \\
&  &\times  \Re \left[ \sum_{f=f_{l}}^{f_{u}}e^{2\pi i f(\hat{\Omega}_{q} - \hat{\Omega}_{p})\cdot \Delta \vec{x}(t)/c} G(t,f) \right] \, ,\nonumber \\
\end{eqnarray}
\noindent where $\Gamma(\hat{\Omega}_{q},t)$ is the time-varying baseline antenna pattern, and $G(t,f)$ is a measure of the spectral strength of the source relative to the baseline's noise PSDs:
\begin{eqnarray}\label{eq3.20}
\Gamma(\hat{\Omega},t) &:=& F_{1}^{+}(\hat{\Omega},t) F_{2}^{+}(\hat{\Omega},t) + F_{1}^{\times}(\hat{\Omega},t) F_{2}^{\times}(\hat{\Omega},t) \, ,\nonumber \\
G(t,f) &:=& \frac{H^{2}(f)}{\xi_{(1)}(t,f)\,\xi_{(2)}(t,f)} \, .  
\end{eqnarray}
In the weak-signal limit the noise-covariance matrix is approximately equal to the beam matrix,
\begin{equation}
N_{p q} \approx \,{\mathcal B}_{p q}\,.
\end{equation}
Its diagonal elements inform us about the sensitivity of 
the network to the different pixels in the sky for an SGWB with PSD $H(f)$.

The statistic $\lambda$ is the maximized (log-)likelihood ratio for a single-baseline SGWB search and is the same statistic introduced in Appendix C of Ref.~\cite{ThraneEtal}. Here, it is expressed specifically in terms of quantities defined in the pixel basis. It has zero mean and unit variance in the absence of a signal. When a signal is present in the data and its parameters are
matched exactly by the template's, the mean of the statistic is
\begin{equation}\label{eq3.21}
\left<\lambda\right>={\mathcal P}\sqrt{ ({\bm{\mathcal B}}\cdot\hat{\bm{\mathcal P}})^{k} ({\bf N}^{-1})_{k k^{\prime}} ({\bm{\mathcal B}}\cdot \hat{\bm{\mathcal P}})^{k^{\prime}} } \; .
\end{equation}
The variance of the statistic remains unchanged. One can extend this single-baseline statistic to the case of a multibaseline network. That statistic arises directly from maximizing the log-likelihood ratio for a network and is given by 
\begin{eqnarray}\label{eq3.22}
\lambda_{\mathcal N}  &=& \frac{\sum_{{\mathcal I}=1}^{N_{b}}S_{\mathcal I}^k ({\bf N}_{\mathcal I}^{-1})_{k k^{\prime}} ({\bf{\bm{\mathcal B}_{\mathcal I}} \cdot \hat{\bm{\mathcal P}}})^{k^{\prime}}} {\sqrt{\sum_{{\mathcal I}=1}^{N_{b}} ({\bm{\mathcal B}}_{\mathcal I} \cdot \hat{\bm{\mathcal P}})^{q}({\bf N}_{\mathcal I}^{-1})_{q r} ({\bm{\mathcal B}}_{\mathcal I} \cdot \hat{\bm{\mathcal P}})^{r} } } \; ,\nonumber \\
&=& \frac{\sum_{{\mathcal I}=1}^{N_{b}}S^{\mathcal I}_{k}\hat{\mathcal P}^{k}}{\sqrt{\sum_{{\mathcal I}=1}^{N_{b}}\hat{\mathcal P}^{q}{\mathcal B}^{\mathcal I}_{qr}\hat{\mathcal P}^{r}}}\; ,
\end{eqnarray}
where $\mathcal{I}$ is the baseline index and the subscript $\mathcal{N}$ highlights that this MLR statistic is for a \textit{network} of baselines. 


The MLR statistic is a detection statistic for SGWBs in the same manner as the standard matched-filter statistic is for deterministic GW sources. The latter is also obtained by maximizing the likelihood ratio with respect to the strength of the deterministic source. Searching for a signal from a deterministic source involves maximizing the matched-filter statistic over a bank of templates defined on the signal's parameter space. For SGWBs, as well, the detection statistic can be the MLR, maximized further with respect to different SGWB models given by $\hat{\bm{\mathcal P}}$, perhaps parametrized by a smaller number of parameters than the number of components of $\hat{\bm{\mathcal P}}$. This is in contrast to the existing searches for anisotropic GW backgrounds. 
Past dirty-map-based searches precluded the presence of a signal by 
demonstrating that the map is consistent with a Gaussian distribution, up to statistical fluctuations allowed by the number of independent ``samples'' on the sky~\cite{S4Radiometer}. However, they did not provide a confidence level for the presence or absence of a broadband or spatially extended signal. A better approach is to solve the inverse problem in an orthogonal basis, namely, the pixel~\cite{Mitra} or spherical harmonic~\cite{ThraneEtal} basis.
This yields an estimate of the background, i.e., a ``clean'' (deconvolved) map and the corresponding noise-covariance matrix. However, this approach depends heavily on how well-posed the inverse problem is and how accurately it can be implemented numerically. Consequently, a detection statistic constructed on the deconvolved data can be affected by similar maladies. To work well, the inverse problem requires that the network of interferometers is sufficiently nondegenerate, which is not always the case. Indeed, the deconvolution procedure can enhance spatial noise correlations and, sometimes, even introduce artifacts, thereby adversely affecting parameter estimation and signal detection by such a procedure. 

As we prove here, the detection problem does not require a well-posed inverse problem
and exists even for a degenerate network. A detection statistic is best defined on the dirty map, as opposed to the clean map. As an added advantage, a dirty-map-based statistic is faster to compute, since it obviates the computational overhead required for obtaining the clean map. 
While it is possible to use an arbitrary sky model, such as the {\em one-dimensional} basis $\hat{\bm{\mathcal P}}$, 
and estimate the strength of the SGWB for that particular model, the MLR statistic in Eq.~(\ref{eq3.17}) provides a well-understood construct that can be maximized over a set of parameters for selecting the model that best fits the data.

To elaborate further on the way the new statistic works, let us consider the example of a directed pixel-space search, which is performed for only one source and assumes that the angular extent of the source is, at most, one pixel. In the standard radiometer search~\cite{Ballmer, Mitra}, the dirty map $S^k$ is computed for each pixel $k$ in the sky.
The (signal part of the) dirty map is generally peaked at the source pixel and has broad structures, including large negative patches, around it. One way of inferring the presence or absence of a source in this image requires deconvolving it.
However, as we show later, deconvolution of a relatively weak source can result in a clean map with significant errors, especially when the sky is divided into around 3000 pixels or more. (A network resolution of several square degrees requires a few thousand pixels across the sky.) Also, computing the noise-covariance matrix can be numerically challenging. The MLR is a good choice in this situation, since it combines all the pixel values to provide a single number for the detection statistic that is simple to use in drawing inferences on the presence or absence of a signal in the network data.

If a parametrized model of the background is available, one can construct the likelihood-ratio statistic from the dirty map and maximize it over the parameter space. The maximized likelihood-ratio statistic can also be used to perform a more advanced blind search, 
where no prior information is available about angular distribution of the power in the SGWB. For each basis component, one can assert that only that basis component is present in the signal and compute the statistic with the corresponding sky model. Thus, using the dirty maps of ``point estimate'' or SNR obtained by the existing radiometer search, our prescription takes one step forward and can provide a map of likelihood ratios, which is statistically a more robust and meaningful quantity, given a set of highly correlated observations.

The construction of the MLR statistic on a dirty map is simple. Equation (\ref{eq3.17}) shows that it is the scalar product of the observed map $S^k$ and a sky-model-dependent normalized ``template.'' The template is proportional to ${\bm{\mathcal B}} \cdot {\bm{\mathcal P}}$, which is the expected signal in the dirty map for a sky model $\hat{\bm{\mathcal P}}$.
The inverse of the noise-covariance matrix is the metric in the pixel space. The sky model can be defined in a straightforward way. For instance, to search for a point source localized to a single pixel, one would use a $\hat{\bm{\mathcal P}}$ with all but one component, namely, the component corresponding to that pixel, set to zero. Indeed, ${\bm{\mathcal B}} \cdot {\bm{\mathcal P}}$ is simply the point-spread function of the pixel with the nonzero component of $\hat{\bm{\mathcal P}}$. Also notice that the inverse of the noise-covariance matrix, being proportional to the beam matrix, cancels out algebraically in the expression for the MLR. Therefore, unlike for deconvolution, for MLR construction the computation of this matrix is not needed. Otherwise, the latter procedure would have been computationally similar to solving the inverse problem, avoiding which is one of the main motivations for this work.

To complete the discussion, we note that the construction of a MLR statistic is not limited to dirty maps and can be implemented for clean maps. A clean map can be expressed as
\begin{equation}
\widetilde{\bm{\mathcal P}} \ = \ {\bm{\mathcal P}} + {\bf n}_c \; ,
\end{equation}
where ${\bm{\mathcal P}}$ is the true sky map, and ${\bf n}_c$ is Gaussian noise with covariance ${\bf \Sigma}$, which is related to the dirty-map noise-covariance matrix through the relation ${\bf \Sigma} = ({\bm{\mathcal B}}^T {\bf{N}}^{-1} {\bm{\mathcal B}})^{-1}$. Therefore, following the same procedure as that for the dirty map, one can write the MLR statistic for a clean map as
\begin{equation}\label{eq:cleanMapStat}
\lambda_c  = \frac{\widetilde{\bm{\mathcal P}} \cdot {\bf \Sigma}^{-1} \cdot {\bm{\mathcal P}}} {\sqrt{\bm{\mathcal P} \cdot {\bf \Sigma}^{-1} \cdot \bm{\mathcal P} } } \; ,
\end{equation}
and, thereby, obtain model-based or blind likelihood-ratio maps.

\subsection{Performance of optimal detection statistic}
\label{subsec:detStatPerform}

We numerically study the performance of the optimal statistic and compare with the existing method. We use the LIGO 4km detectors located in Hanford (H1) and Livingston (L1). Unless otherwise stated, the noise PSDs of all detectors are taken to be their (smoothed) first-generation design sensitivities~\cite{Abbott2009rpp, Acernese2007}. The frequency band considered here spans $40-1024$Hz, with a bin size $\Delta f = 1$Hz. The source PSD is taken to be a constant, $H(f)=1.516\times 10^{-48}$/Hz. Note that the spectral index of the source PSD has a significant effect on the resolution of the network. Predictions from astrophysical and cosmological models suggest the nominal range of the spectral index to be between $-3$ and $1$. The higher the spectral index, the higher the resolution, and the more computationally expensive the directed search.
We take the spectral index to be zero here by setting $H(f)$ as a constant.

The directed search is performed by dividing the (simulated) strain data from all detectors into segments with a duration of $192$ sec. The noise is taken to be stationary. The sky is tessellated into $3072$ pixels by using the Hierarchical Equal Area isoLatitude Pixelization~\cite{healpix, healpix-paper}. The choice of the signal integration duration is taken to be a sidereal day, which leads to the azimuthal symmetry of the baseline sensitivities and sky resolutions. The justification for choosing the above parameter values can be found, e.g., in \citet{Mitra}.

\begin{figure}[h!]
\begin{center}
\subfigure[~Injected map]
{\includegraphics[width=0.238\textwidth]{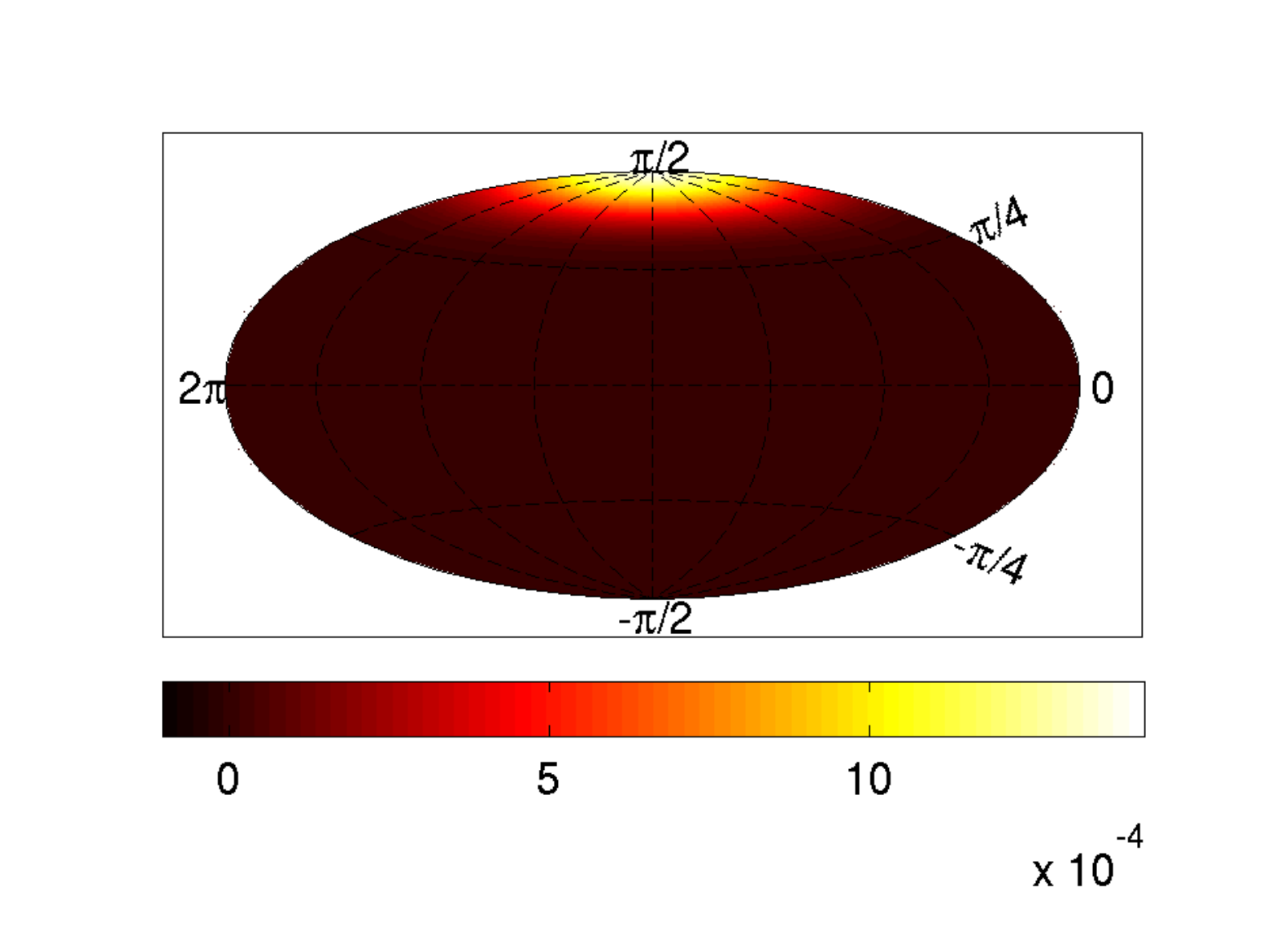}}
\subfigure[~Dirty map]
{\includegraphics[width=0.238\textwidth]{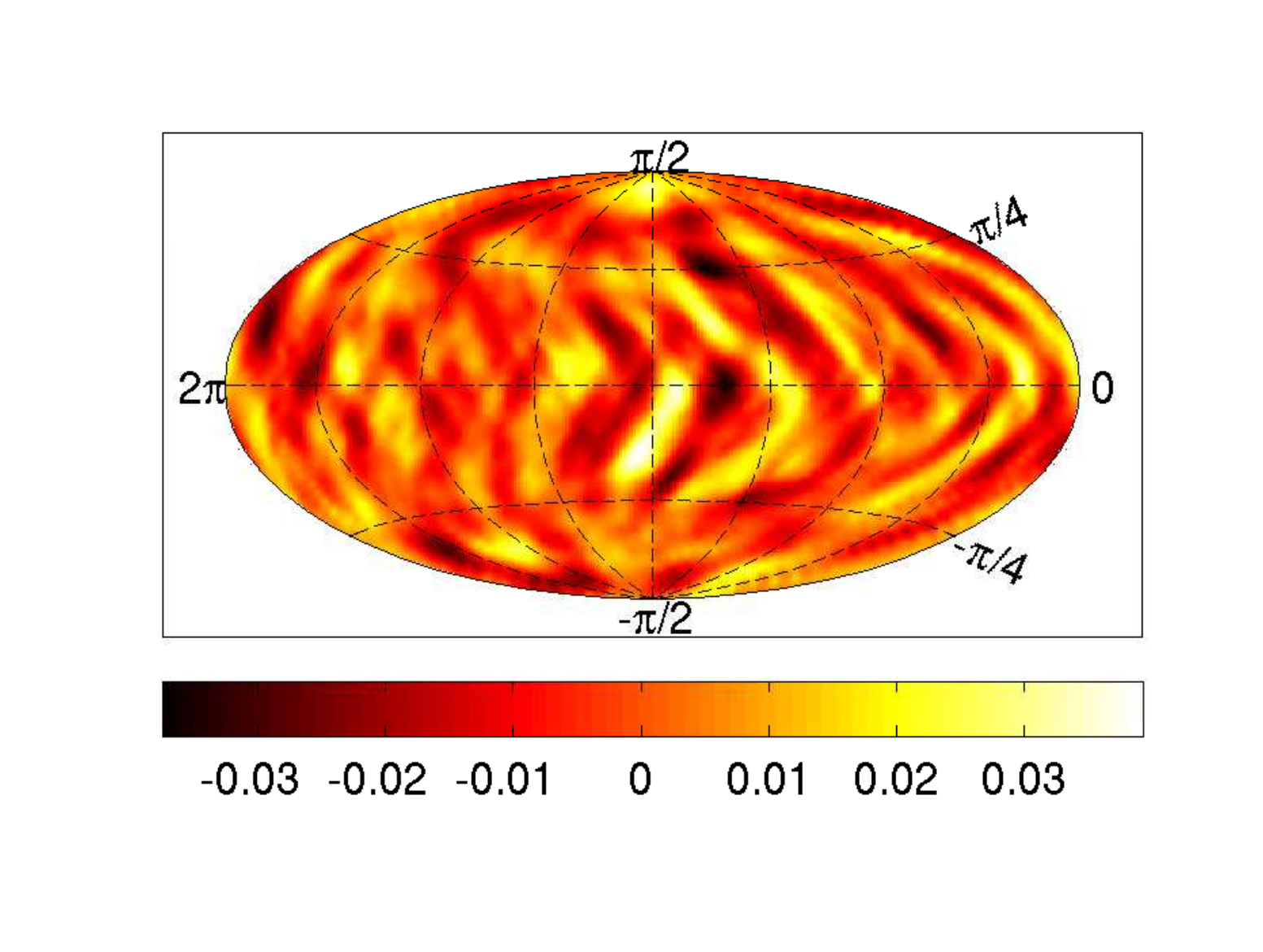}}
\subfigure[~Clean map]
{\includegraphics[width=0.238\textwidth]{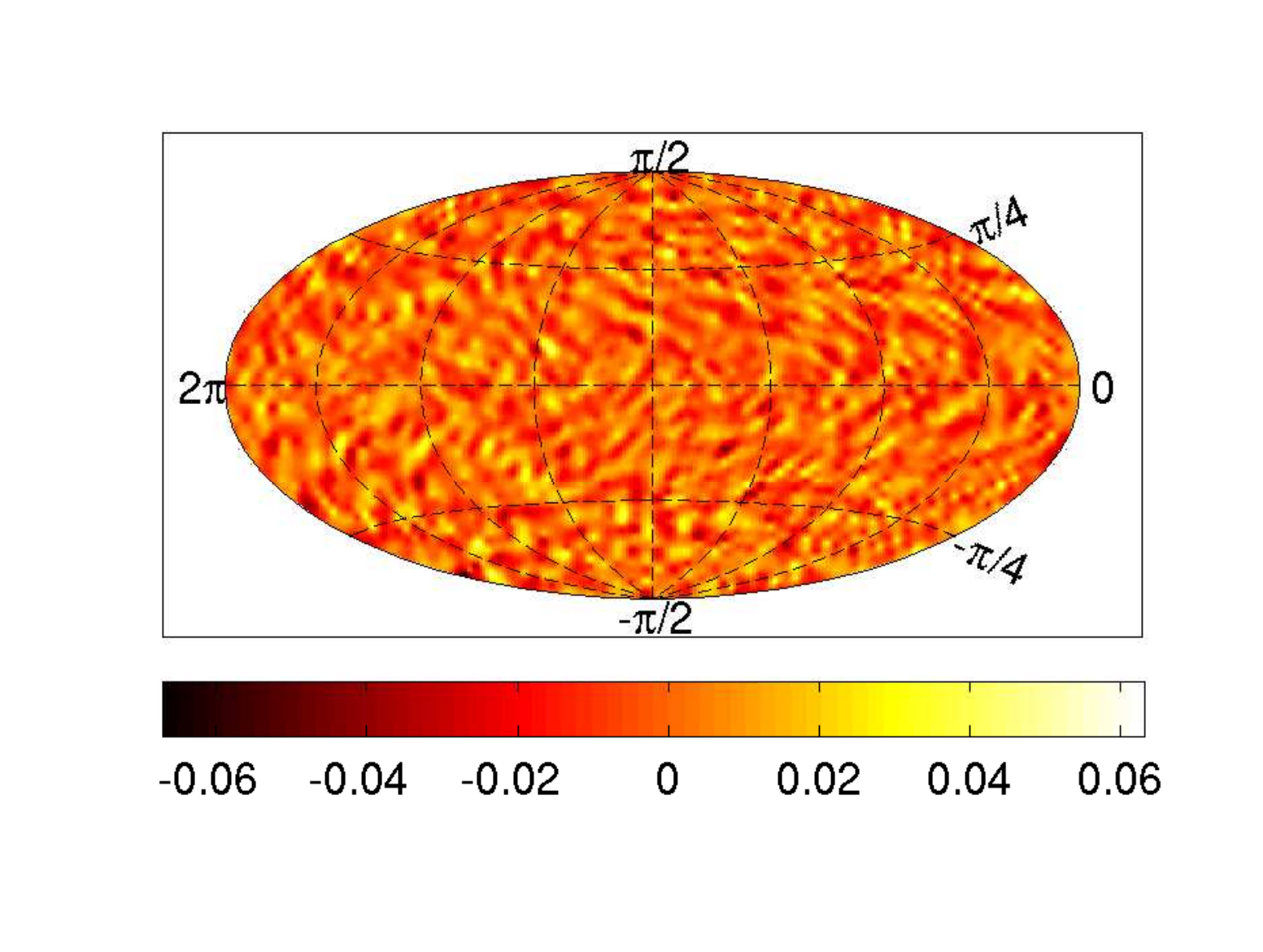}}
\caption{\label{fig:statmapCAP} The ${\mathcal P}^k$ sky map of a weak polar-cap GWB source is shown in (a). The dirty (b) and clean (c) maps for this source were constructed using the radiometer algorithm for the LIGO H1L1 baseline. The last two maps for this weak source are visually very similar to those for the noise-only case (which is not shown here).}
\end{center}
\end{figure}

\begin{figure}[h!]
\begin{center}
\subfigure[~Noise]
{\includegraphics[width=0.45\textwidth]{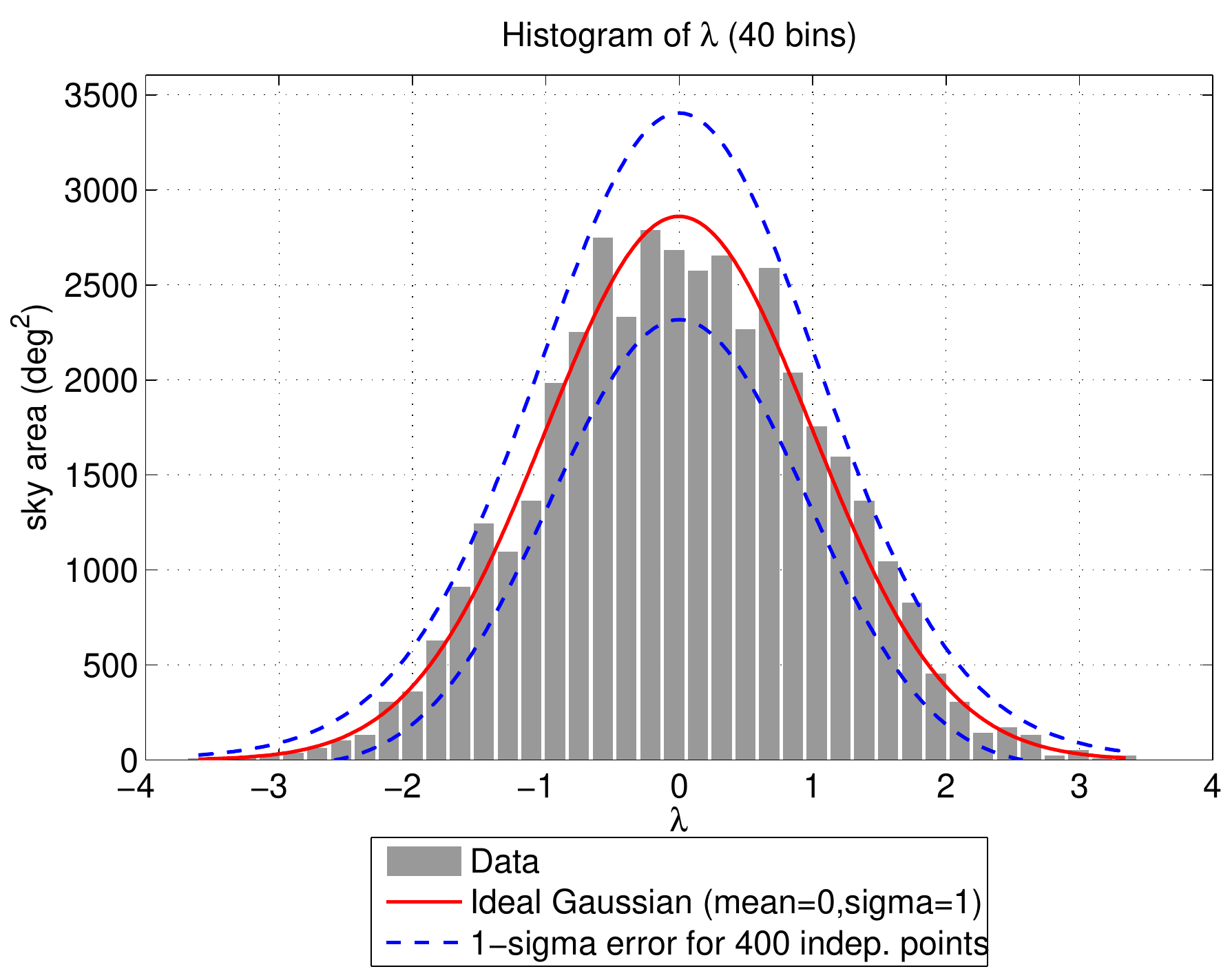}}
\subfigure[~CAP]
{\includegraphics[width=0.45\textwidth]{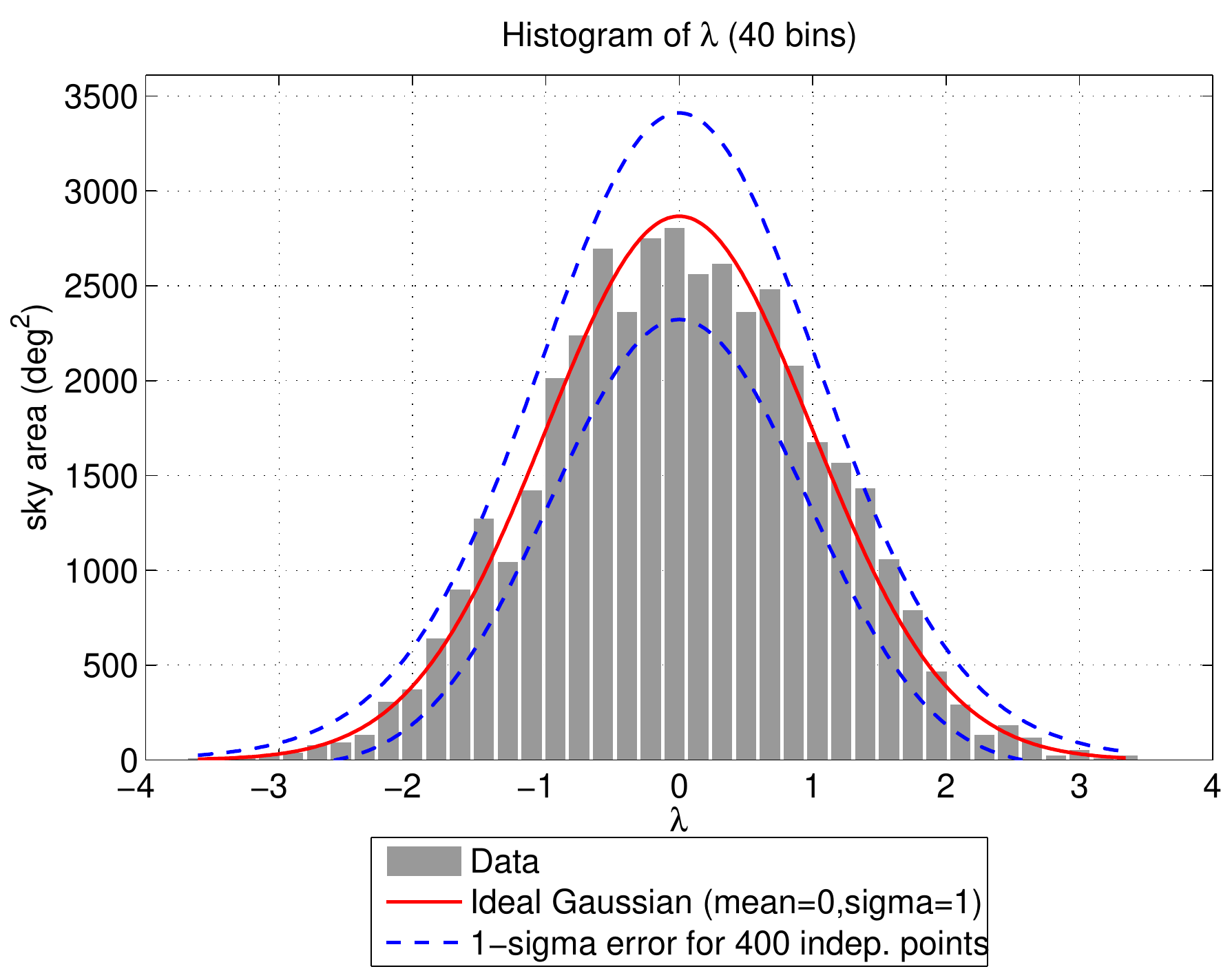}}
\caption{\label{fig:histRadio} Histograms of the dirty maps for the noise-only data set (top) and weak polar-cap signal (bottom), which is depicted in Fig.~\ref{fig:statmapCAP}, are shown here. These two histograms are consistent with that of zero-mean Gaussian data (solid curved line), up to $1\sigma$ errors: Following \citet{S4Radiometer}, the $1\sigma$ error boundaries for $400$ degrees of freedom have also been overlaid for a consistency check. 
}
\end{center}
\end{figure}

We first construct simulated data sets of two kinds, one with only noise and the other with a weak signal from a ``polar-cap'' source added to that noise. The sky map of the latter case is shown in the first plot in Fig.~\ref{fig:statmapCAP}. We make dirty maps for these two cases using single-pixel source templates for each of the 3072 pixels. These maps are essentially the maps of SNR for the directed search, as can be seen by substituting
\begin{equation}
\hat{\mathcal P}^{k} \ = \ \left\{ \begin{array}{cl}1 & \text{for target pixel}\,, \\ 0 & \text{for remaining pixels}\,,\end{array}\right.
\end{equation}
into Eq.~(\ref{eq3.17})~\cite{Mitra}. The dirty maps for both cases look very similar, and only one of them, namely, the one for the polar-cap source, is shown in the second plot of Fig.~\ref{fig:statmapCAP}. Not surprisingly, it is also similar to the dirty map presented in \citet{S4Radiometer} for real data.
Indeed, the MLR values over the dirty-map pixels appear to follow a normal distribution, as shown in Fig.~\ref{fig:histRadio}. This is consistent with the distribution presented in \citet{S4Radiometer}. Following that reference, we also plot the $1\sigma$ error envelope around the Gaussian fit for $400$ degrees of freedom and observe that the tops of every bar in the histogram lie within that envelope.
The important point to note here is that, for both noise-only and weak-signal (polar-cap) data sets the distributions in Fig.~\ref{fig:histRadio} are very similar and consistent with a normal distribution. 

\begin{figure}[h!]
\begin{center}
\subfigure[~Noise]
{\includegraphics[width=0.45\textwidth]{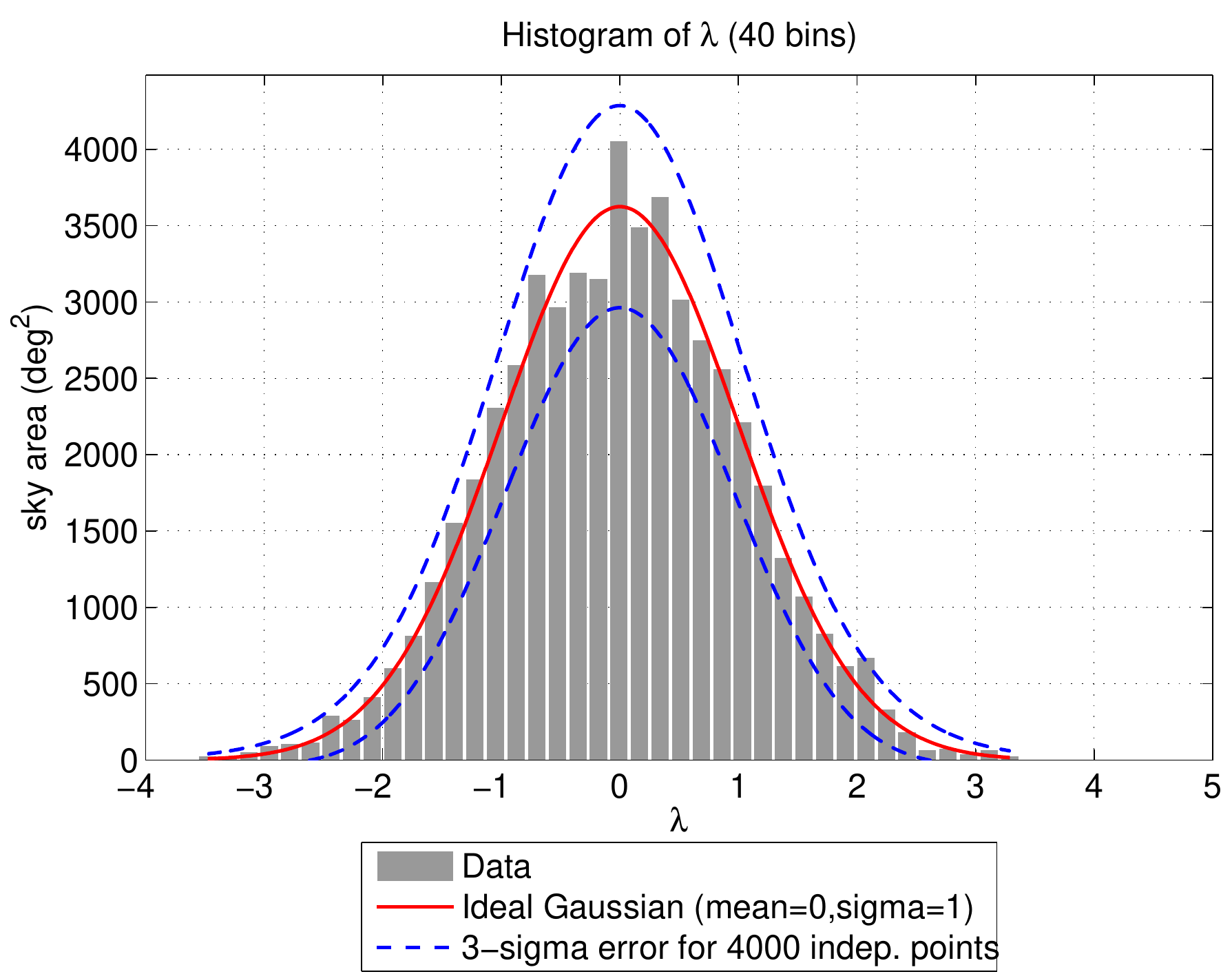}}
\subfigure[~CAP]
{\includegraphics[width=0.45\textwidth]{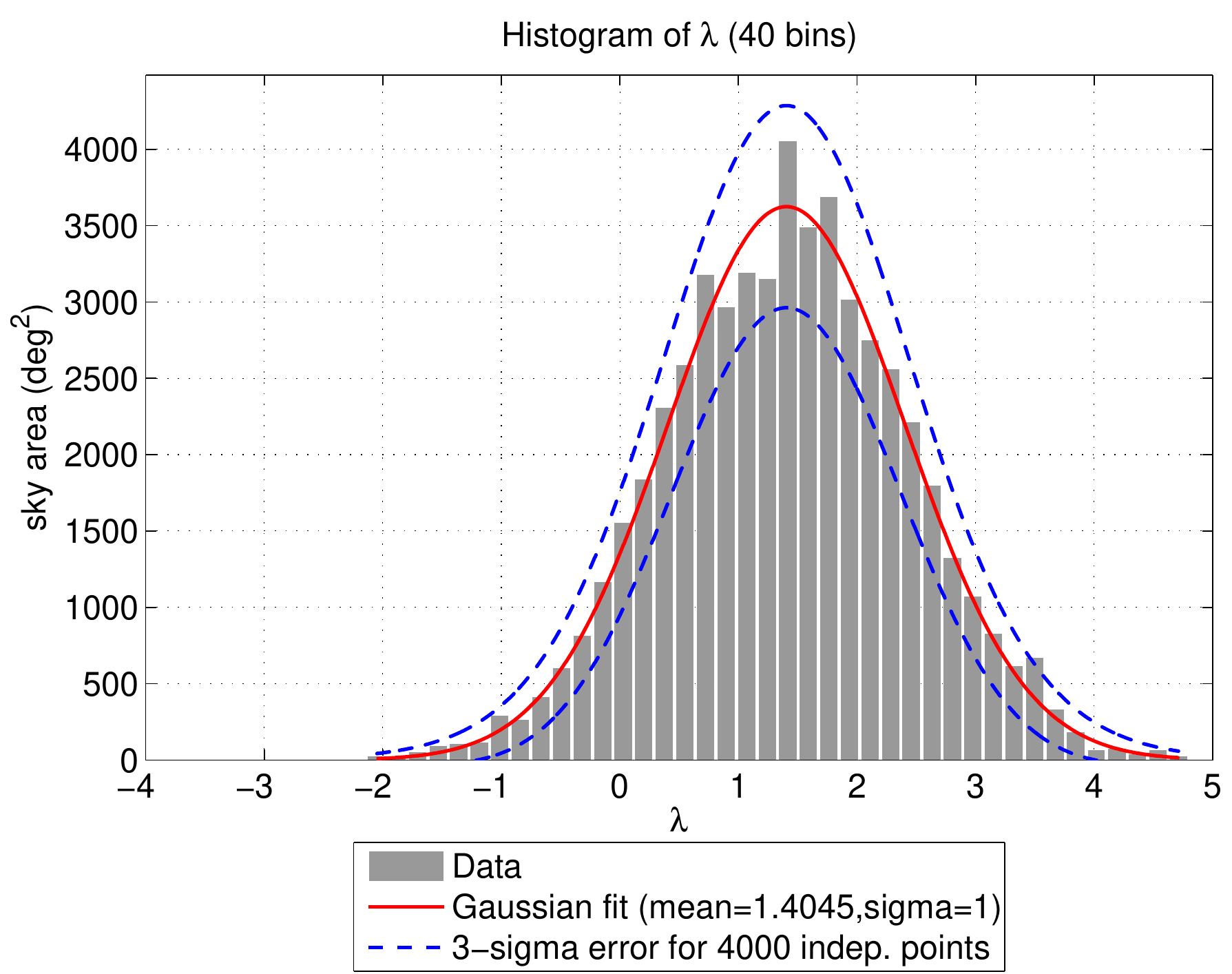}}
\caption{\label{fig:histMLR} Monte Carlo simulations with 4000 noise realizations were performed to study the performance of the MLR statistic. The distribution of the MLR statistic obtained for dirty maps generated from noise-only (top) and weak polar-cap injection (bottom) are shown. Clearly, the MLR statistic detects the signal at $\sim 1.4\sigma$ level.}
\end{center}
\end{figure}

If we now pretend that we know the broad shape of the GWB sky and use the $\hat{\mathcal P}^{k}$ of the polar-cap signal as our template for computing the MLR statistic, 
we find that the above two cases can be distinguished better: In the noise-only case, the MLR statistic equals $-0.023$, while with the weak polar-cap signal it is $1.400$, which is significantly larger than the former, as explained below. 
To corroborate this claim, we computed the MLRs on an ensemble of $4000$ realizations of noise, with and without the weak polar-cap signal. In Fig.~\ref{fig:histMLR}, we show the distribution of the MLR statistic ($\lambda$) for noise-only (top) and weak-signal (bottom) cases. Clearly, the noise-only $\lambda$ values are normally distributed with a zero mean, and the weak-signal $\lambda$ values are normally distributed with a mean of $\sim 1.4\sigma$, where $\sigma \approx 1$.
%
This experiment confirms that, given our assumptions on the detector noise and the signal, the MLR statistic can considerably enhance the detectability of a weak diffuse stochastic background, if a reasonable model of the background is available. This is true even when the distribution of the dirty-map pixel values for that source is close to zero-mean Gaussian.

\section {Performance of Multibaseline Radiometers}
\label{sec:performance}

In this section we define a set of figures of merit to compare the performance of a network of baselines with that of its individual baselines for a directed search of a SGWB.

A single figure of merit may not suffice in capturing all the attributes of a baseline or a network of detectors benefiting a SGWB search. A certain baseline or network configuration can have good sensitivity if the detectors are optimally orientated, but have poor resolution if they are proximally located. Optimally oriented detectors may be very sensitive to certain anisotropy modes, but insensitive to others, making the estimation problem highly degenerate. On the other hand, a network of detectors that are oriented differently may have moderate, yet uniform, sensitivity to all spherical harmonic modes of a SGWB, thereby mitigating the ill-posedness of the estimation problem. Such a network, however, will perform worse than one where all the detectors are aligned similarly
in a low-frequency, all-sky isotropic search. Therefore, the relevance of a figure of merit is determined by the kind of search one is undertaking. Here, we propose a set of figures of merit that are relevant to current searches of anisotropic stochastic background and that are special cases of the general ML framework presented in~\citet{ThraneEtal}.

For the numerical simulation studies below, we use the same detector characteristics as mentioned in Section~\ref{subsec:detStatPerform}, but we now include the Virgo detector (V1) in Cascina, Italy to construct a three-baseline network. The baselines and their network are named by concatenating the symbols for the participating detectors; e.g., the Hanford-Livingston baseline is termed as H1L1, and the network of the above three detectors is termed as H1L1V1.

\subsection{Sensitivity}

The first figure of merit is the ``sensitivity'' of a network and is motivated by a similar quantity defined in \citet{Cella} for the all-sky isotropic search. In practice, a greater sensitivity implies a better confidence level, at which detection can be made or upper limits can be inferred.

We define the single-baseline sensitivity for a directed search as the expectation value of the MLR in Eq.~\eqref{eq3.21} for a SGWB source with ${\mathcal P}$ set to unity,
\begin{eqnarray}\label{eq4.1}
{\text{Sensitivity}}& = & \sqrt{ ({\bm{\mathcal B}}\cdot\hat{\bm{\mathcal P}})^{k} ({\bf N}^{-1})_{k k^{\prime}} ({\bm{\mathcal B}}\cdot \hat{\bm{\mathcal P}})^{k^{\prime}} } \; ,  \nonumber \\
& = & \sqrt{ \hat{\mathcal P}^{k} {\mathcal{B}}_{k k^{\prime}} \hat{\mathcal P}^{k^{\prime}} } \; .
\end{eqnarray}
The sensitivity can be expressed in the spherical harmonic basis as follows:
\begin{equation}\label{eq4.2}
{\text{Sensitivity}} = \sqrt{{\mathcal P}_{l m} {\mathcal{B}}^{l m \,l^{\prime}m^{\prime}} {\mathcal P}_{l^{\prime}m^{\prime}} } \; ,
\end{equation}
where
\begin{equation}\label{eq4.3}
{\mathcal P}_{l m} = \int d\hat{\Omega}\; \hat{\mathcal P}(\hat{\Omega})\, Y_{l m}^{*}(\hat{\Omega}) \; ,
\end{equation}
\begin{equation}\label{eq4.4}
{\mathcal{B}}_{l m \,l^{\prime} m^{\prime}} = \int \int d\hat{\Omega} \,d{\hat{\Omega}}^{\prime}\; Y_{l m}(\hat{\Omega}) \,\mathcal{B}(\hat{\Omega}, {\hat{\Omega}}^{\prime})\, Y_{l^{\prime} m^{\prime}}({\hat{\Omega}}^{\prime}) \; .
\end{equation}

Owing to the statistical independence of the baselines, the multibaseline sensitivity squared is the sum of squares of the individual baseline sensitivities, as was also noted for the isotropic-background baseline sensitivities in \citet{Cella}:
\begin{equation}\label{eq4.5}
{\text{Sensitivity}}^{2}_{\mathcal{N}} =  \sum_{\mathcal{I}}{\text{Sensitivity}}^{2}_{\mathcal{I}} \; .
\end{equation}
For an unpolarized background from a single pixel, say, labeled $k$, and with $\hat{{\mathcal P}}^{r} = \delta^{r(k)}$, the sensitivity expression simplifies to
\begin{equation}\label{eq4.6}
{\text{Sensitivity}_{(k)}}=\sqrt{ {\mathcal B}^{q(k)} ({\bf N}^{-1})_{q r} {\mathcal B}^{r(k)} } = \sqrt{\mathcal{B}^{(k)(k)}  } \; .
\end{equation}
Unless otherwise mentioned, there is no sum over the repeated parenthetic indices in this paper.\\

\begin{figure}[h!]
\begin{center}
\includegraphics[width=0.45\textwidth]{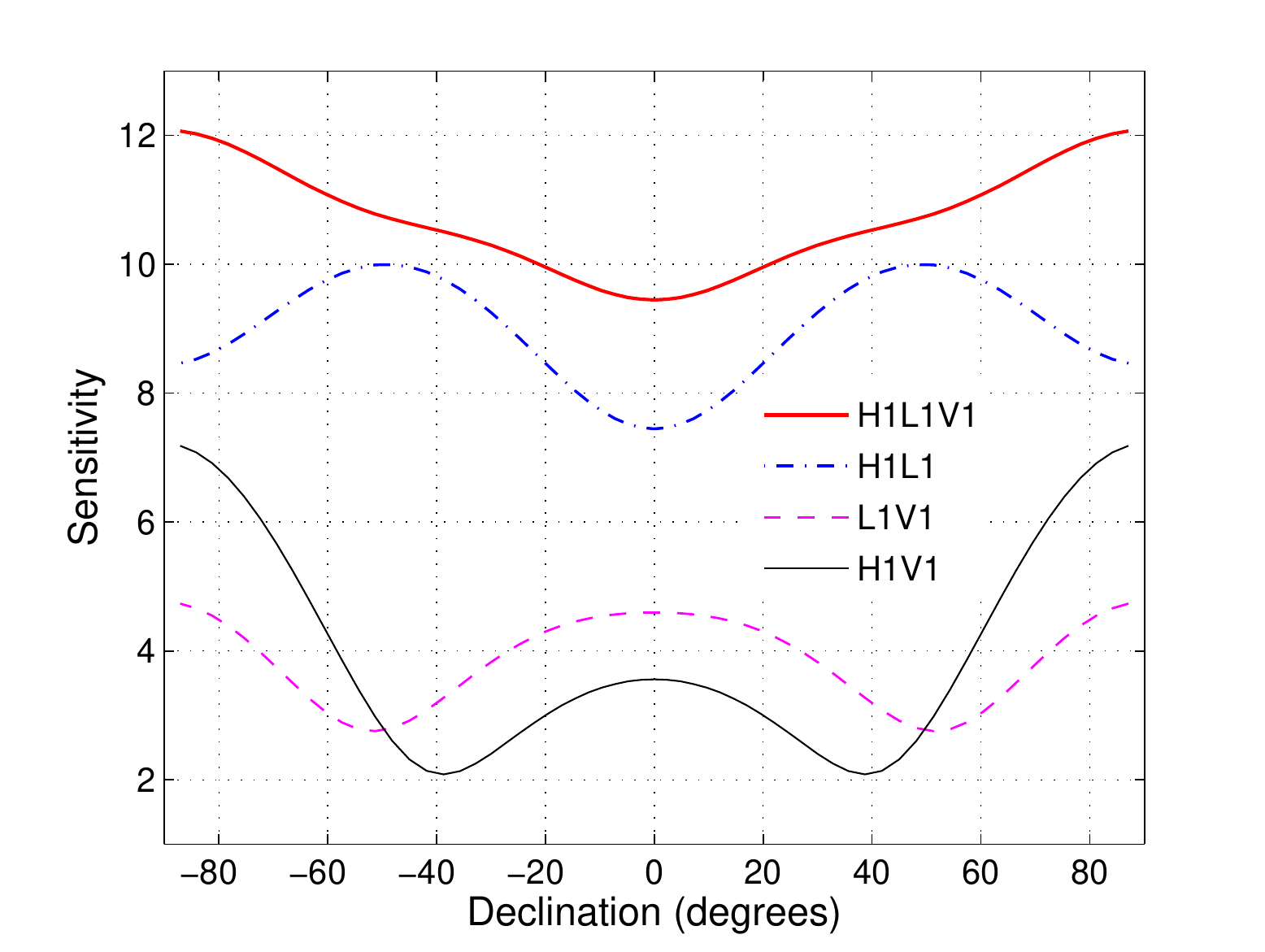}
\includegraphics[width=0.45\textwidth]{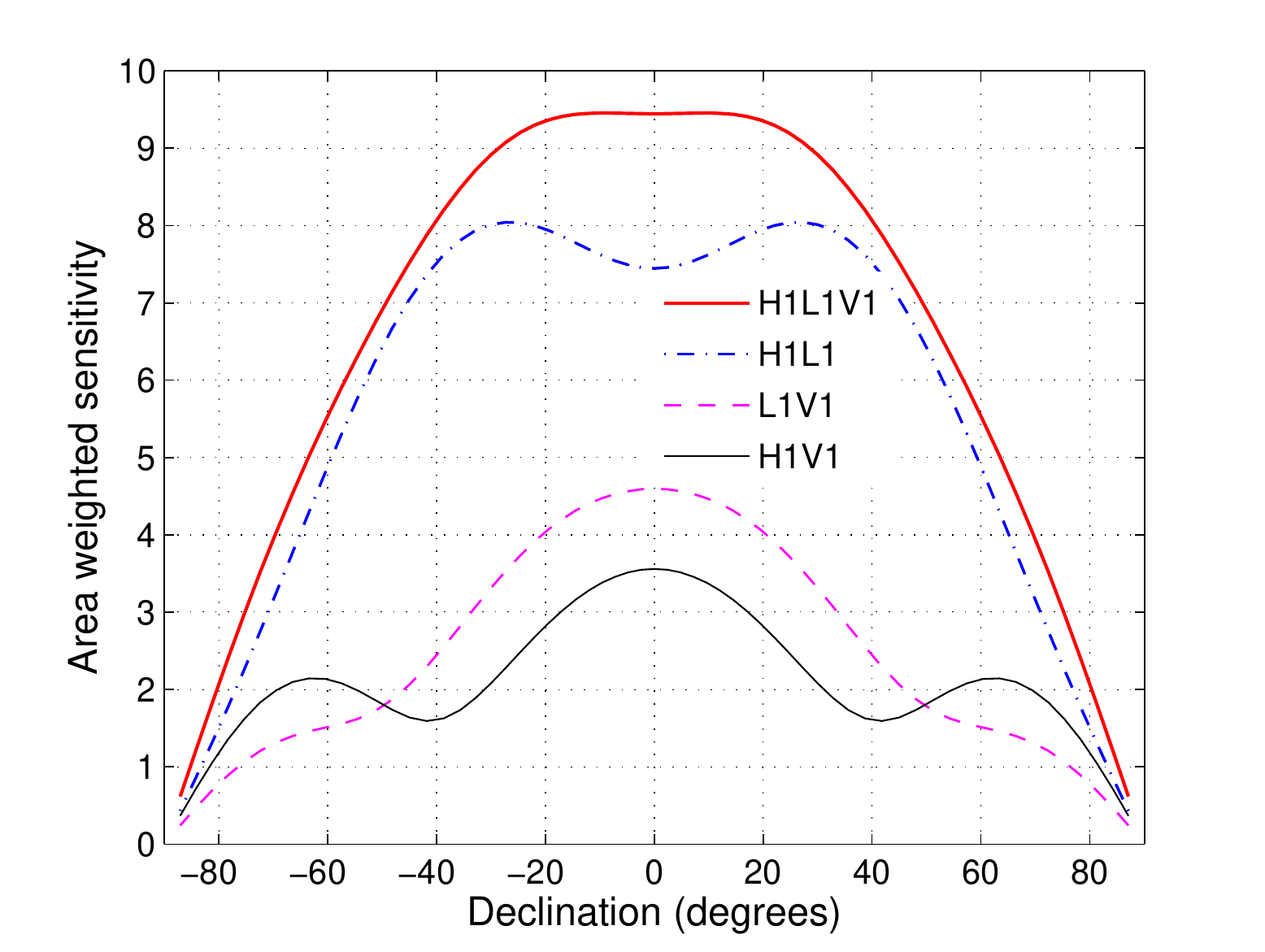}
\caption{\label{fig:sensitivity}The sensitivities (top) and their area-weighted counterparts (bottom) of three different baselines and their network are plotted as functions of the declination of a single-pixel SGWB source. The source PSD ($H(f)=1.516\times 10^{-48}$strain/Hz) is chosen such that it has maximum SNR$=10$ in the H1L1 baseline. (Note that the source parameter ${\mathcal P}$ is set to unity for these plots.) The signal band considered here is $40$-$1024$~Hz.}
\end{center}
\end{figure}

In the top panel of Fig.~\ref{fig:sensitivity}, we compare the sensitivities of the baselines and the whole network as a function of declination. (As noted above, the sensitivities are azimuthally symmetric.) For a fair comparison, we also replot them after weighting them with the cosine of the latitude, in effect, to assign equal weight to every pixel on the sky. It is clear that the H1L1 baseline has much better sensitivity due to the 
similar orientations of the two detectors. Still, inclusion of Virgo, which is oriented quite differently relative to H1 and L1, improves the sensitivities of the network by $\sim 10$\% (which corresponds to an increase in the observational volume by $\sim 30$\%), especially in the regions where the H1L1 baseline does not perform well. However, this network improvement is highly superseded by all other performance improvements indicated by corresponding figures of merit introduced in this section.

\begin{figure}[h!]
\begin{center}
\includegraphics[width=0.45\textwidth]{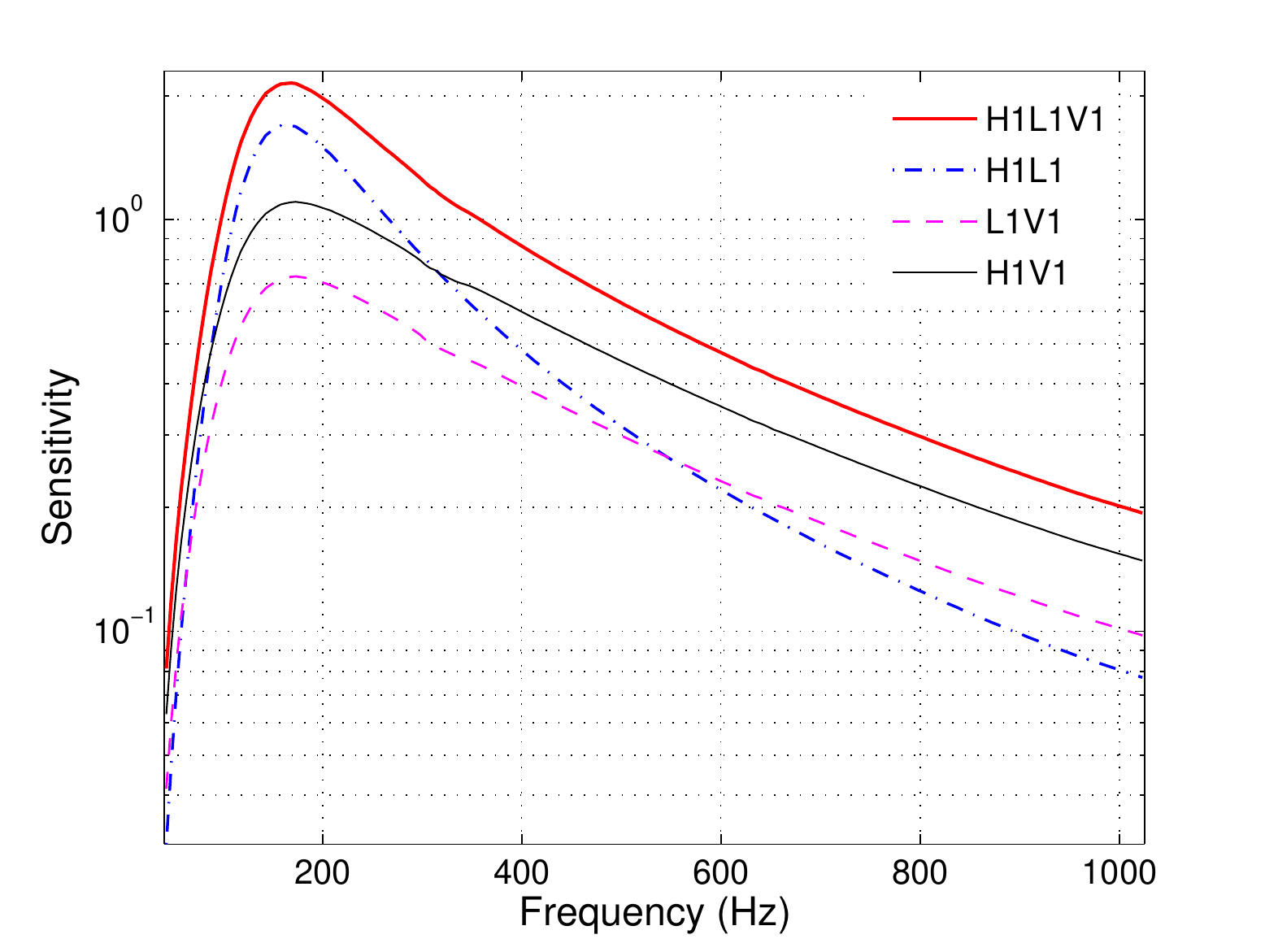}
\includegraphics[width=0.45\textwidth]{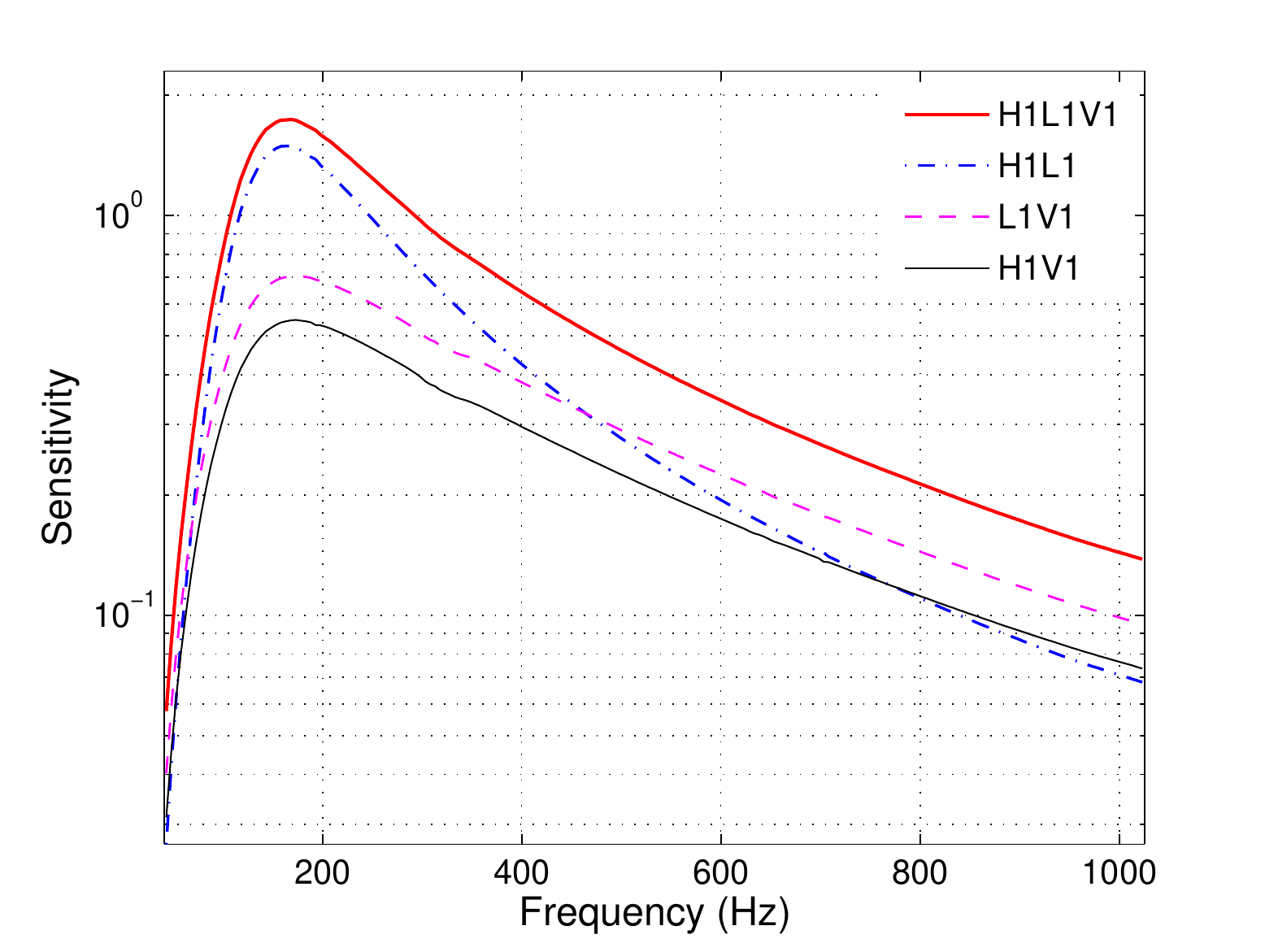}
\caption{\label{fig:narrowBandSens}The sensitivities of three single baselines and their multibaseline network plotted as functions of the central frequency of the source band. The source is chosen to have a constant $H(f)=1.516\times 10^{-48}$strain/Hz and a band width of 5Hz. The top panel represents the sensitivities at the celestial poles, and the bottom panel represents those at the celestial equator.}
\end{center}
\end{figure}

In Fig.~\ref{fig:narrowBandSens} we plot the narrowband ($5$Hz) sensitivities at two locations, namely, the Celestial North Pole (top) and the equator (bottom).  Performance improvement of a network for a narrow band search at high frequencies is better than the (frequency integrated) broadband search.

\subsection{Sky coverage}

In a directional search, the main advantage of a network lies in the fact that it vastly improves sky coverage, which, in turn, leads to better parameter estimation, including localization and sky-map reconstruction considered later in this section. In this subsection, we 
illustrate the advantage of using a network of detectors, as compared to using its individual baselines, to this end.

\begin{figure}[h!]
\begin{center}
\subfigure[~H1L1]
{\includegraphics[width=0.238\textwidth]{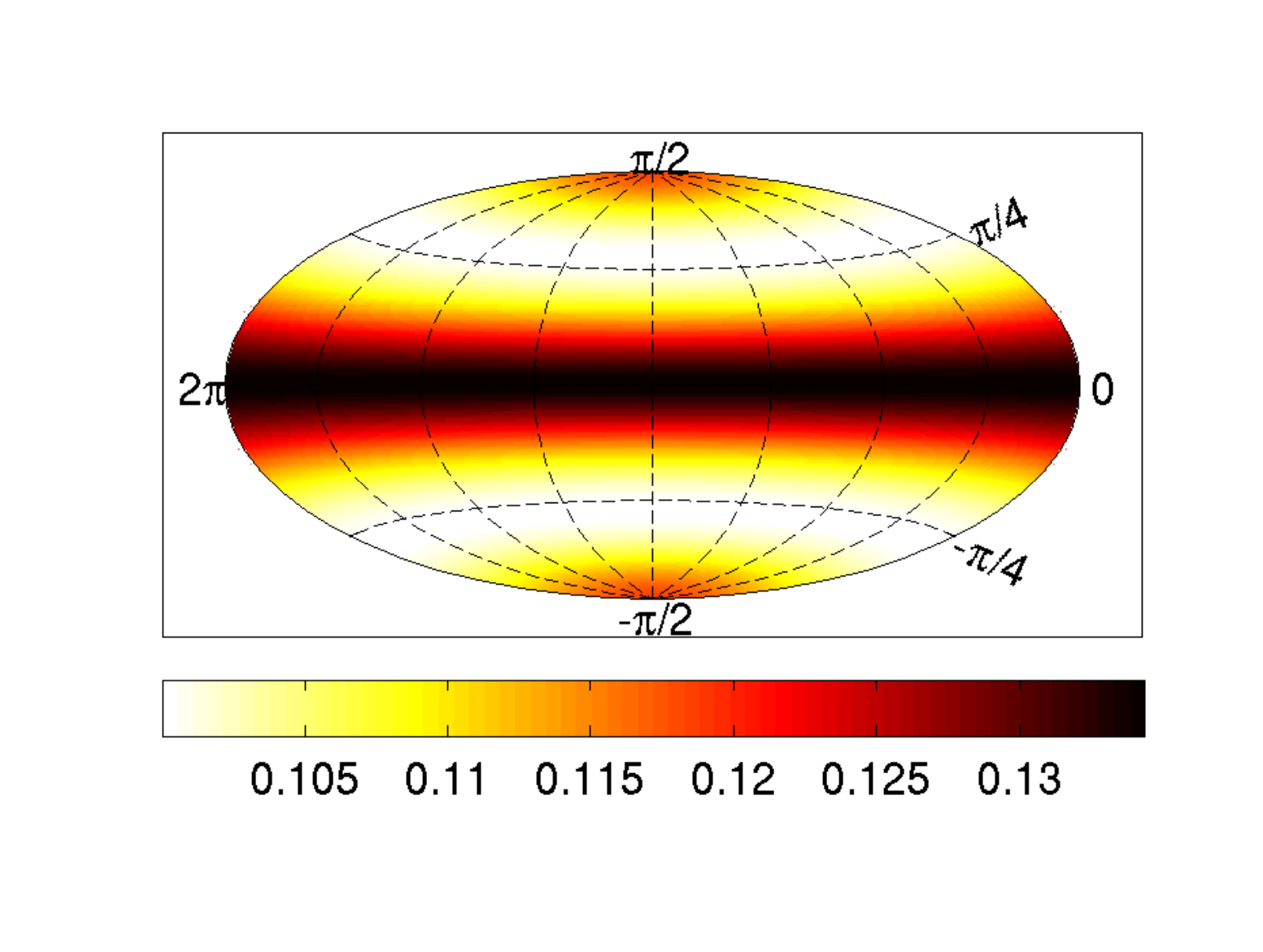}}
\subfigure[~L1V1]
{\includegraphics[width=0.238\textwidth]{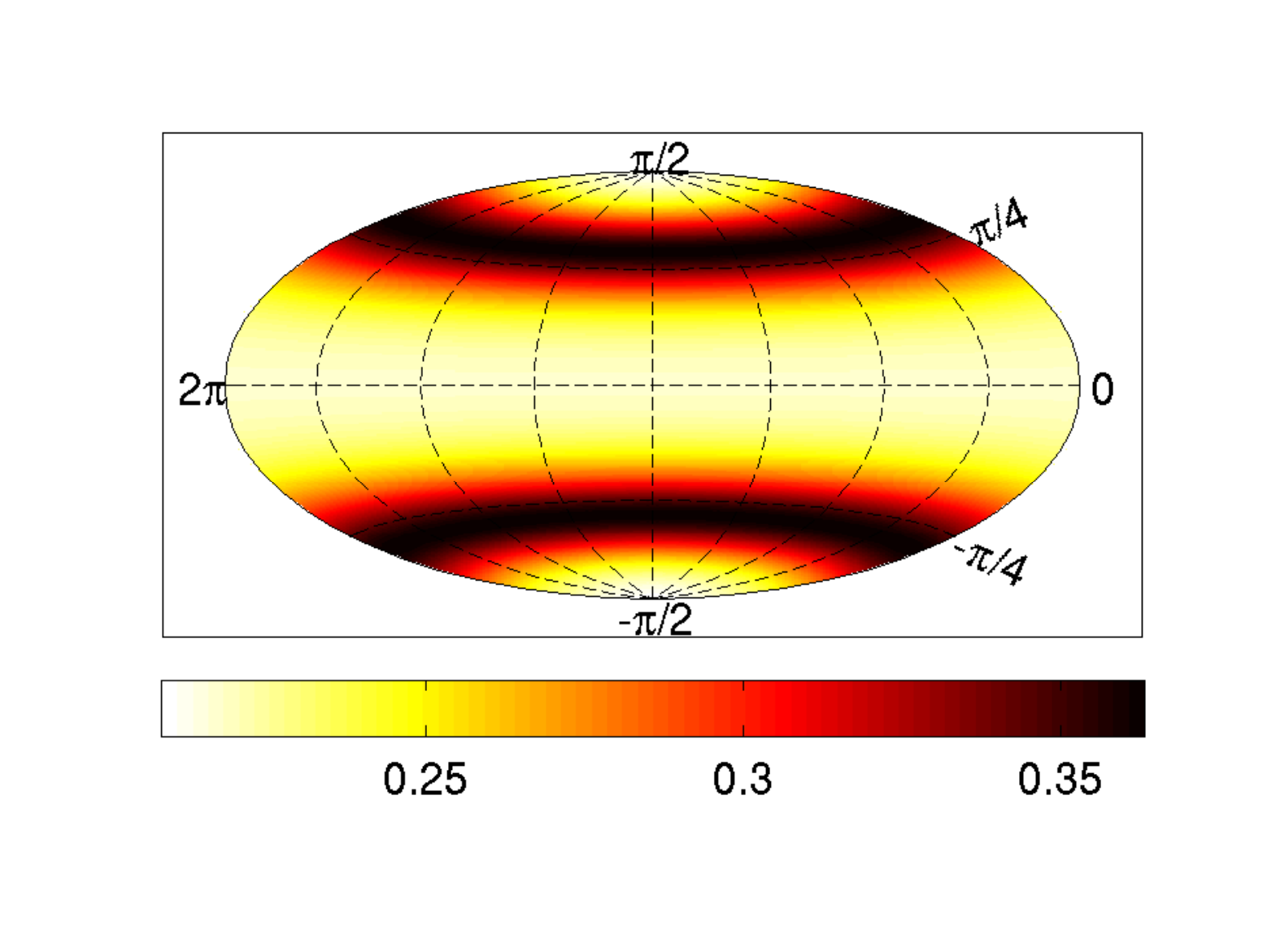}}
\subfigure[~H1V1]
{\includegraphics[width=0.238\textwidth]{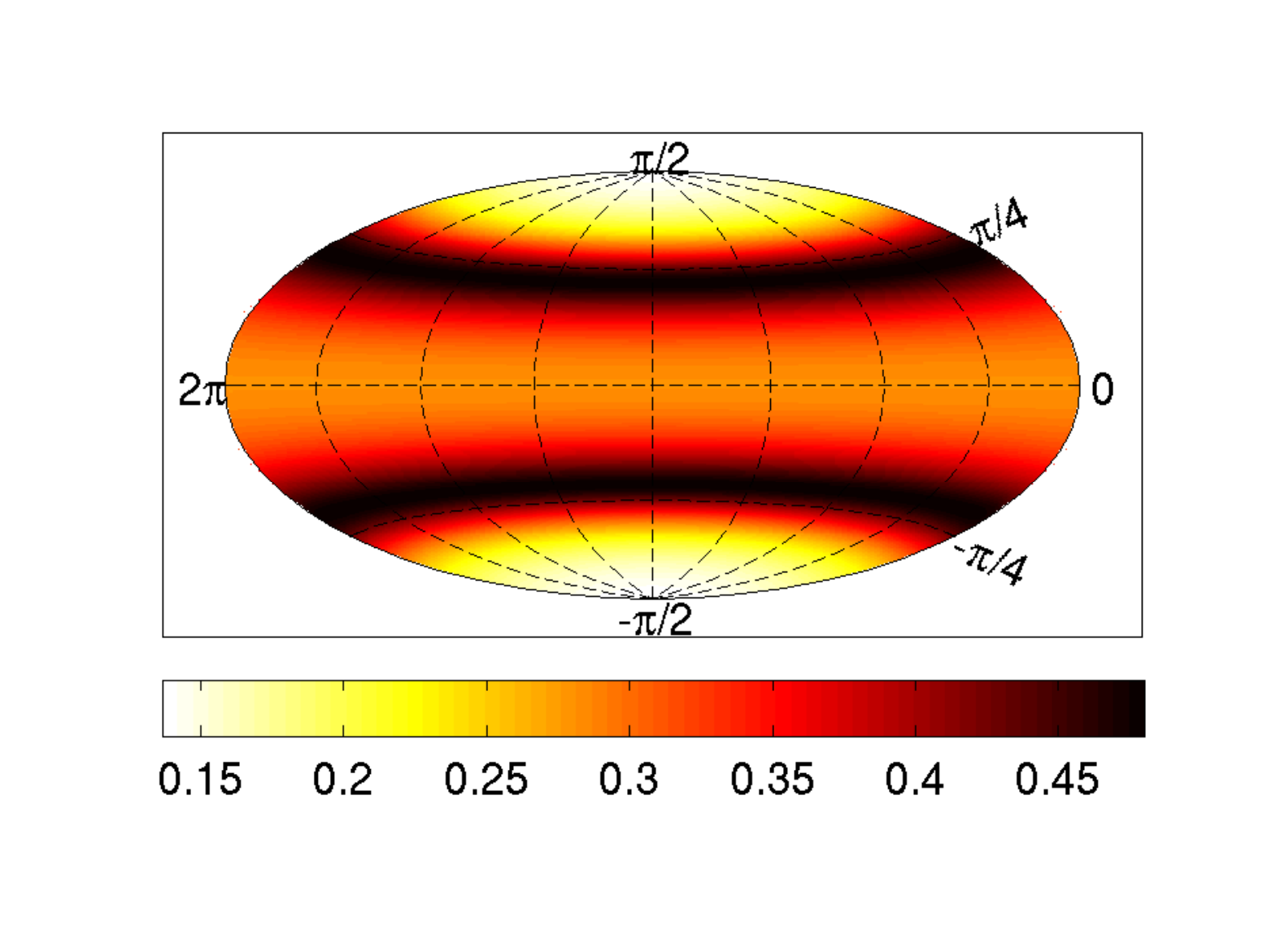}}
\subfigure[~H1L1V1]
{\includegraphics[width=0.238\textwidth]{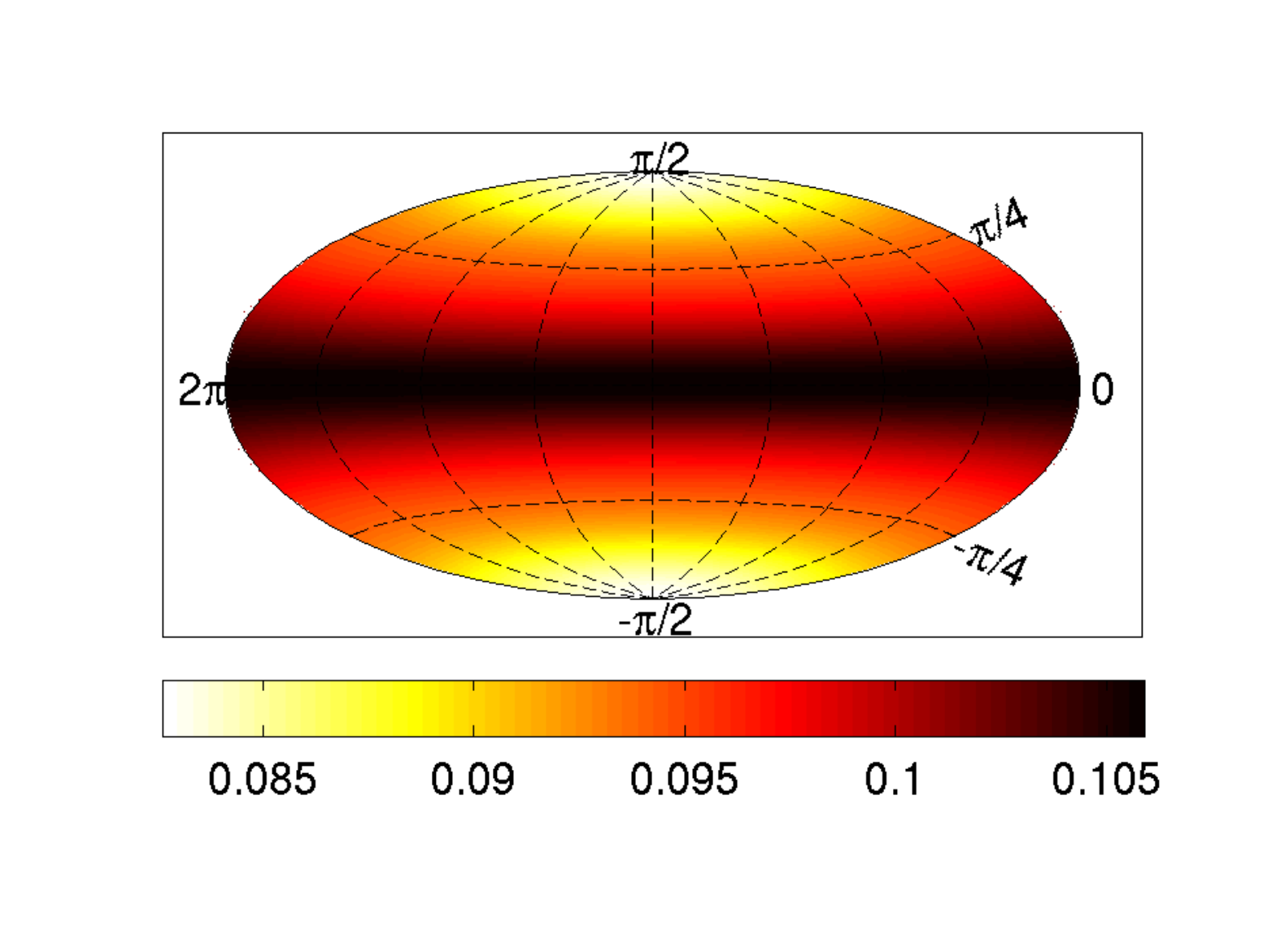}}
\caption{\label{fig:compSD}The standard deviation of dirty maps measured by the three LIGO-Virgo baselines and the full network for a constant $H(f)$
are plotted here. The aim of this figure is to show how individual baselines complement each other, thereby making the ``effective deviation'' (i.e., the square root of the harmonic mean of variances of the individual baselines) of the combined map observed by the network nearly uniform. Note that the color scale in the network plot has lesser spread than the individual baselines. 
The absolute scale of the maps depends on the normalization of the filter, and only the relative scale is important here. The azimuthal symmetry is present because we are considering a whole sidereal day's observation, with stationary noise.}
\end{center}
\end{figure}

In general, one radiometer baseline cannot sample the whole sky uniformly; the measurement errors in some parts of the sky are much worse than those in the other parts. Introduction of new baselines with different orientations improves filling in these ``holes'' by scanning the sky with different antenna-pattern functions. In the first three plots in Fig.~\ref{fig:compSD}, we show the standard deviation in measuring the dirty map by the three individual baselines, namely, H1L1, H1V1, L1V1, and their network H1L1V1~\footnote{The absolute scales of the plots are not important since the emphasis here is on relative performance of the different baselines and their network.}. The azimuthal symmetry mentioned before is explicitly observed here. The H1L1 baseline has the least deviation at most declinations, due to the optimal orientations of the H1 and L1 detectors. Again, H1V1 and L1V1 baselines have low deviation in the regions where H1L1 does not perform well. Since the dirty maps from different baselines are operationally combined with an inverse-noise-variance weight, the harmonic mean of the variances provides the effective variance of the combined dirty map. The last plot in Fig.~\ref{fig:compSD} shows the effective deviation for a network of detectors. Clearly, the deviation now has smaller spread and also, by construction (harmonic mean), the deviations are smaller than those for the individual baselines.

\begin{figure*}[h!t]
\begin{center}
\includegraphics[width=0.30\textwidth]{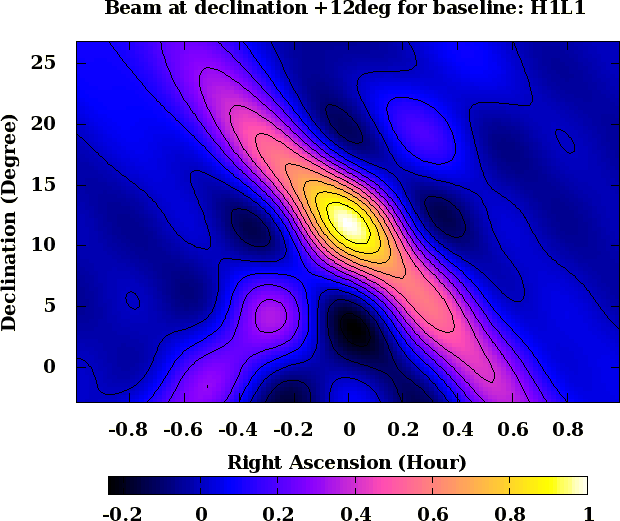}\quad
\includegraphics[width=0.30\textwidth]{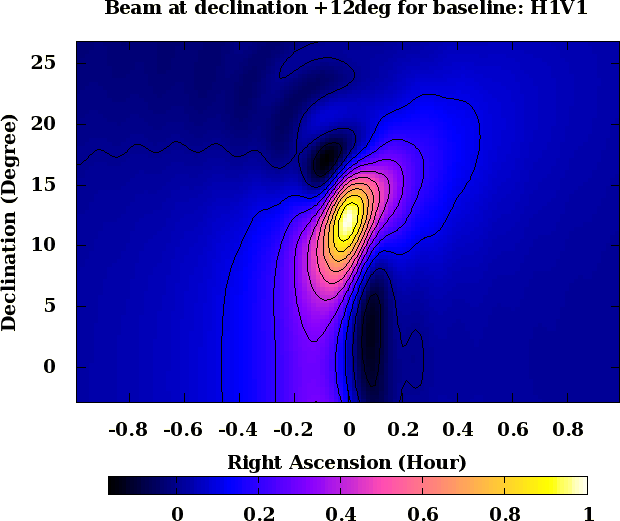}\quad
\includegraphics[width=0.30\textwidth]{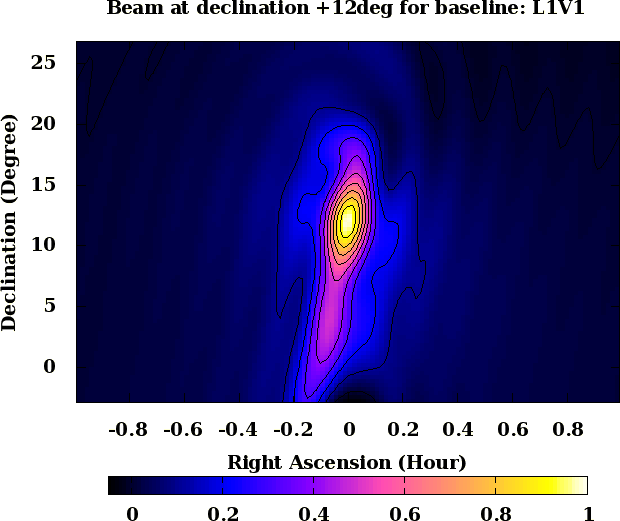}
\caption{\label{fig:compBeams}The beam functions for the three LIGO-Virgo baselines for $H(f) = {\rm constant}$  are shown here. This figure illustrates that different baselines also complement each other in terms of angular resolution along different tangential sky directions.}
\end{center}
\end{figure*}

Most importantly, a network also complements the single-baseline observations in terms of angular resolution. The beam functions for each radiometer baseline are highly asymmetric, which means that a given position on the sky is probed with quite different angular resolutions in the tangential directions. To illustrate this aspect, the typical beam functions for the three LIGO-Virgo baselines in the direction of the Virgo cluster are shown in Fig.~\ref{fig:compBeams}. If we consider the beam for the H1L1 baseline, the sensitive part of the beam is similar to a highly eccentric ellipse, suggesting that the angular resolution along the minor axis is much finer than that along the major axis. The beams for the baselines in a network involving the Virgo detector provide better resolution due to the longer baselines: The beams are finer along the major axis of the H1L1 beam, thus complementing the H1L1 observation, which is a major motivation for using a network.
This, in turn, improves the 
condition number of the Fisher information matrix, thereby, reducing the numerical errors in the anisotropy estimation problem at ``high'' resolution, i.e., near or beyond the diffraction limited resolution, and significantly improves source localization accuracy.

\begin{figure}[h!]
\begin{center}
\includegraphics[width=0.45\textwidth]{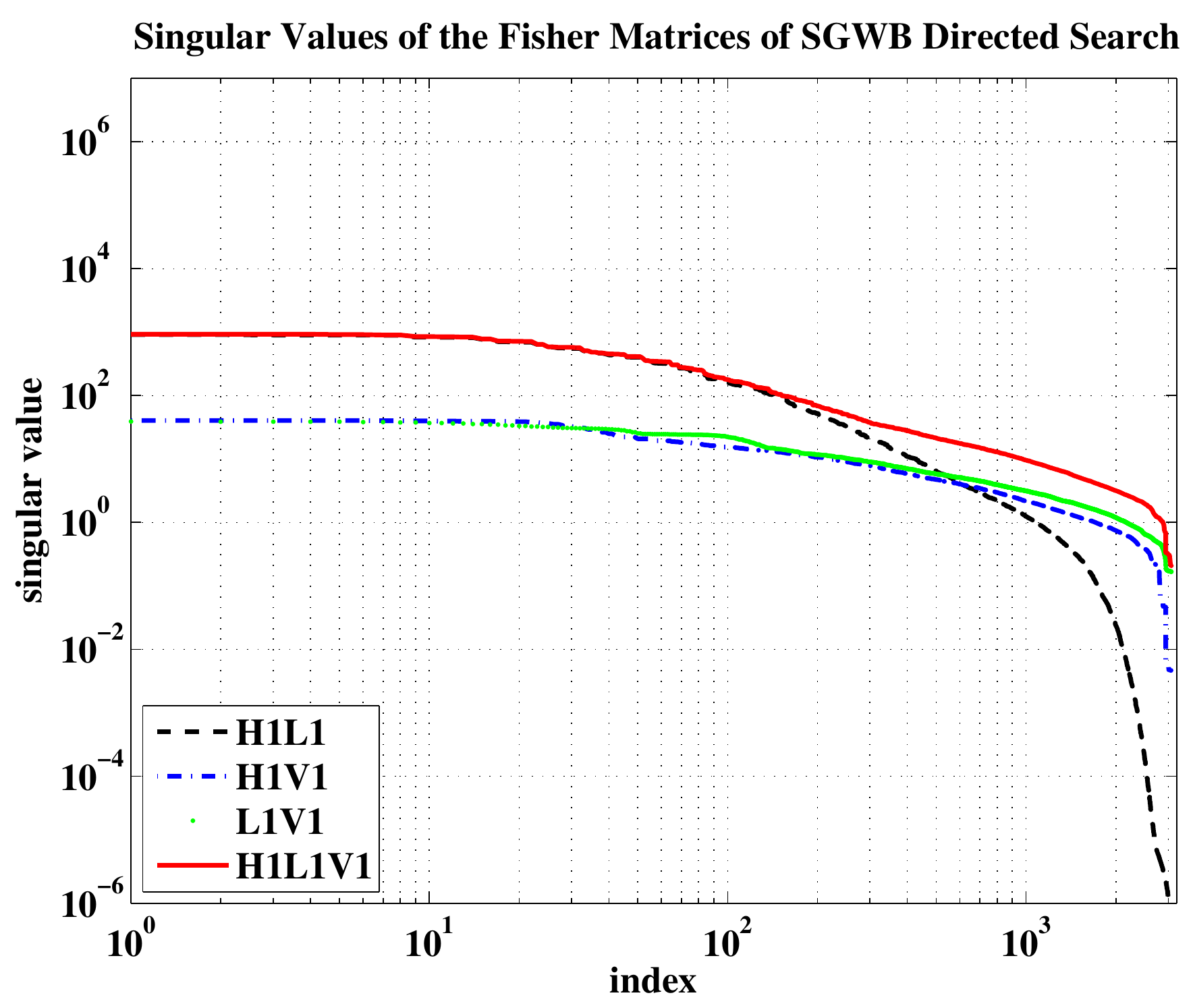}
\caption{\label{fig:SVD}Singular values of the Fisher matrices for individual baselines and the whole network are plotted in this figure. The LIGO H1L1 baseline (dashed curved line) has very small singular values at high resolutions; therefore, estimation of anisotropy at those resolutions is an ill-defined problem. The network (solid curved line) makes the singular values much more uniform, thereby regularizing the inverse problem.}
\end{center}
\end{figure}

Singular value decomposition of the Fisher information matrices provides a more quantitative verification of the above claim. Figure~\ref{fig:SVD} shows the singular values of the Fisher matrices for the individual baselines and the whole network. The LIGO baseline has very small singular values at higher resolutions (dashed curved line), which implies that estimation of anisotropy at those resolutions is an ill-defined problem. The network reduces the difference between high and low singular values and regularizes the inverse problem at high resolution (solid curved line).

\subsection{Parameter accuracy }

An important figure of merit for a directed search is how well a point source can be localized or its other parameters be constrained. In a noise-dominated mapping experiment, it is not easy to identify sources in the observed images. If there are candidate sources that have been modeled by other astronomical observations,
one can utilize that information to detect or constrain parameters of such sources. If the source was very accurately modeled, the optimal strategy would be to design a specific search focused on that source. But, in practice, with very limited knowledge of sources, the optimal strategy would be to vary the parameters within a reasonable range and maximize the log-likelihood ratio.

One of the main advantages of the MLR statistic is that it allows estimation of parameters of the source, given a model. For example, if there is an extended source, such as a cluster of galaxies, with an angular scale comparable to the resolution of the radiometer, and there exists a reasonable model for its mass distribution, one can maximize the log-likelihood ratio to find the center of the cluster~\footnote{Note that the image of the source may be too faint to be visually prominent in the dirty or clean maps.}. 
Even for a blind search, this method may prove to be advantageous to perform a finer search around the poorly estimated parameters of a potential candidate source.

In this section, we assess the accuracy with which a pointlike (single-pixel) source can be located using a network of GW detectors as compared to its individual baselines. The parameter estimation accuracy is deduced from the elements of the Fisher information matrix \cite{Helstrom}. 

For an unpolarized background from a single pixel, labeled $k$, and with $\hat{{\mathcal P}}^{r} = \delta^{r(k)}$, the single-baseline detection statistic follows from Eq.~\eqref{eq3.17} to be
\begin{equation}\label{eq5.1}
\lambda_{(k)}  = \frac{S^p ({\bf N}^{-1})_{p q} {\bf{\mathcal B}}^{q}_{(k)}} {\sqrt{ {\bf{\mathcal B}}^{r}_{(k)}({\bf N}^{-1})_{r s} {\bf{\mathcal B}}^{s}_{(k)} } } \;,
\end{equation}
which can be interpreted as the inner product of the data, ${\bf S}$, and a unit-norm template $\hat{\bm{\mathcal B}}_k$. Hence, the \textit{match} \cite{Owen} between the unit-norm templates for the $k^{\text{th}}$ and the $k^{\prime~\text{th}}$ pixels is
\begin{eqnarray}\label{eq5.2}
M & = & \frac{\mathcal{B}^{p}_{(k)}({\bf N}^{-1})_{p q}\mathcal{B}^{q}_{(k^{\prime})}} {\sqrt{\mathcal{B}^{r}_{(k)}({\bf N}^{-1})_{r s}\mathcal{B}^{s}_{(k)}}\;\sqrt{\mathcal{B}^{r^{\prime}}_{(k^{\prime})}({\bf N}^{-1})_{r^{\prime} s^{\prime}}\mathcal{B}^{s^{\prime}}_{(k^{\prime})}}} \; , \nonumber \\
& = & \frac{\mathcal{B}_{(k) (k^{\prime})}}{\sqrt{\mathcal{B}_{(k)(k)}}\sqrt{\mathcal{B}_{(k^{\prime}) (k^{\prime})}}} \, ,
\end{eqnarray}
where the inner products are all defined in terms of $\bf{N^{-1}}$. Define ${\bf{\Theta}}_{(k)}\equiv\{\mu_{k} ,\phi_{k}\}$ as the pixel coordinates, where $\mu_{k} \equiv \cos {\theta}_{k}$, with ${\theta}_{k}$ and $\phi_{k}$ being the declination and right ascension of the $k^{\text{th}}$ pixel, respectively. Since the \textit{match} has a maximum value of unity at $k^{\prime}=k$, one can expand $M$ in a Taylor series about $\Delta\mu_{(k)}=0$ and $\Delta \phi_{(k)}=0$ as
\begin{eqnarray}\label{eq5.3}
M &\approx& 1+\frac{1}{2}\left(\frac{\partial^{2}M}{\partial\Theta_{(k^{\prime})}^{\mu}\partial\Theta_{(k^{\prime})}^{\nu}}\right)\Bigg|_{{\bf{\Theta}}_{(k^{\prime})}={\bf{\Theta}}_{(k)}}\Delta\Theta_{(k)}^{\mu}\Delta\Theta_{(k)}^{\nu} \, , \nonumber \\
&\approx& 1-\frac{\Gamma_{(k)\mu\nu}}{(\text{SNR})^{2}_{(k)}}\Delta\Theta_{(k)}^{\mu}\Delta\Theta_{(k)}^{\nu} \, ,
\end{eqnarray}
where 
$\Gamma_{(k)\mu\nu}\equiv\Gamma_{\mu\nu}({\bf{\Theta}}_{(k)})$ are the components of the Fisher information matrix ${\bf\Gamma}_{(k)}$, and
$(\text{SNR})_{(k)}$ is the signal-to-noise ratio in the $k^{\text{th}}$ pixel.
%

For large $\text{SNR}$, the error variance-covariance matrix obeys
\begin{equation}\label{eq5.4}
\left({\bf\Gamma}_{(k)}^{-1}\right)^{\mu\nu} \approx \left<\Delta\Theta_{(k)}^{\mu}\Delta\Theta_{(k)}^{\nu}\right> \,.
\end{equation}
The estimation error in the measurement of the sky-position solid angle (in steradians) is given by \cite{Cutler}
\begin{eqnarray}\label{eq5.5}
&&\Delta\Omega_{(k)} \ = \nonumber\\
&& \ \ 2\pi\sqrt{\left<(\Delta\cos\theta_{(k)})^{2}\right>\left<(\Delta\phi_{(k)})^{2}\right> - \left<\Delta\cos\theta_{(k)}\Delta\phi_{(k)}\right>^{2}} \, .\nonumber\\
\end{eqnarray}
The Fisher information matrix for multiple baselines is just the sum of the Fisher matrices for the individual baselines,
\begin{equation}\label{eq5.6}
\left[\Gamma_{(k)\mu\nu}\right]_{\mathcal{N}}=\sum_{\mathcal{I}}\Gamma_{\mathcal{I}(k)\mu\nu} \, ,
\end{equation}\
where $\mathcal{I}$ is the baseline index, and $\Gamma_{\mathcal{I}(k)\mu\nu}$ is the Fisher information matrix of the $\mathcal{I}^{\text{th}}$ baseline, as given in Eq.~\eqref{eq5.3}. Hence, the error variance-covariance matrix for the network is
\begin{equation}\label{eq5.7}
\left(\left[{\bf\Gamma}_{(k)}\right]_{\mathcal{N}}^{-1}\right)^{\mu\nu} \approx \left<\Delta\Theta_{(k)}^{\mu}\Delta\Theta_{(k)}^{\nu}\right>_{\mathcal{N}} \,,
\end{equation}
for large SNR. Therefore, the 1$\sigma$ estimation error in solid angle for locating a pixel source with the multibaseline network is expressed as
\begin{widetext}
\begin{equation}\label{eq5.8}
\left[\Delta\Omega_{(k)}\right]_{\mathcal{N}} = 2\pi\sqrt{\left<(\Delta\cos\theta_{(k)})^{2}\right>_{\mathcal{N}}\left<(\Delta\phi_{(k)})^{2}\right>_{\mathcal{N}} - \left<\Delta\cos\theta_{(k)}\Delta\phi_{(k)}\right>_{\mathcal{N}}^{2}} \, .
\end{equation}
\end{widetext}
Note that this error diminishes with SNR as $1/\text{SNR}^2$, i.e., localization is more accurate at higher SNR.

\begin{figure}[h!]
\begin{center}
\includegraphics[width=0.45\textwidth]{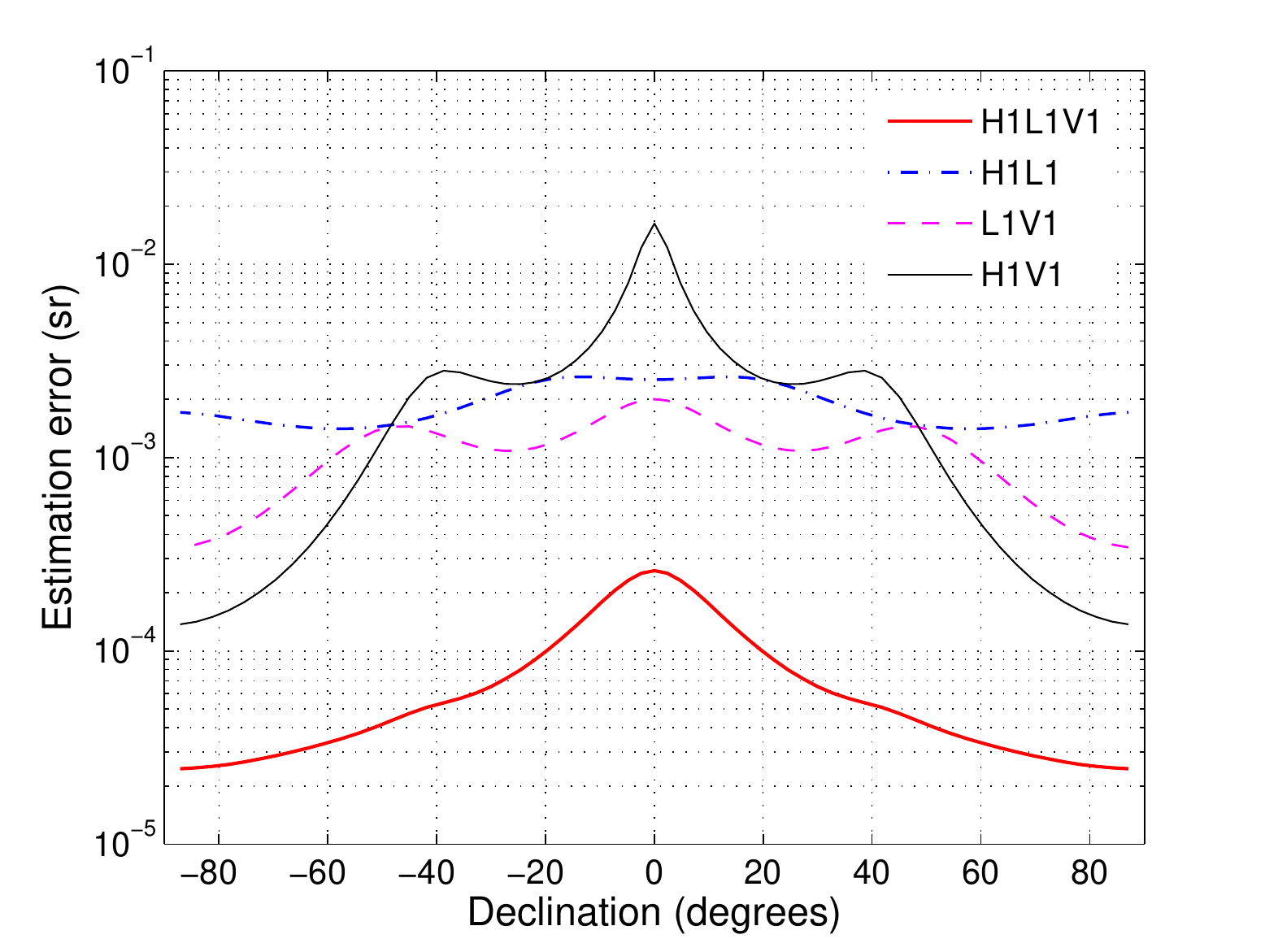}
\includegraphics[width=0.45\textwidth]{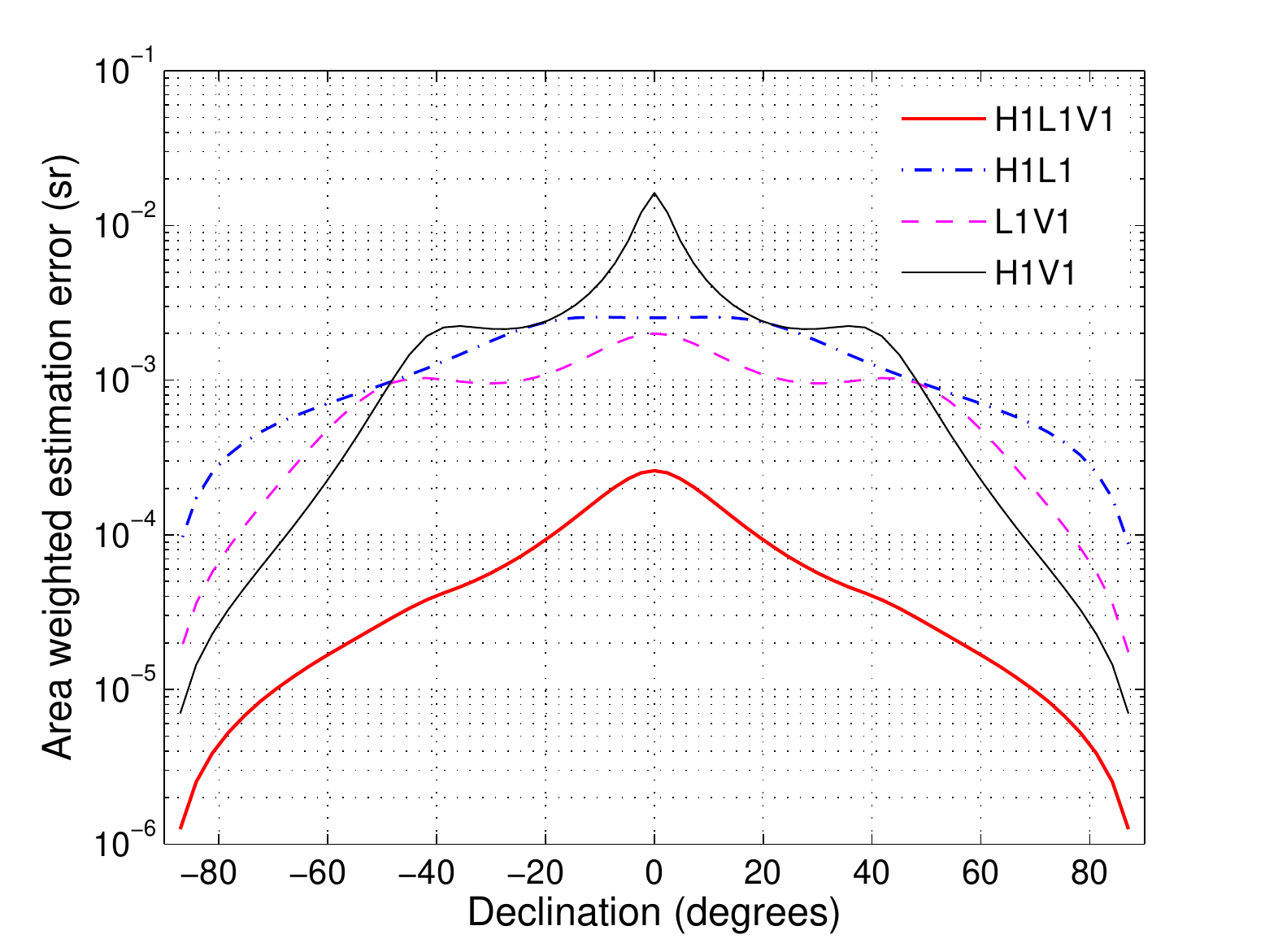}
\caption{\label{fig:locErr}The 1$\sigma$ error (top) and area-weighted 1$\sigma$ error (bottom) in the solid angle for locating a source with three single baselines and the whole network. The network accuracy is better by about 1 order of magnitude or more at most of the declinations. Note that an error of 1sr $\simeq$ 3282.80635 sq-degrees, and that the error here decreases as 1/SNR$^2$.}
\end{center}
\end{figure}

We present the source-localization errors for the individual LIGO-Virgo baselines and the network in the top panel of Fig.~\ref{fig:locErr}. We also show the corresponding area-weighted plots obtained by multiplying these errors with the cosine of the latitudinal angle in the bottom panel of Fig.~\ref{fig:locErr}. 
The network clearly outperforms individual baselines by about 1 order of magnitude or more for almost all declination angles.

The primary focus of this analysis was to obtain statistics that are based on 
dirty-map constructs. 
However, it is straightforward to extend it to be applicable to clean maps. Similarly, although we considered broadband signals, it is possible to easily extend our study to narrow band signals.

\subsection{Map making}

Finally, we compare the quality of sky maps made by the individual baselines and their network since they are among the primary products of anisotropic searches. Here, we consider two figures of merit, namely, the MLR statistic and the normalized mean square error (NMSE) for comparing maps. To compare the dirty maps, we use the MLR statistic introduced earlier, and, to compare clean maps, we use both the MLR statistic and the NMSE (defined below).

For an unpolarized and anisotropic gravitational wave background, the maximum-likelihood estimators of the signal-strength vector are given by
\begin{equation}\label{eq6.1}
{\widetilde{\mathcal{P}}}^{k} = ({\bm{\mathcal{B}}}^{-1})^{k}_{\;k^{\prime}}S^{k^{\prime}}\, ,
\end{equation}
where $S^{k}$ are components of the dirty map ~\eqref{eq3.14} and ${\widetilde{\mathcal{P}}}^{k}$ are components of the deconvolved (clean) map. Note that the clean map, ${\widetilde{\bm{\mathcal{P}}}}$, are the values of ${\bm{\mathcal{P}}}$ that maximize the statistic.  We extend this single-baseline analysis to a multibaseline one by simply adding the dirty maps and beam matrices as $S_{\mathcal{N}}^{k}=\sum_{{\mathcal I}=1}^{N_{b}}S_{\mathcal{I}}^{k}$ and ${\mathcal{B}}_{{\mathcal{N}}k^{\prime}}^{k}=\sum_{{\mathcal I}=1}^{N_{b}}{\mathcal{B}}_{{\mathcal{I}}k^{\prime}}^{k}$. So, the maximum-likelihood estimators for a multibaseline network are given by
\begin{equation}\label{eq6.2}
{\widetilde{\mathcal{P}}}^{k} = ({\bm{\mathcal{B}}_{\mathcal N}}^{-1})^{k}_{\;k^{\prime}}S_{\mathcal{N}}^{k^{\prime}}\,.
\end{equation}
We simulate the data with signal as~\cite{Mitra}
\begin{eqnarray}\label{eq6.3}
\tilde{x}^{*}_{1}(t,f)\tilde{x}_{2}(t,f) & = & \langle \tilde{h}^{*}_{1}(t,f)\tilde{h}_{2}(t,f)\rangle + \tilde{n}^{*}_{1}(t,f)\tilde{n}_{2}(t,f) \, , \nonumber \\
\langle\tilde{h}^{*}_{1}(t,f)\tilde{h}_{2}(t,f^{\prime})\rangle & = & \delta_{f f^{\prime}} H(|f|) \sum_{i}{\mathcal P}^{i}\,\gamma(\hat{\Omega}_{i},t,|f|) \, ,
\end{eqnarray}
where ${\mathcal{P}}^{i}$ is the injected source strength at the $i^{\text{th}}$ pixel, and $\gamma$ is the direction-dependent overlap reduction function. We use a conjugate gradient (CG) method to solve the set of linear equations \eqref{eq6.1} and \eqref{eq6.2}.

The mismatch between two maps, injected and estimated, is measured using the normalized mean square error,
\begin{equation}\label{eq6.4}
\text{NMSE} := \frac{|\widetilde{\bm{\mathcal P}} - \bm{\mathcal P}|^{2}}{|\bm{\mathcal P}|^{2}}\, . 
\end{equation}

\subsubsection{Dirty maps and clean maps}

\begin{figure}[h!]
\begin{center}
\subfigure[~Injected map]
{\includegraphics[width=0.238\textwidth]{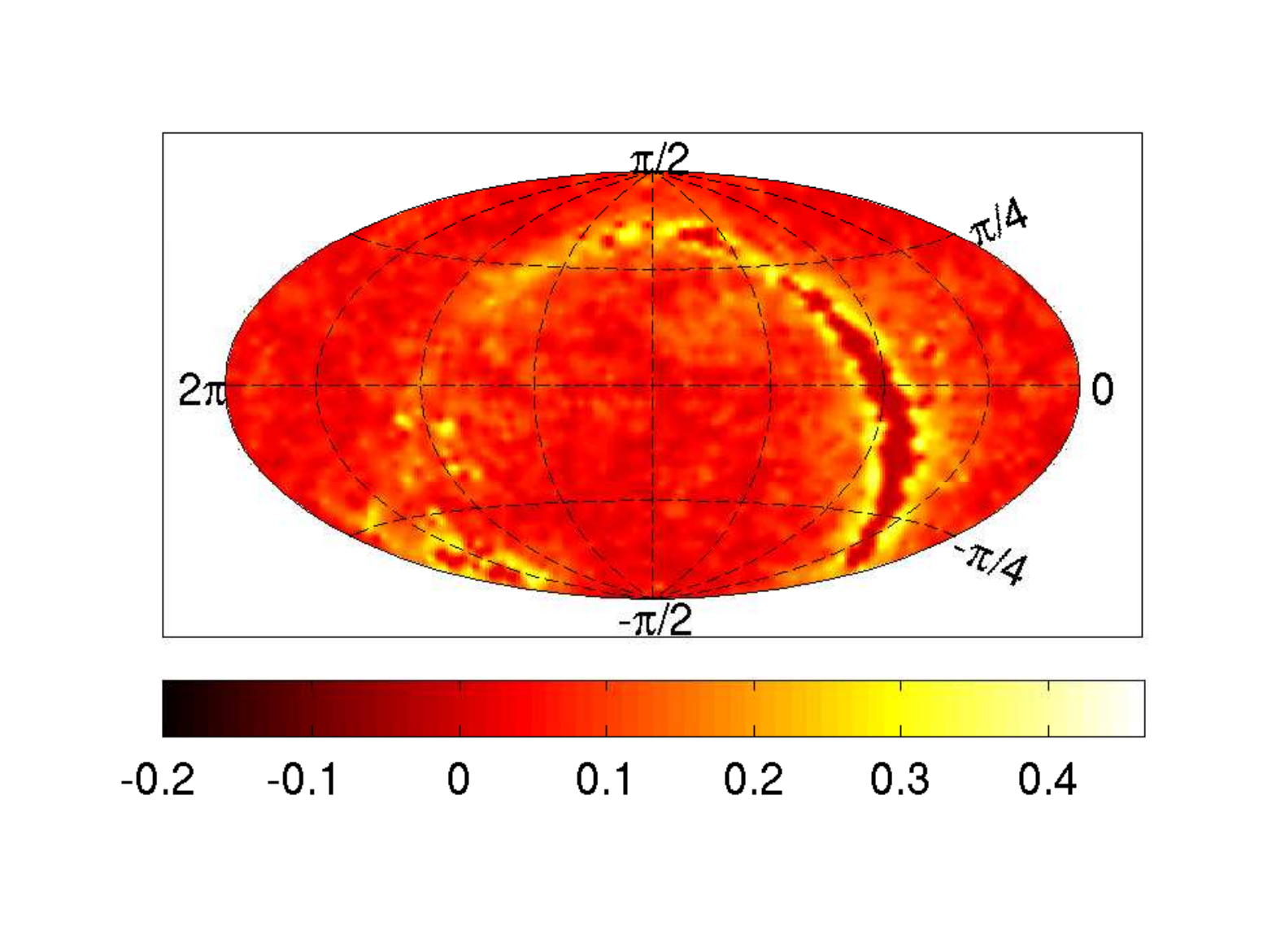}}
\subfigure[~H1L1]
{\includegraphics[width=0.238\textwidth]{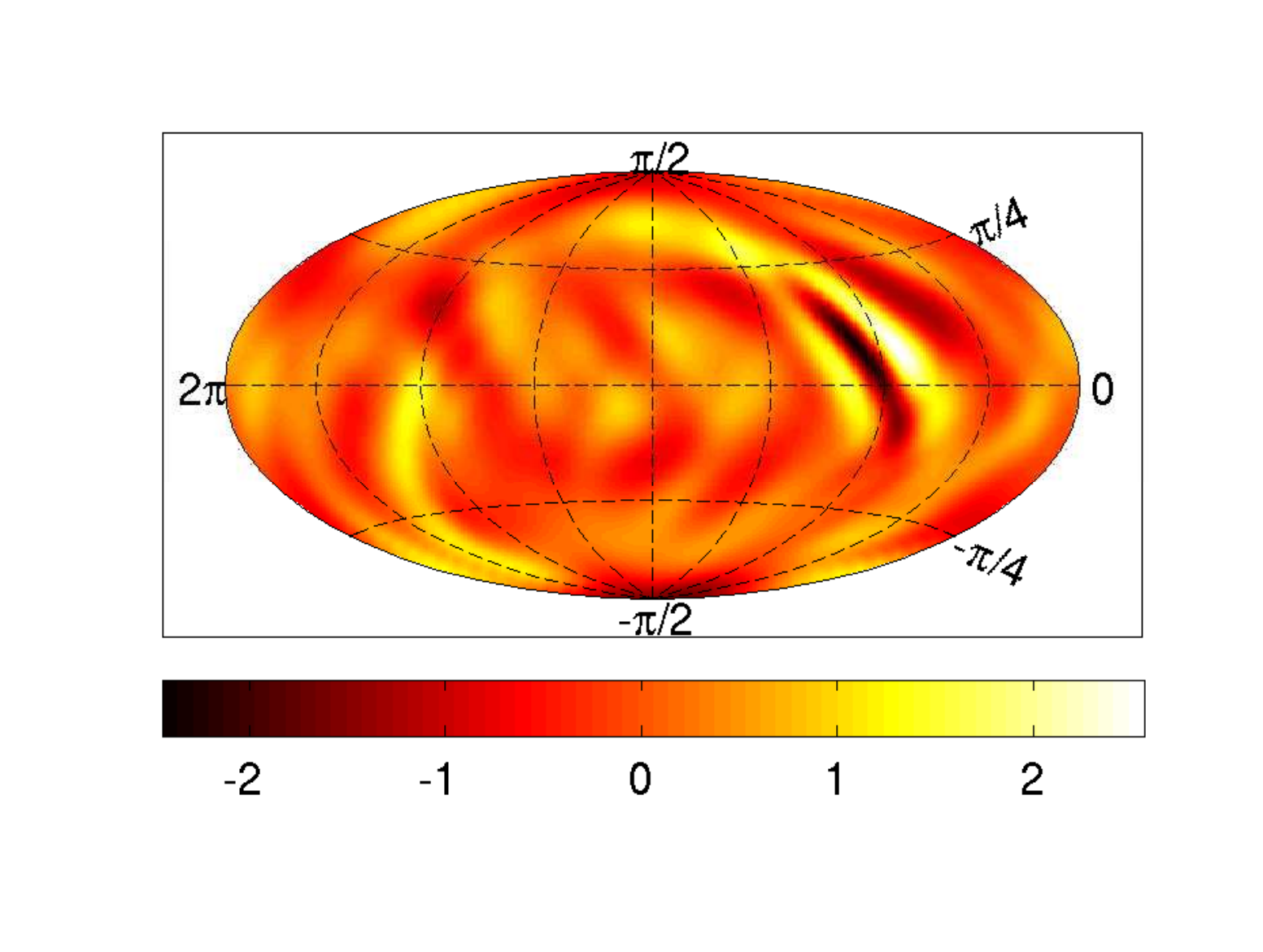}}
\subfigure[~L1V1]
{\includegraphics[width=0.238\textwidth]{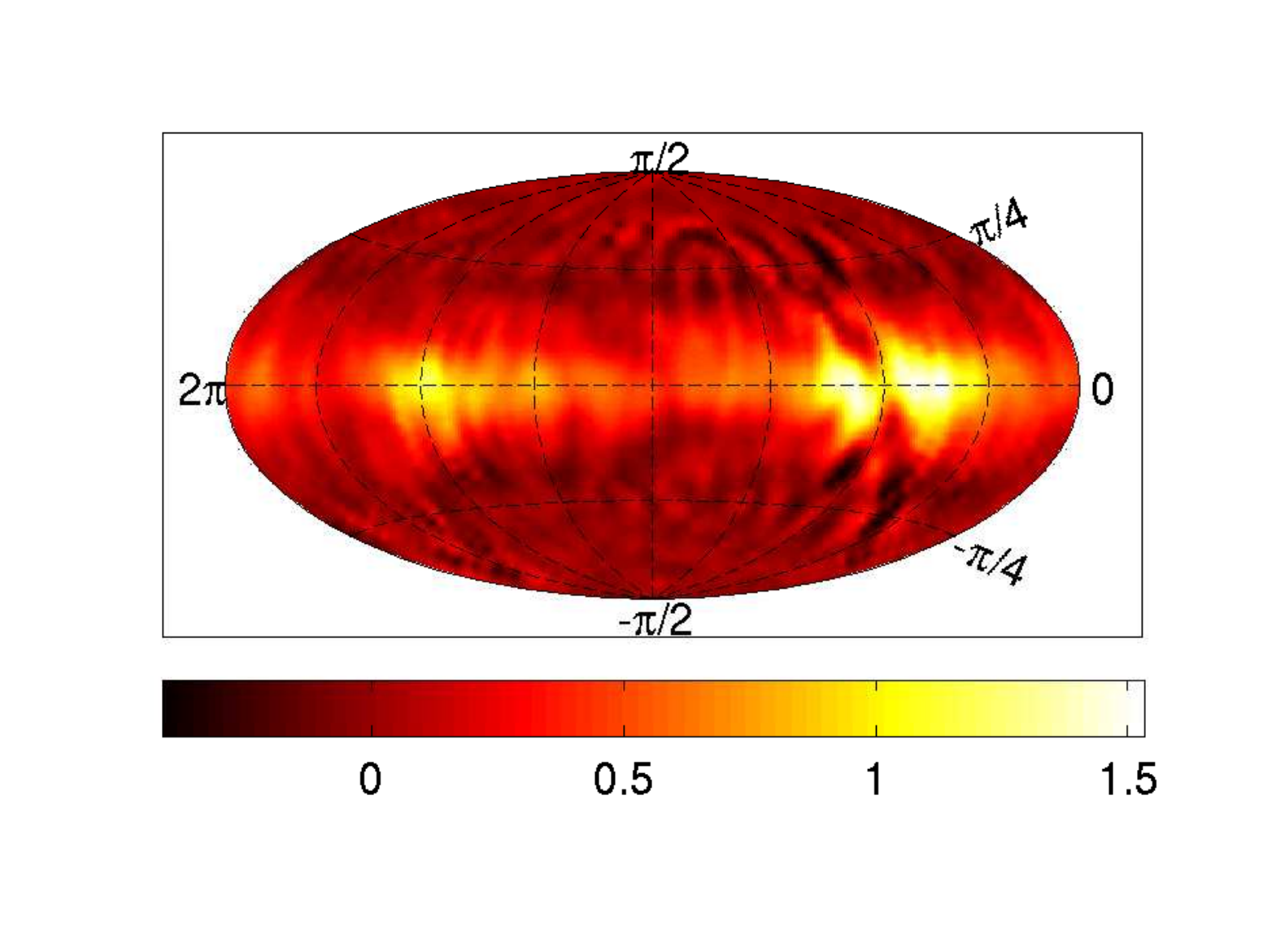}}
\subfigure[~H1V1]
{\includegraphics[width=0.238\textwidth]{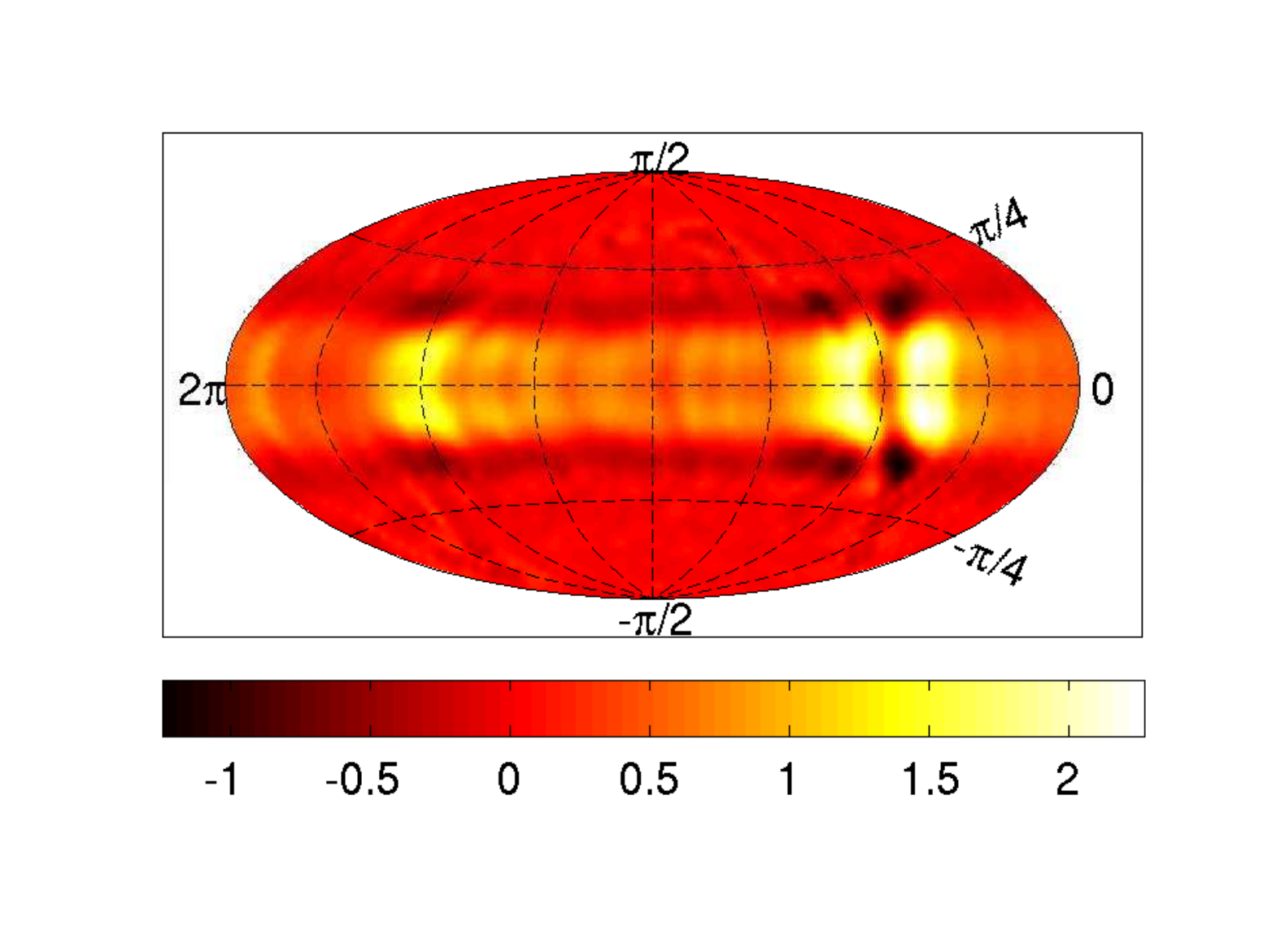}}
\caption{\label{fig:dirtygal}Toy model of an extended source is shown in (a). Dirty maps made from simulated data containing signal from that source are shown in the last three panels for the three LIGO-Virgo baselines.}
\end{center}
\end{figure}

\begin{figure}[h!]
\begin{center}
\subfigure[~H1L1 (NMSE=0.4311)]
{\includegraphics[width=0.238\textwidth]{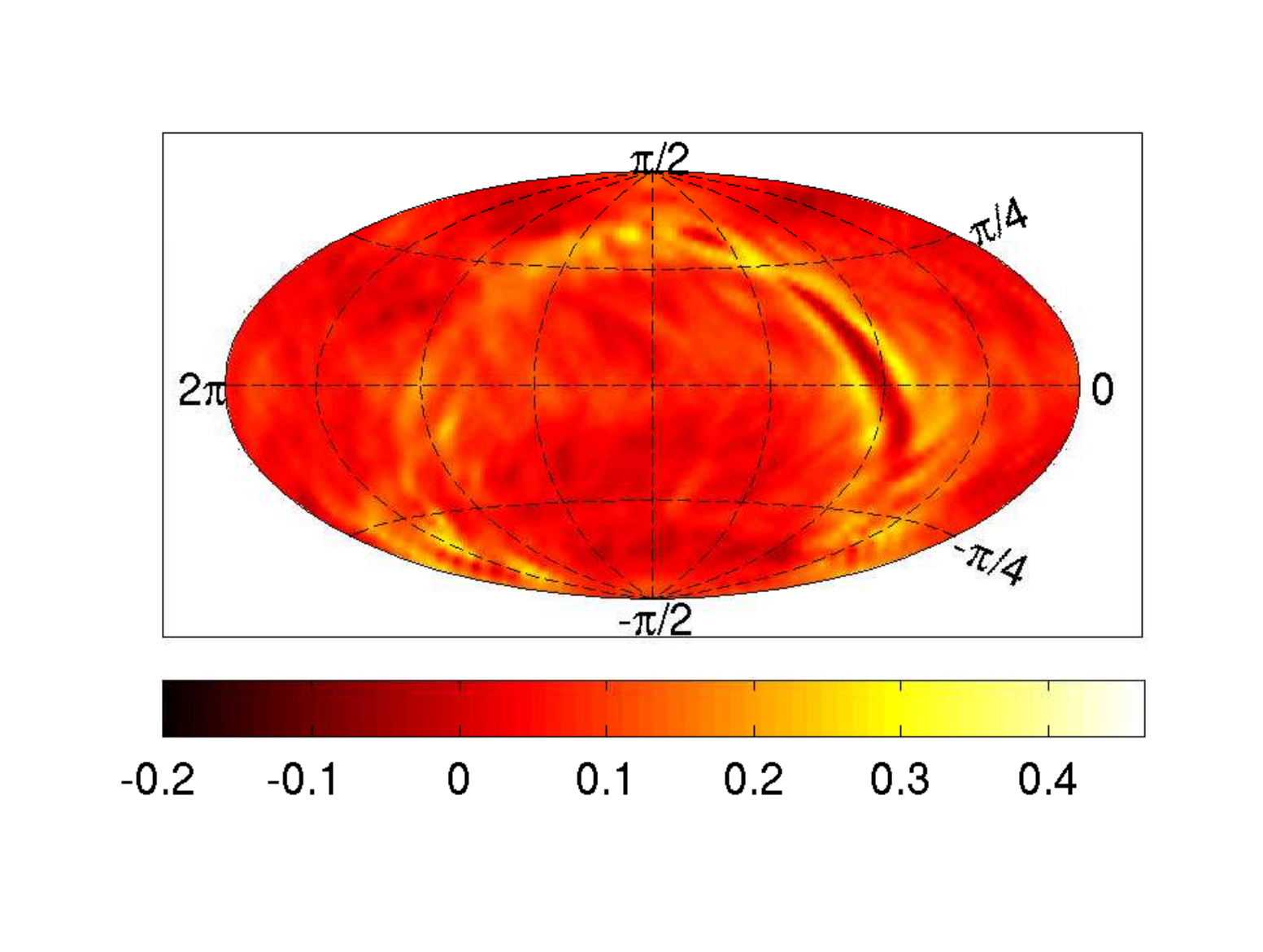}}
\subfigure[~L1V1 (NMSE=0.4171)]
{\includegraphics[width=0.238\textwidth]{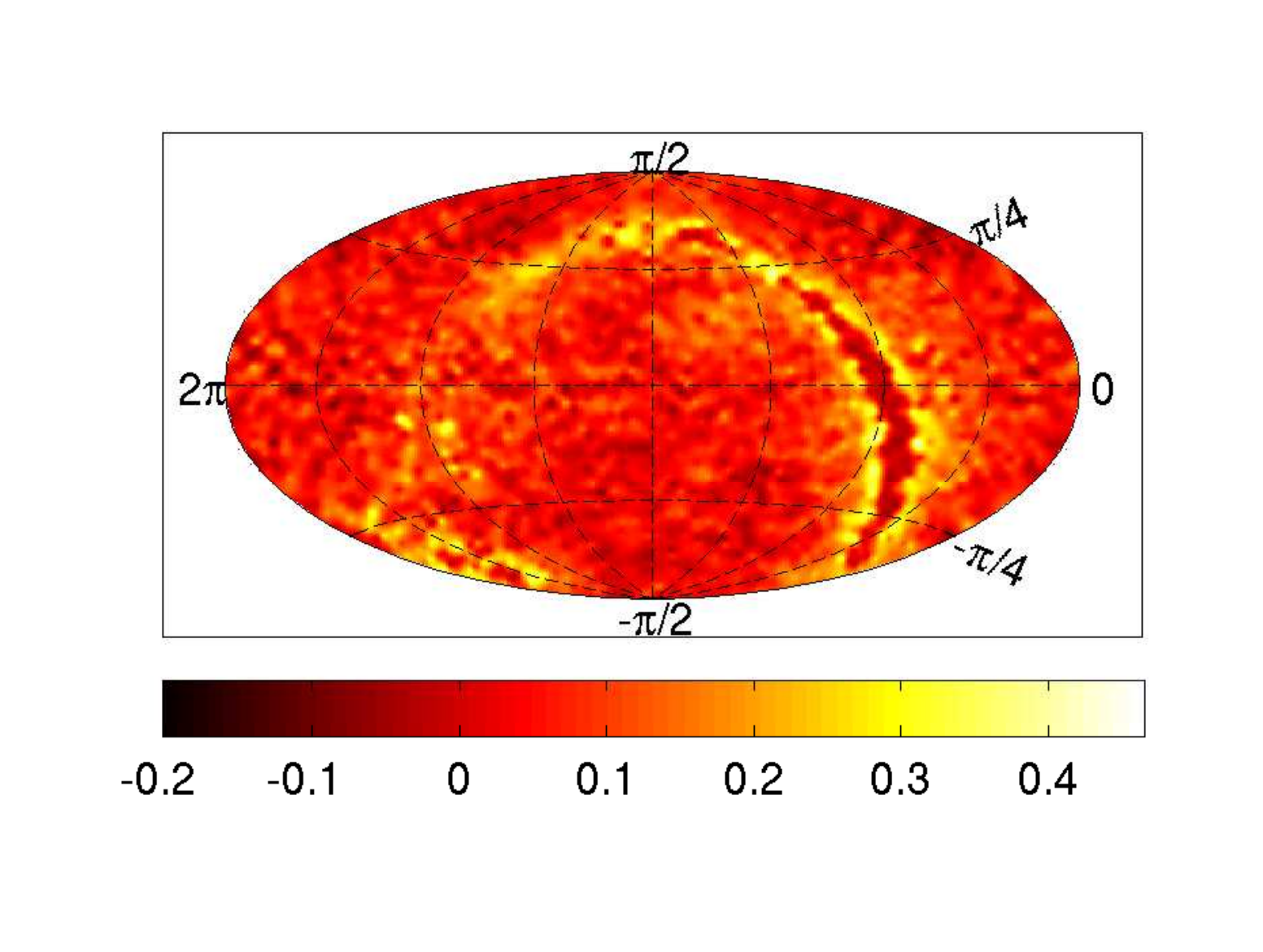}}
\subfigure[~H1V1 (NMSE=0.4735)]
{\includegraphics[width=0.238\textwidth]{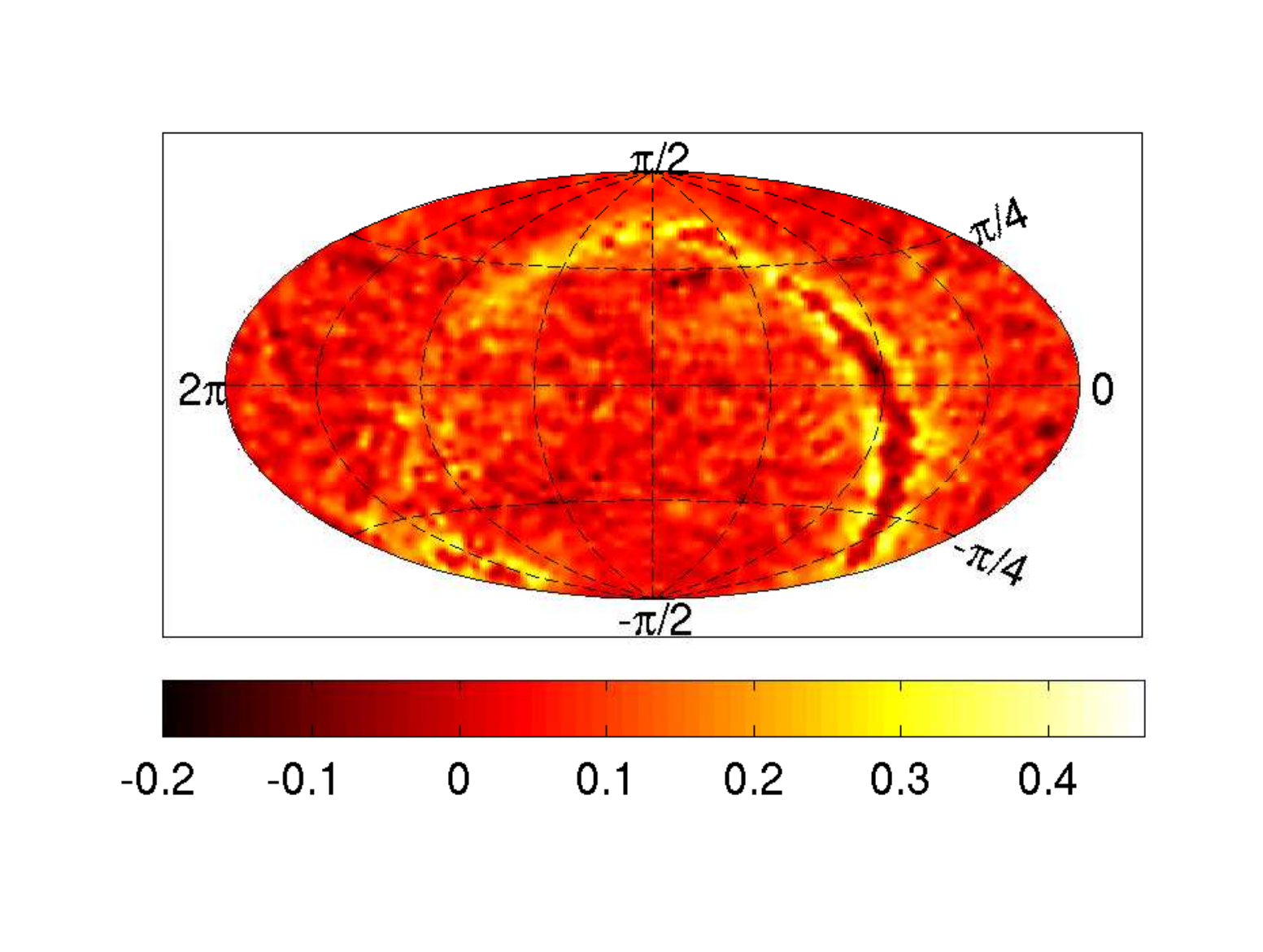}}
\subfigure[~H1L1V1 (NMSE=0.2206)]
{\includegraphics[width=0.238\textwidth]{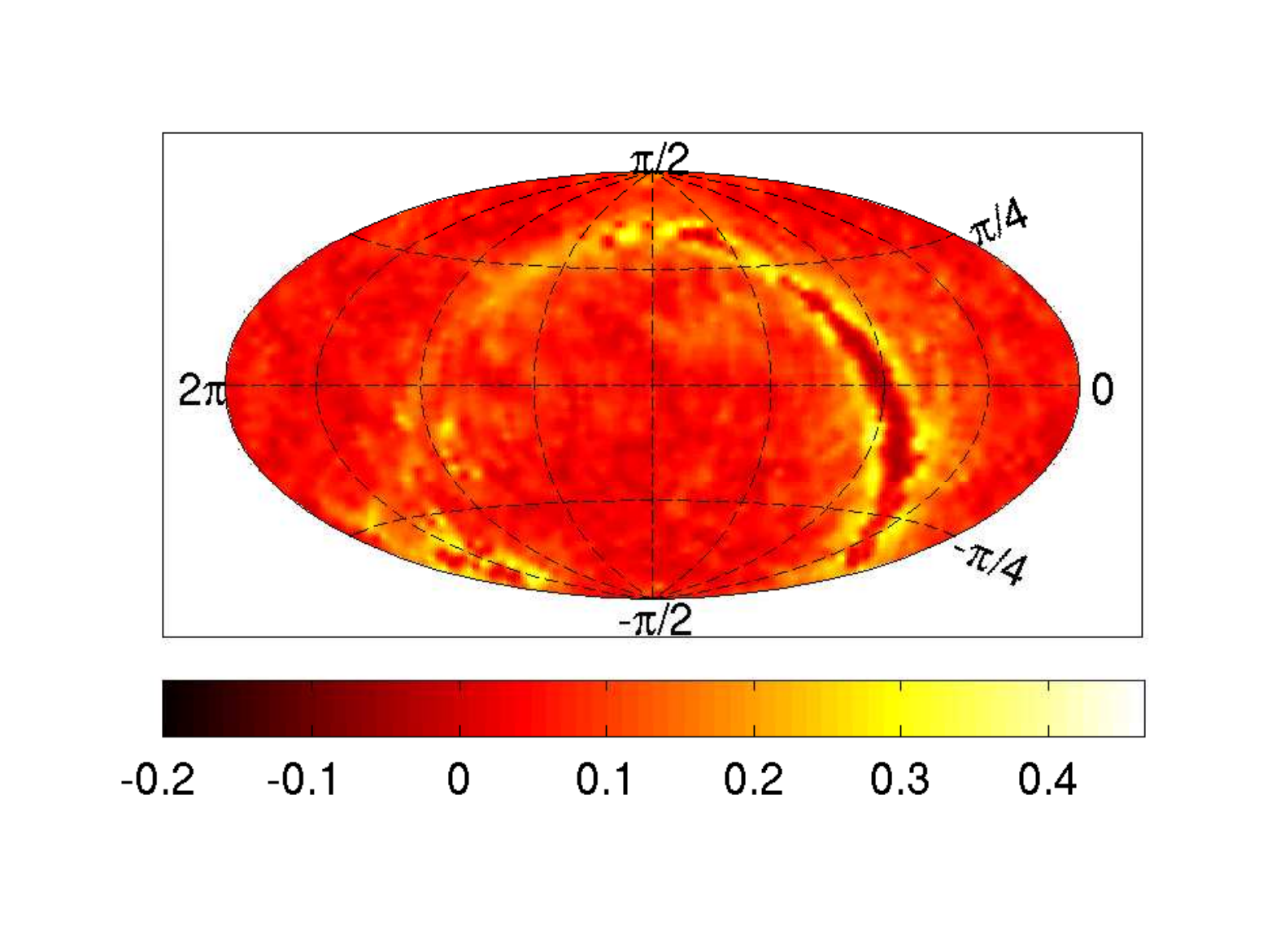}}
\caption{\label{fig:cleangal}Clean maps obtained by the deconvolution of the dirty maps of Fig.~\ref{fig:dirtygal}, using 20 CG iterations, are shown here.}
\end{center}
\end{figure}

\begin{figure}[h!]
\begin{center}
\subfigure[~H1L1]
{\includegraphics[width=0.238\textwidth]{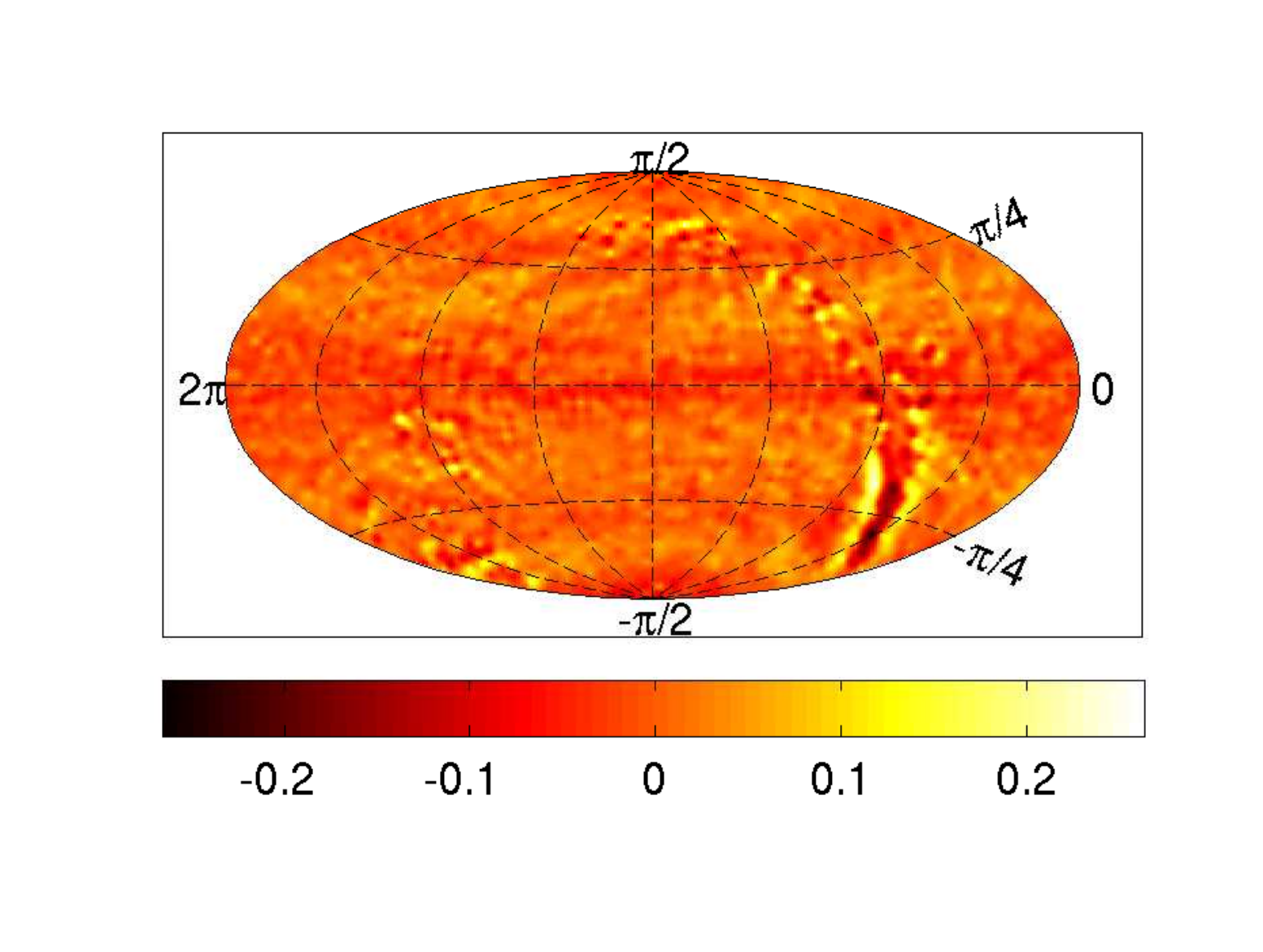}}
\subfigure[~L1V1]
{\includegraphics[width=0.238\textwidth]{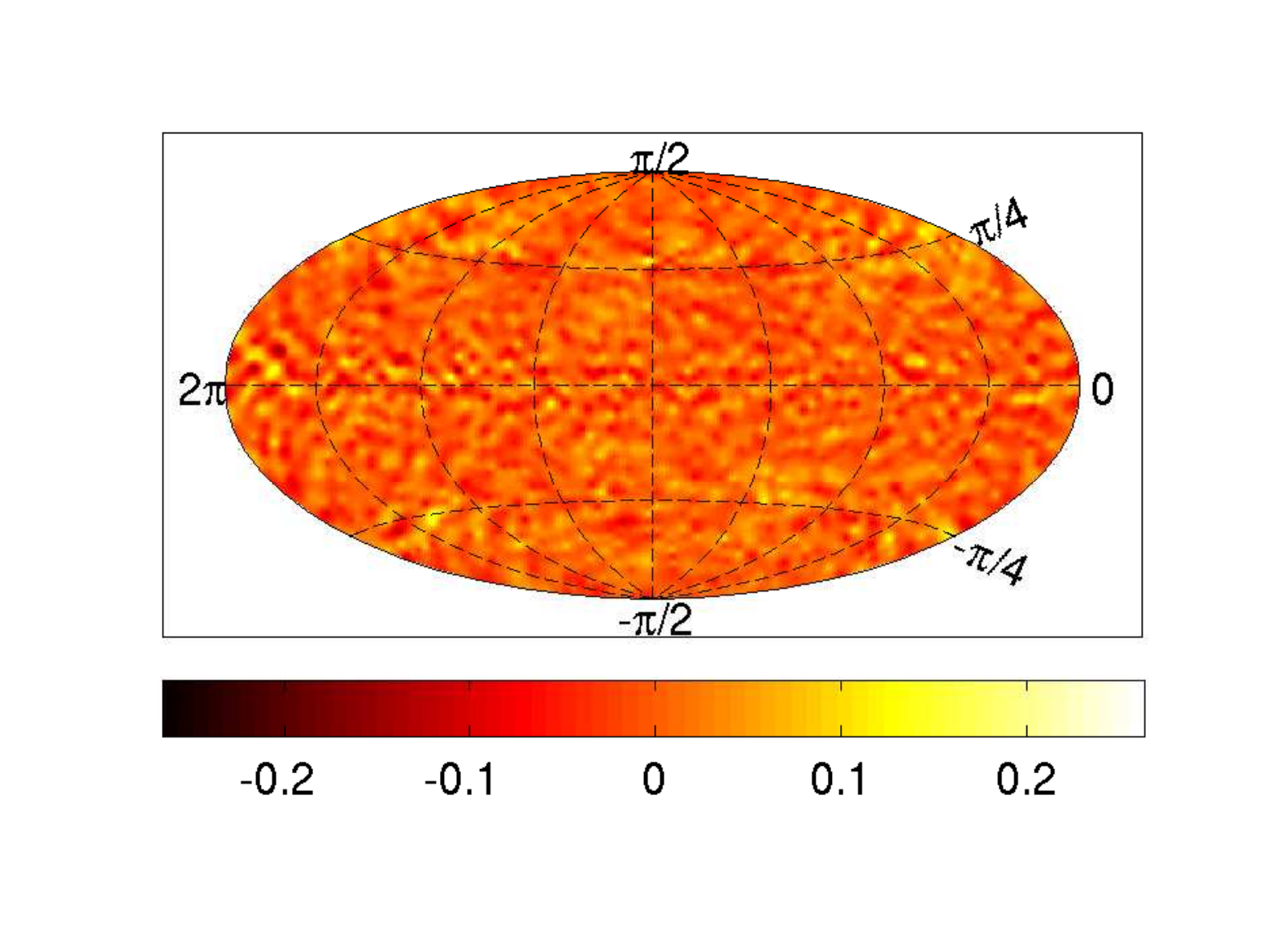}}
\subfigure[~H1V1]
{\includegraphics[width=0.238\textwidth]{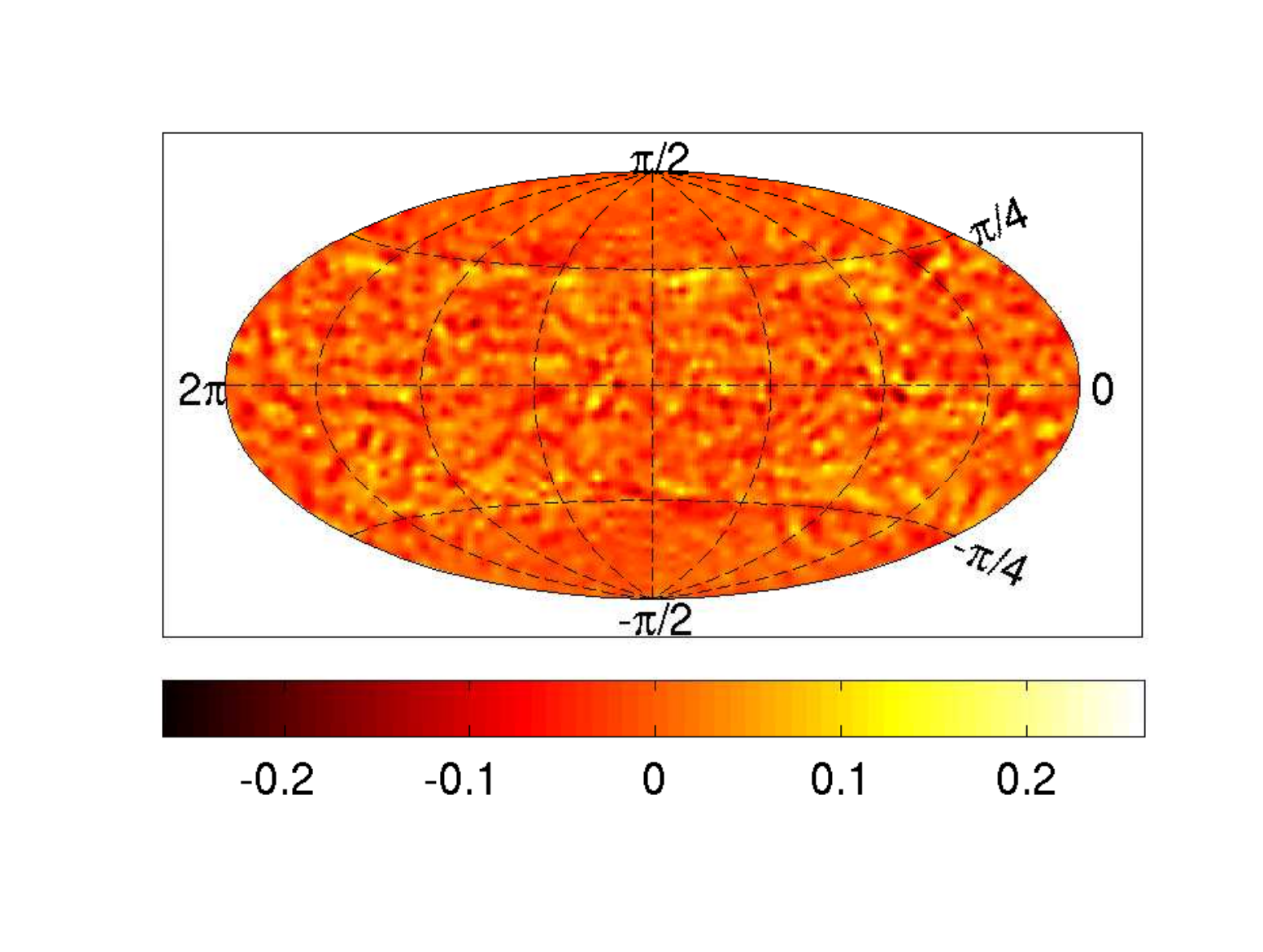}}
\subfigure[~H1L1V1]
{\includegraphics[width=0.238\textwidth]{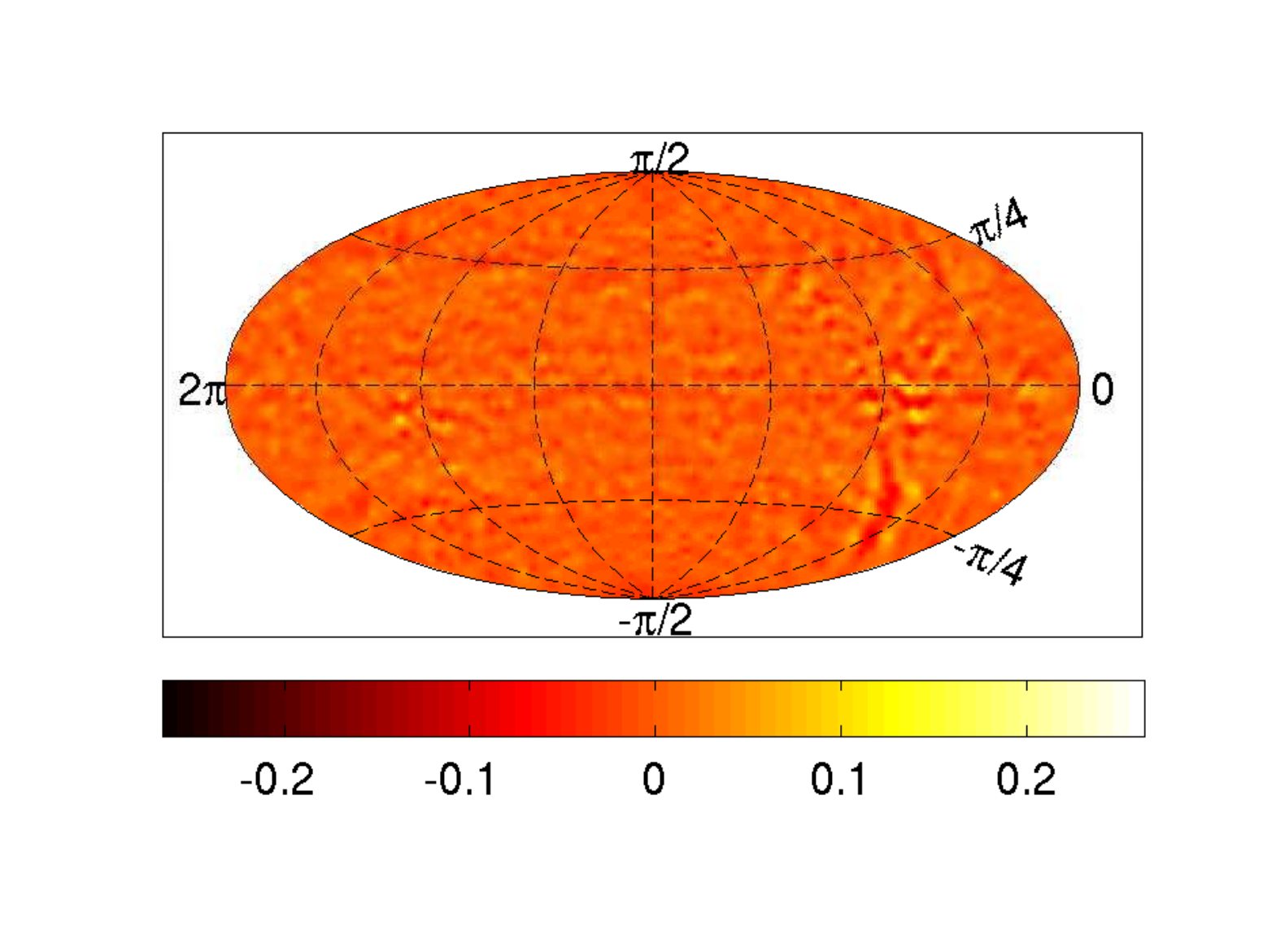}}
\caption{\label{fig:biasgal}Difference between the clean maps of Fig.~\ref{fig:cleangal} and the injected map of Fig.~\ref{fig:dirtygal}(a).}
\end{center}
\end{figure}

\begin{figure}[h!]
\begin{center}
\subfigure[~Injected map]
{\includegraphics[width=0.238\textwidth]{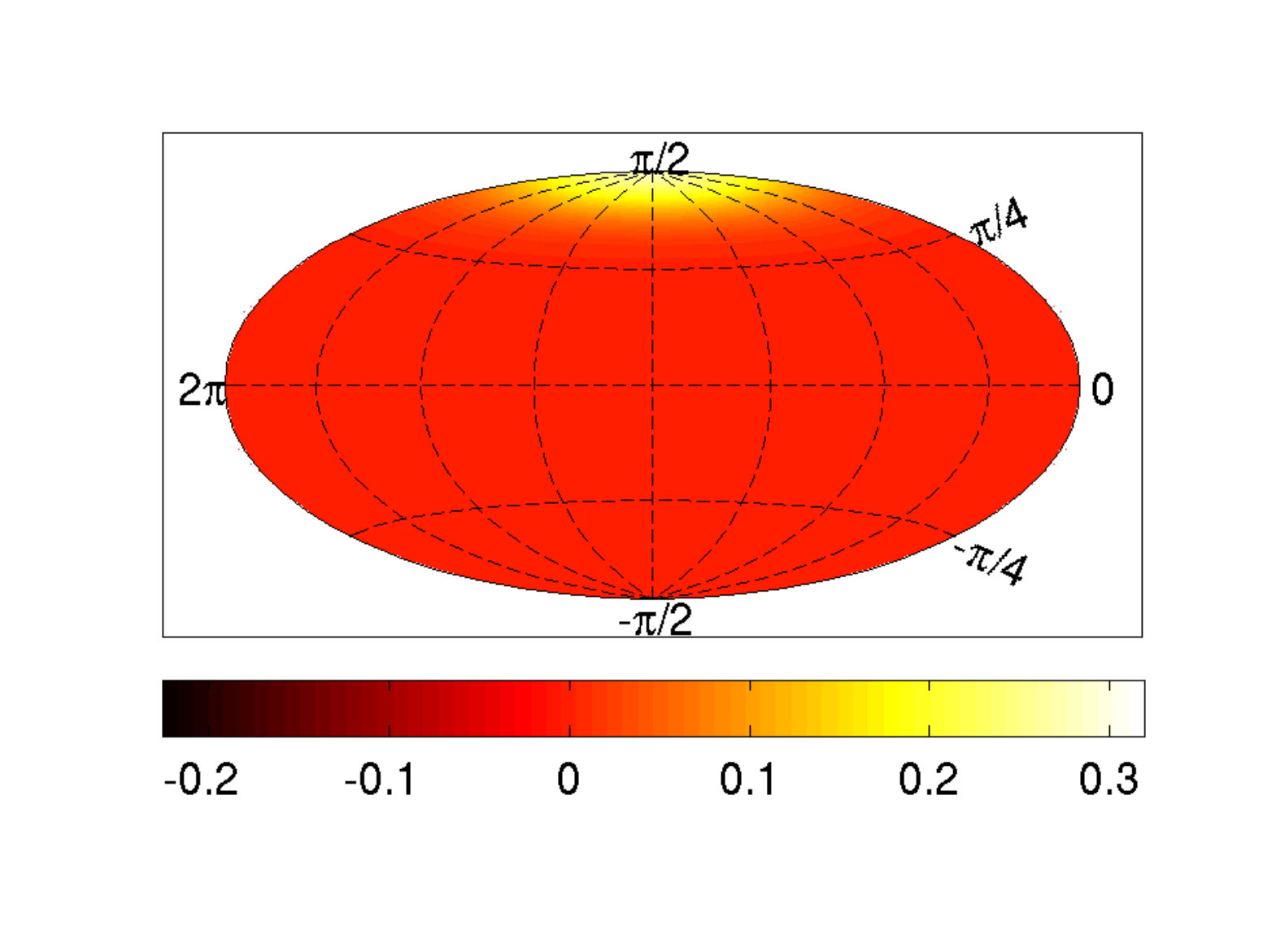}}
\subfigure[~H1L1]
{\includegraphics[width=0.238\textwidth]{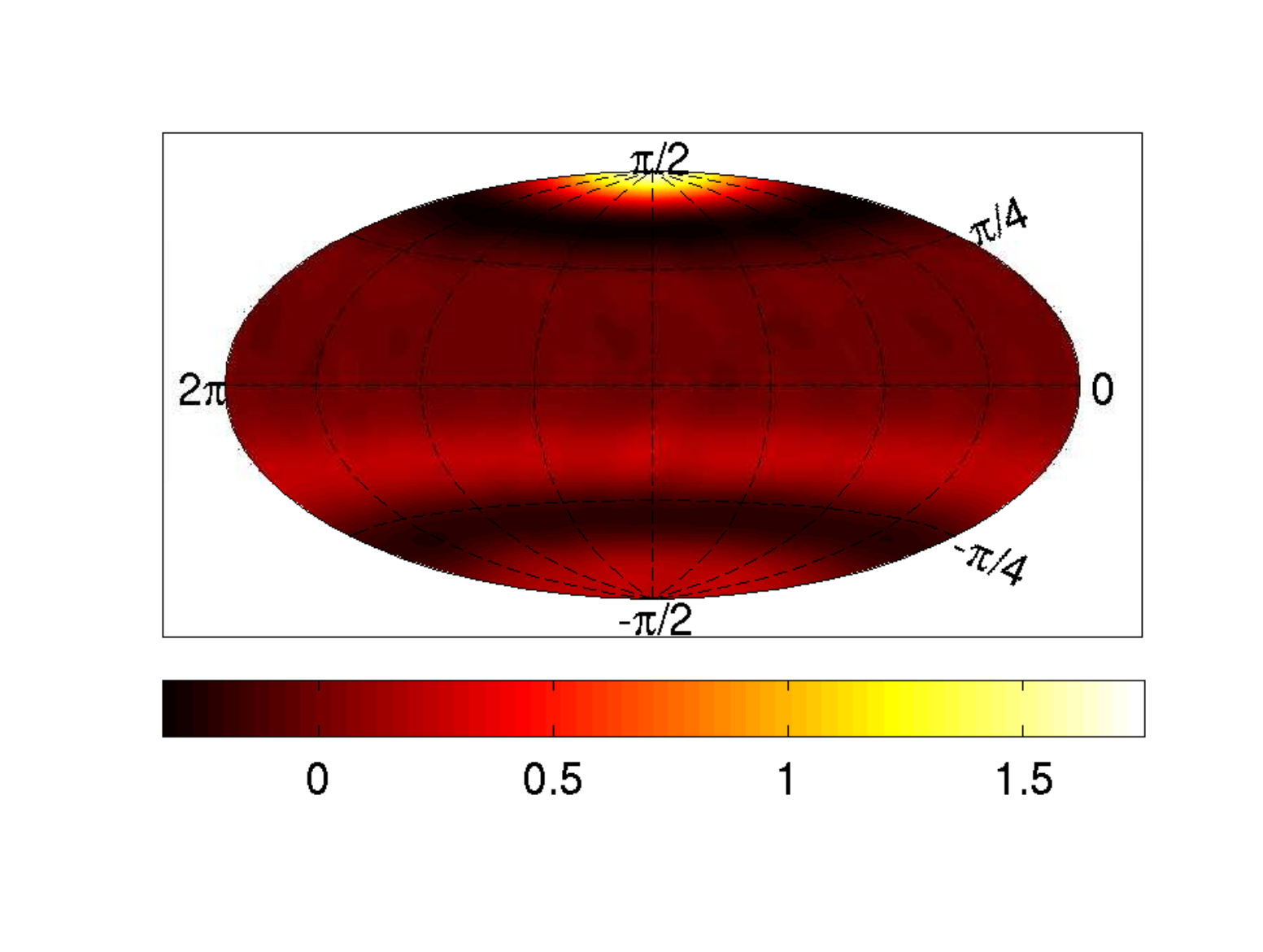}}
\subfigure[~L1V1]
{\includegraphics[width=0.238\textwidth]{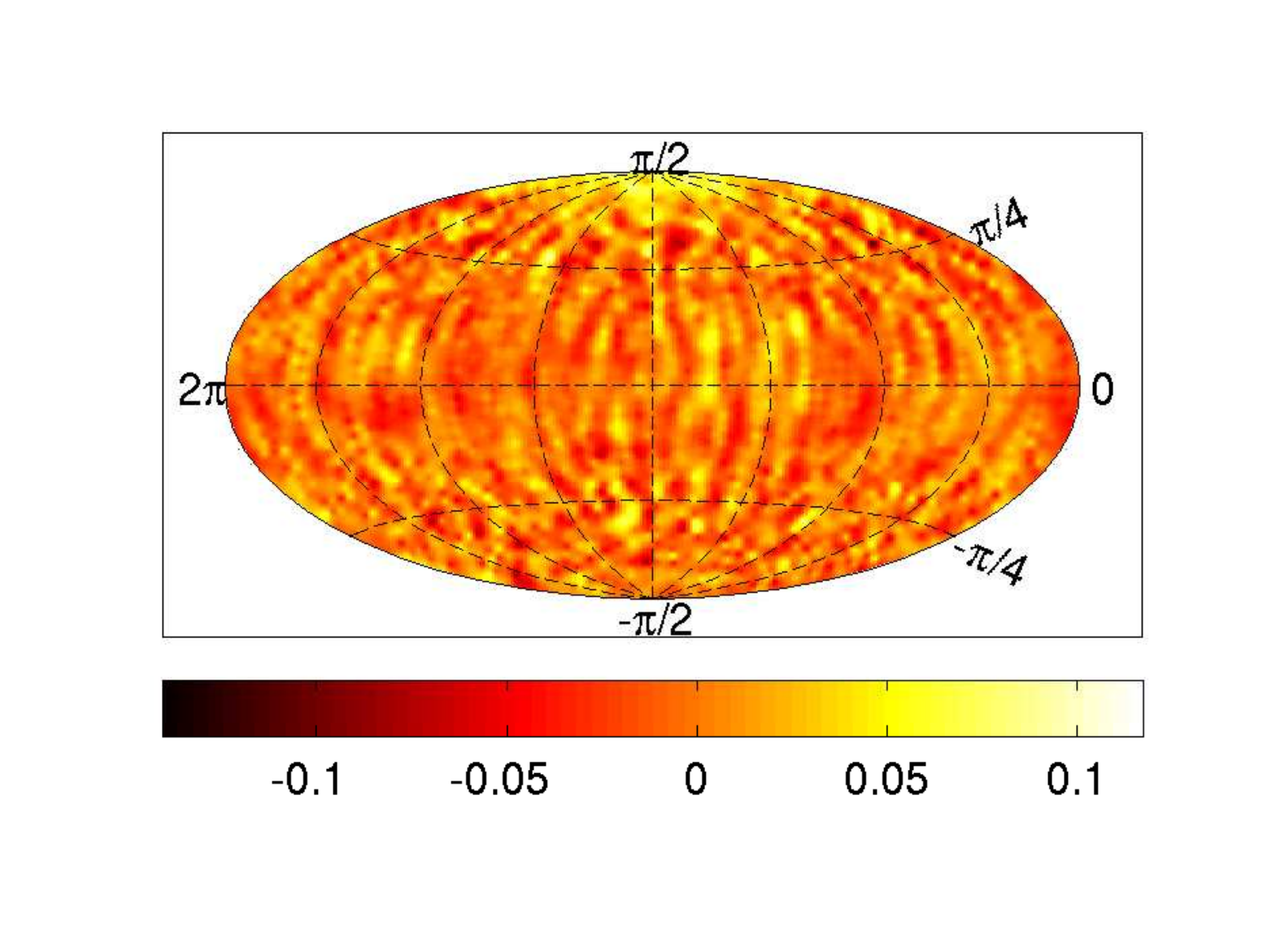}}
\subfigure[~H1V1]
{\includegraphics[width=0.238\textwidth]{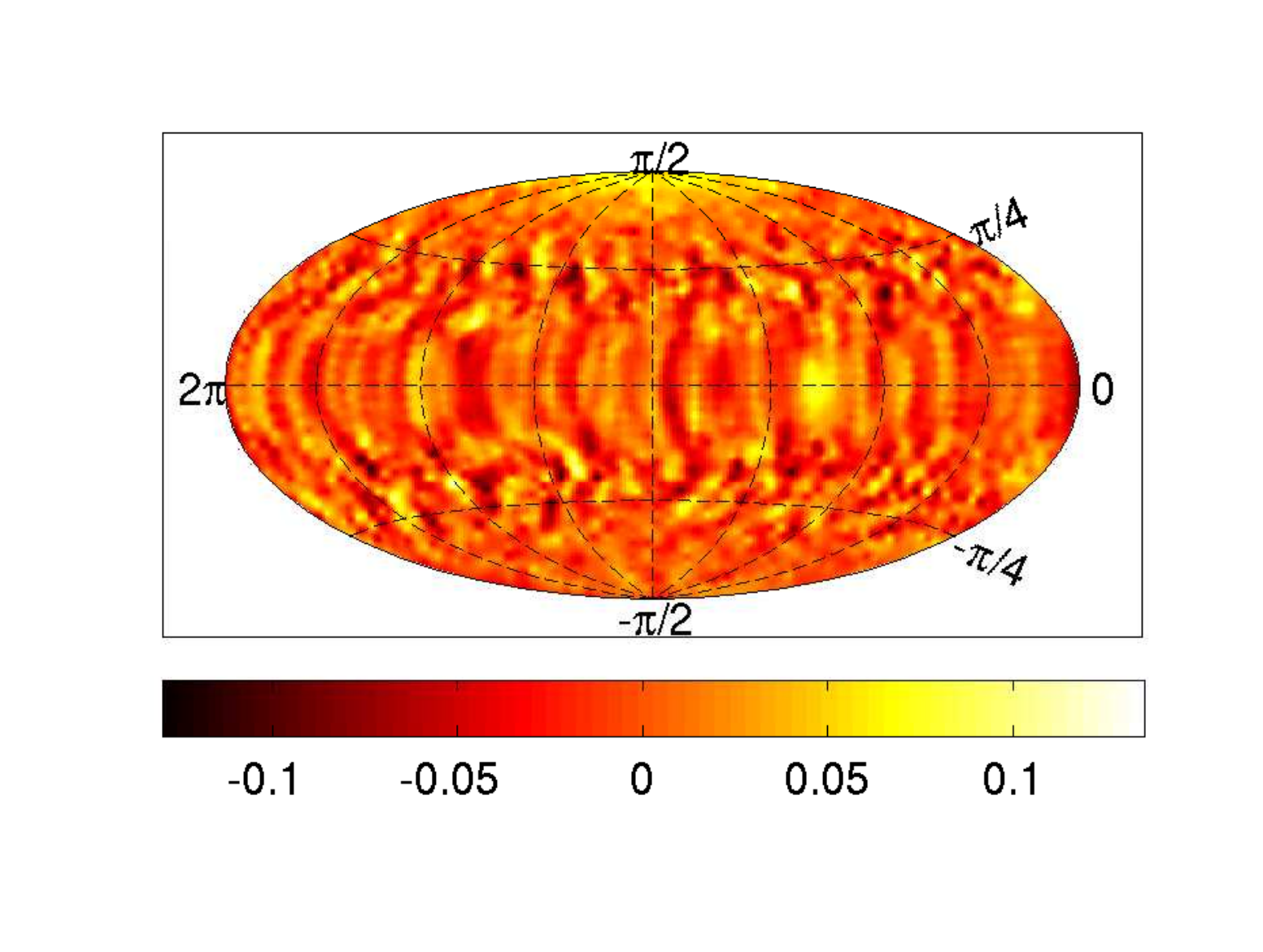}}
\caption{\label{fig:dirtycap}The toy model of a localized source is shown in (a). Dirty maps made from simulated data from three LIGO-Virgo baselines are shown in the last three panels.}
\end{center}
\end{figure}

\begin{figure}[h!]
\begin{center}
\subfigure[~H1L1 (NMSE=0.6642)]
{\includegraphics[width=0.238\textwidth]{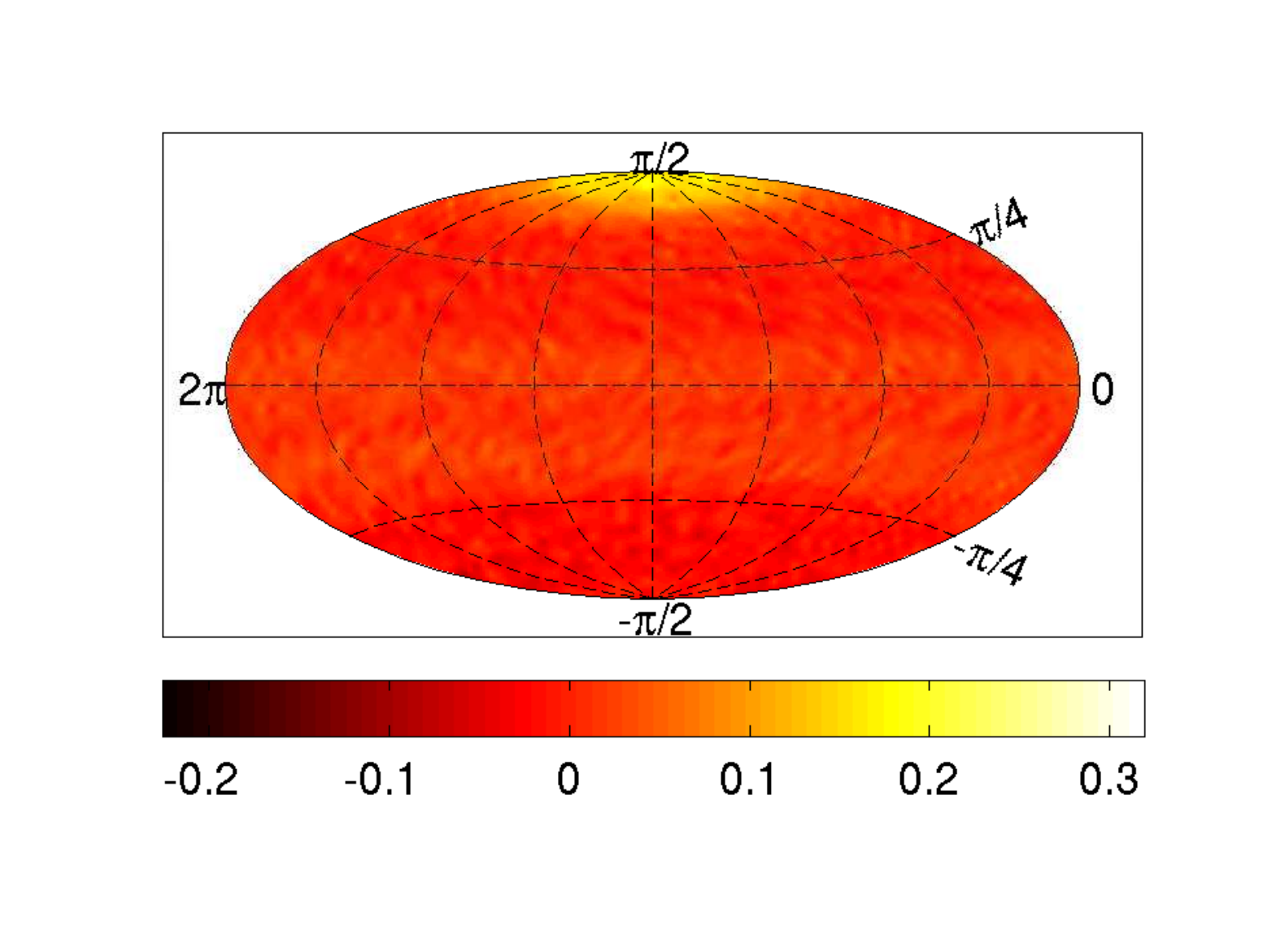}}
\subfigure[~L1V1 (NMSE=1.2939)]
{\includegraphics[width=0.238\textwidth]{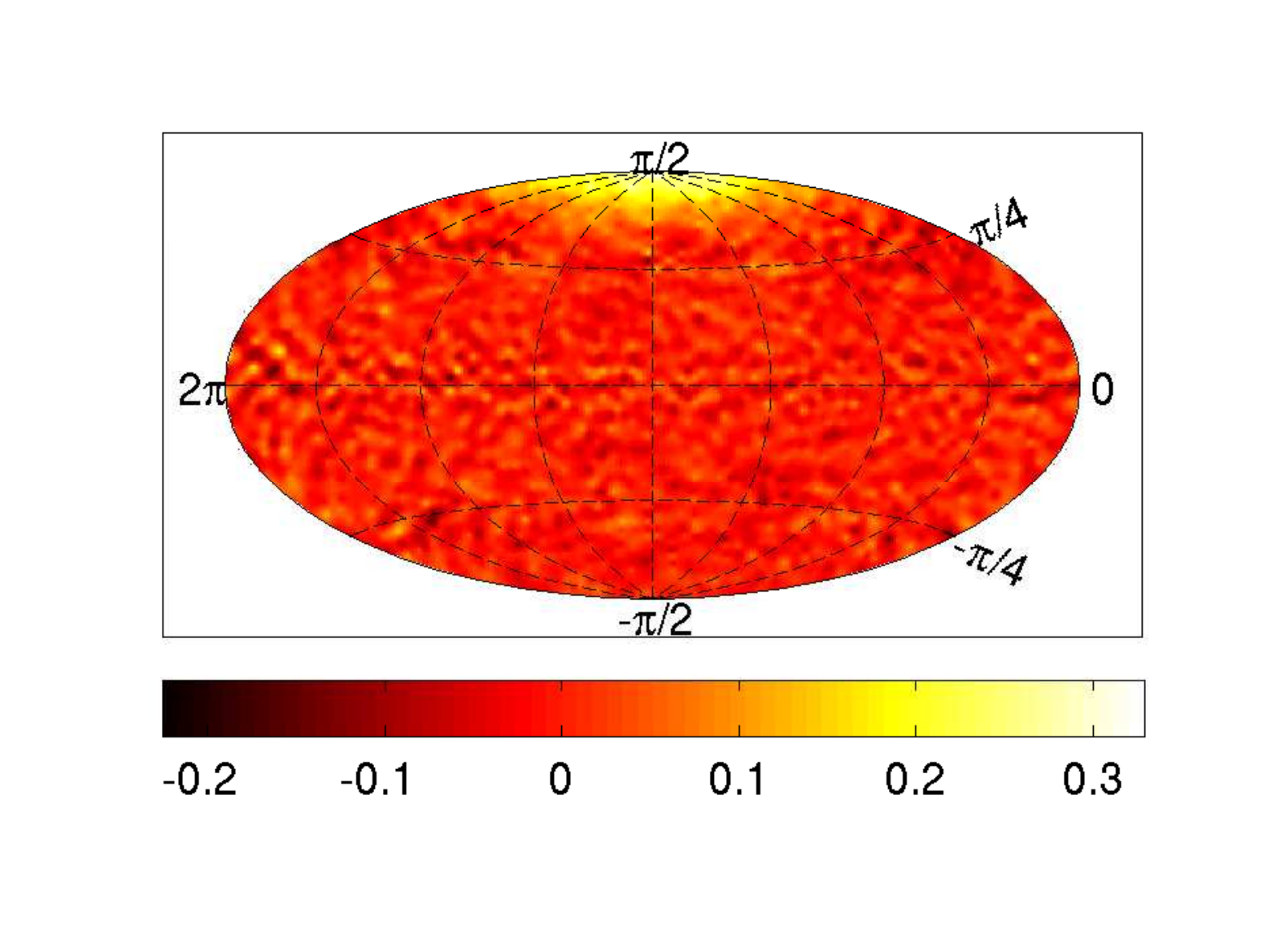}}
\subfigure[~H1V1 (NMSE=1.4750)]
{\includegraphics[width=0.238\textwidth]{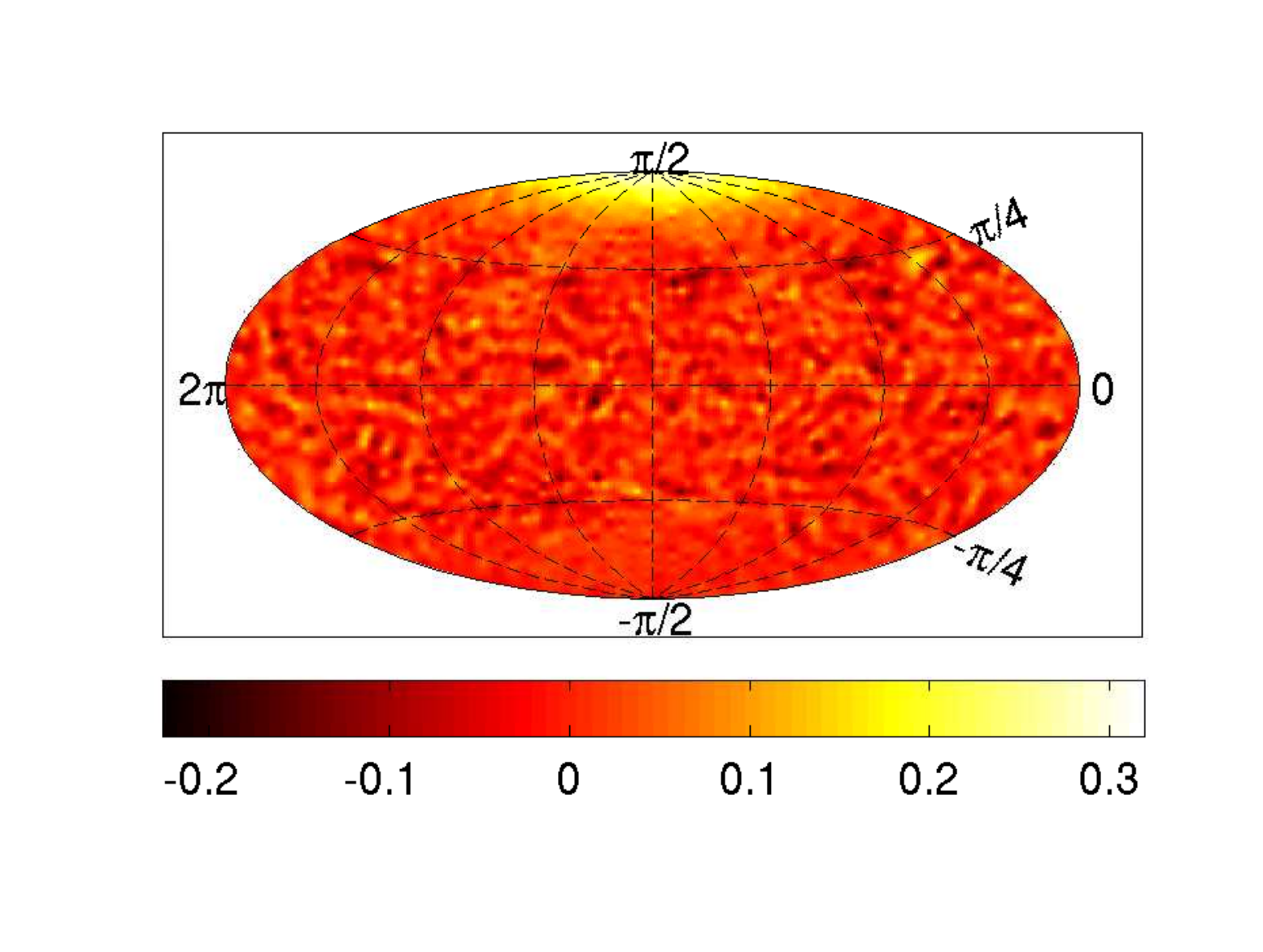}}
\subfigure[~H1L1V1 (NMSE=0.4961)]
{\includegraphics[width=0.238\textwidth]{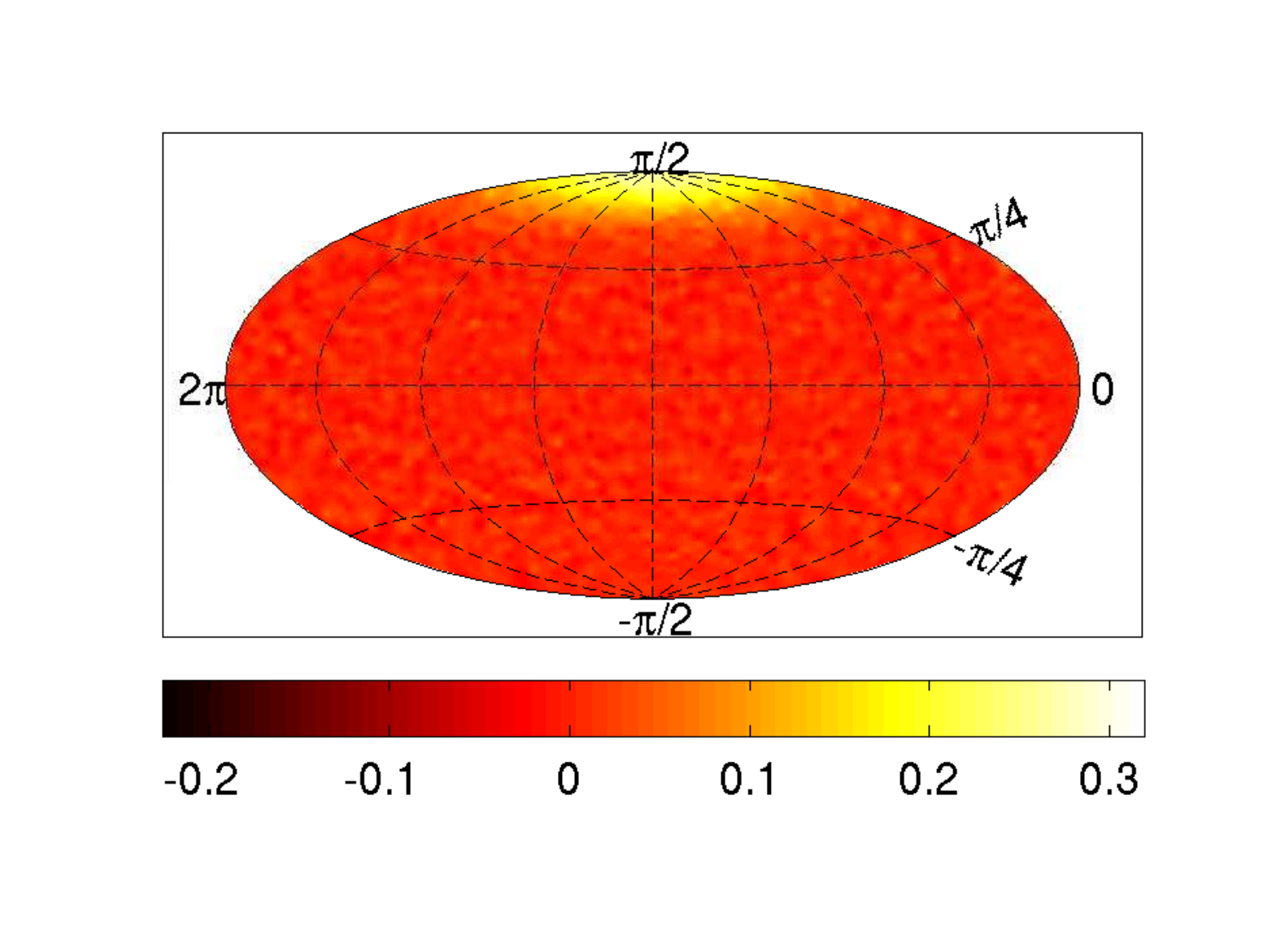}}
\caption{\label{fig:cleancap}Clean maps obtained by the deconvolution of the dirty maps of Fig.~\ref{fig:dirtycap}, using 20 CG iterations, are shown here.}
\end{center}
\end{figure}

\begin{figure}[h!]
\begin{center}
\subfigure[~H1L1]
{\includegraphics[width=0.238\textwidth]{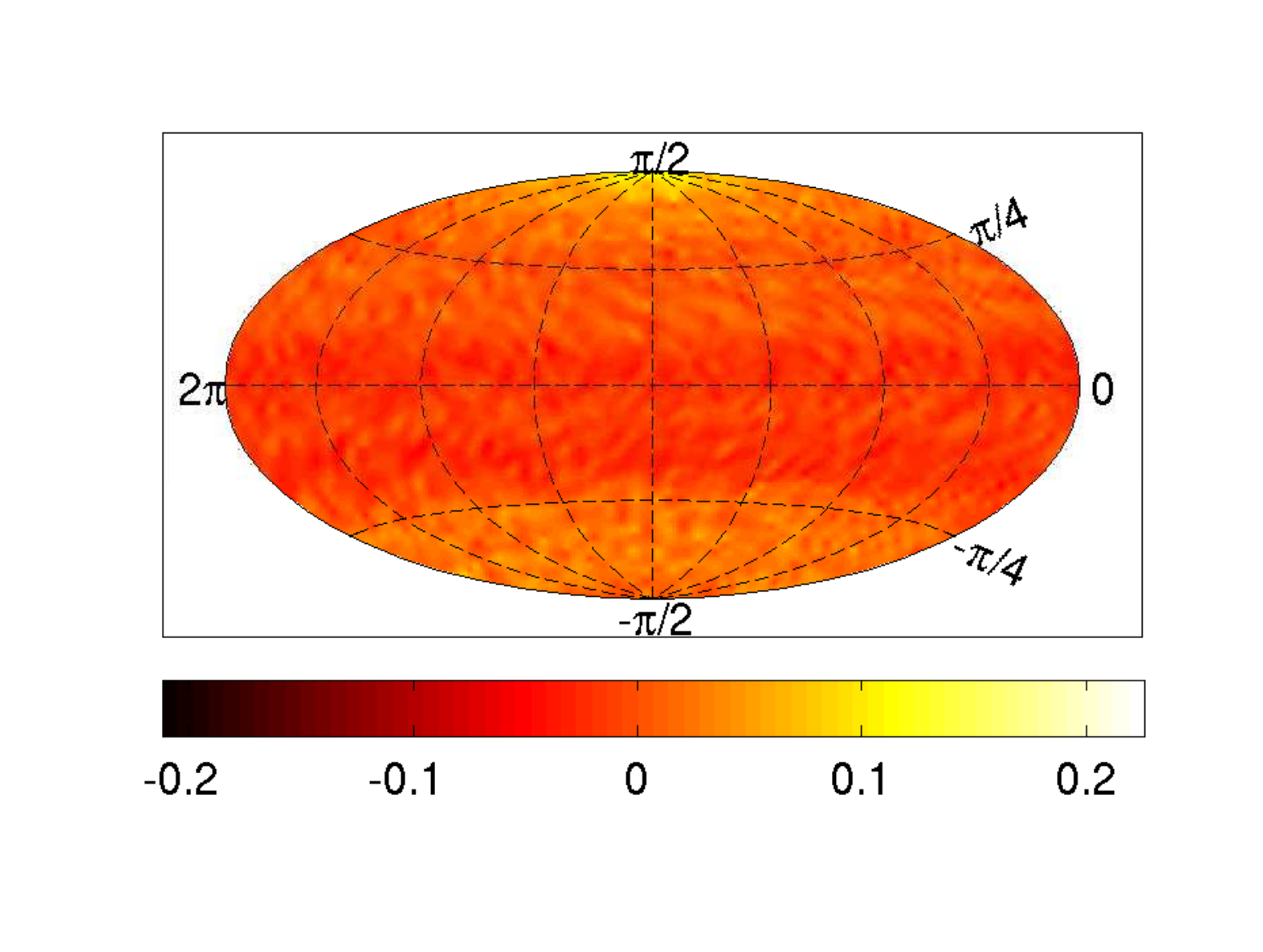}}
\subfigure[~L1V1]
{\includegraphics[width=0.238\textwidth]{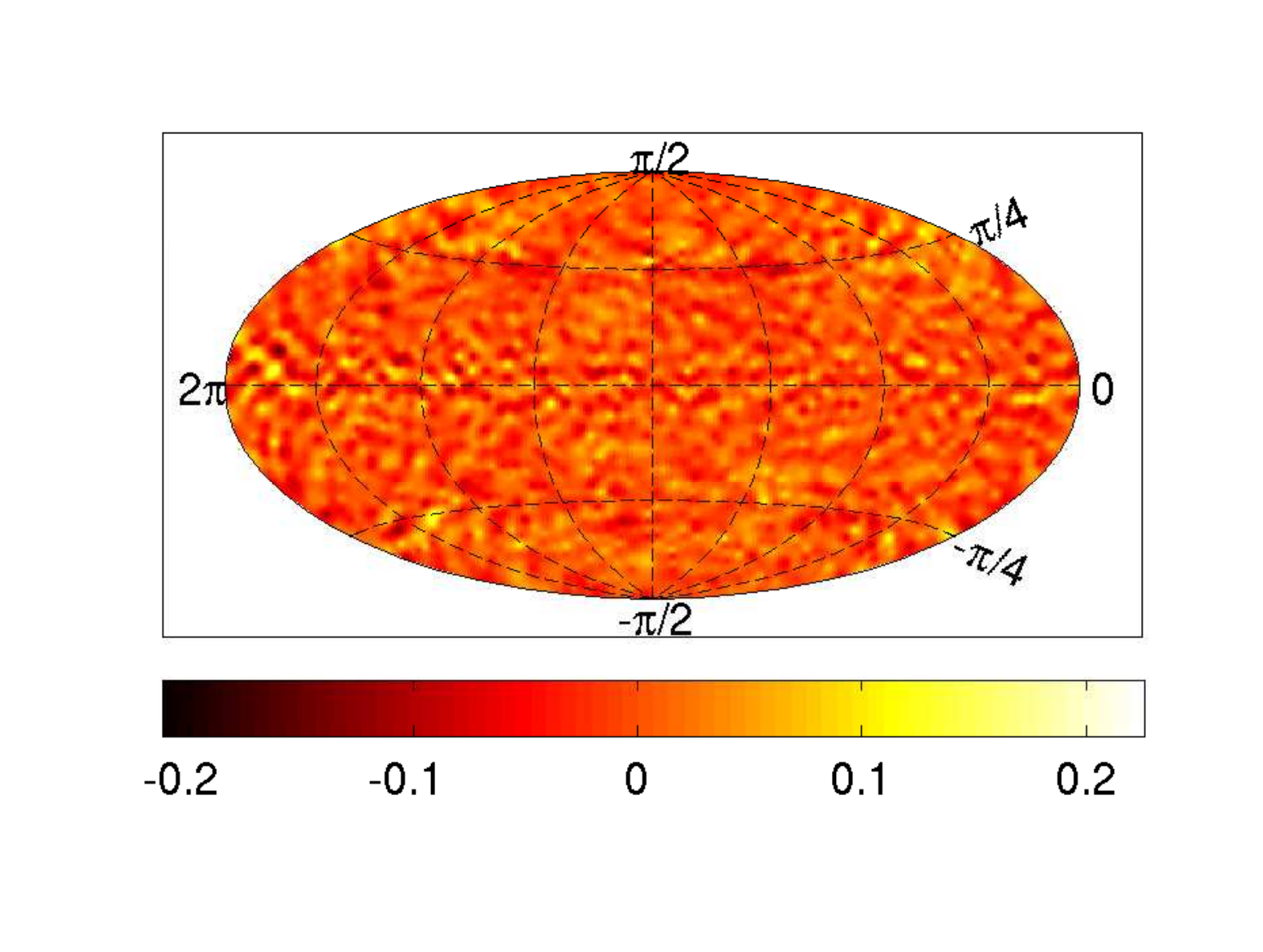}}
\subfigure[~H1V1]
{\includegraphics[width=0.238\textwidth]{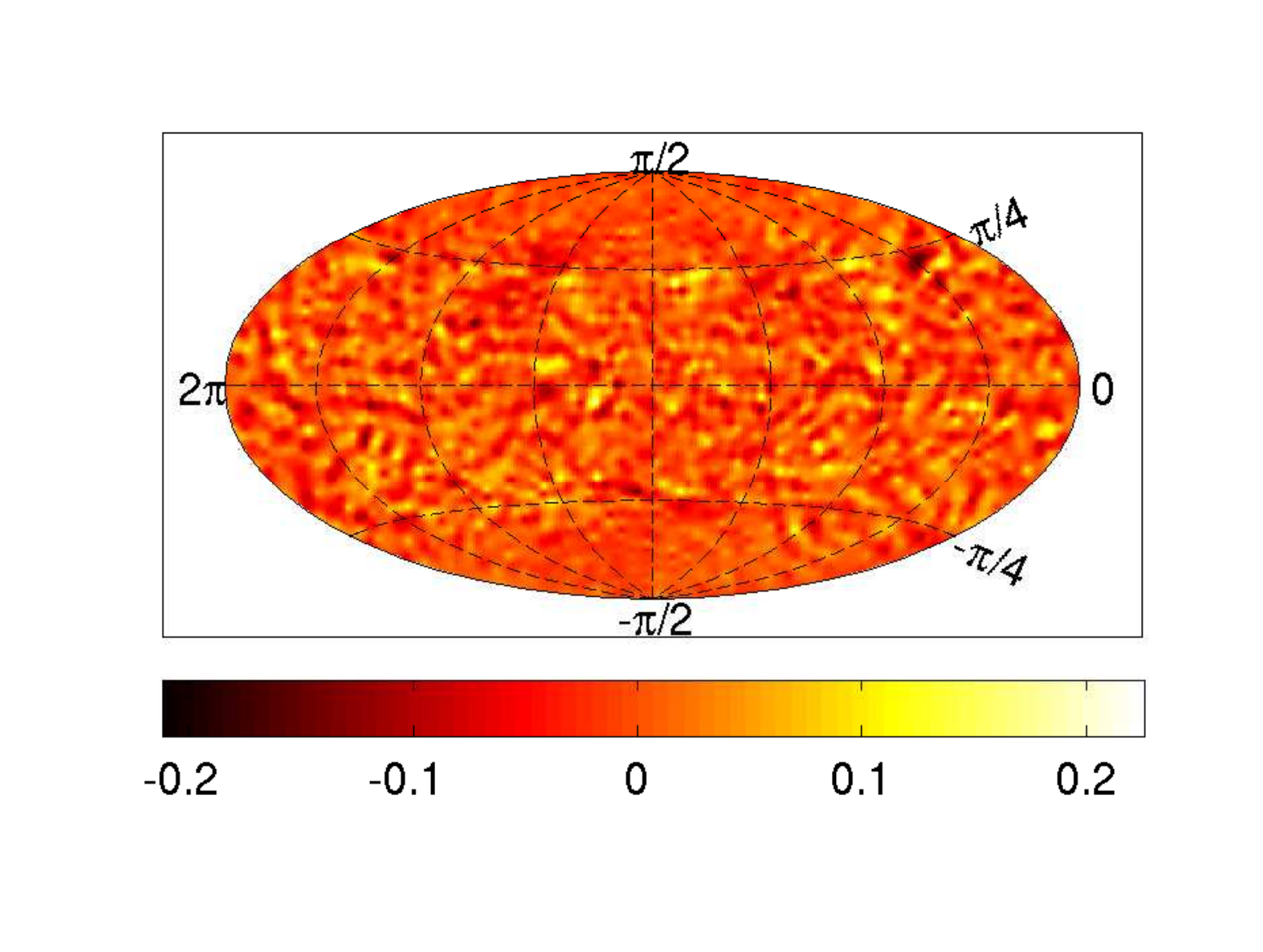}}
\subfigure[~H1L1V1]
{\includegraphics[width=0.238\textwidth]{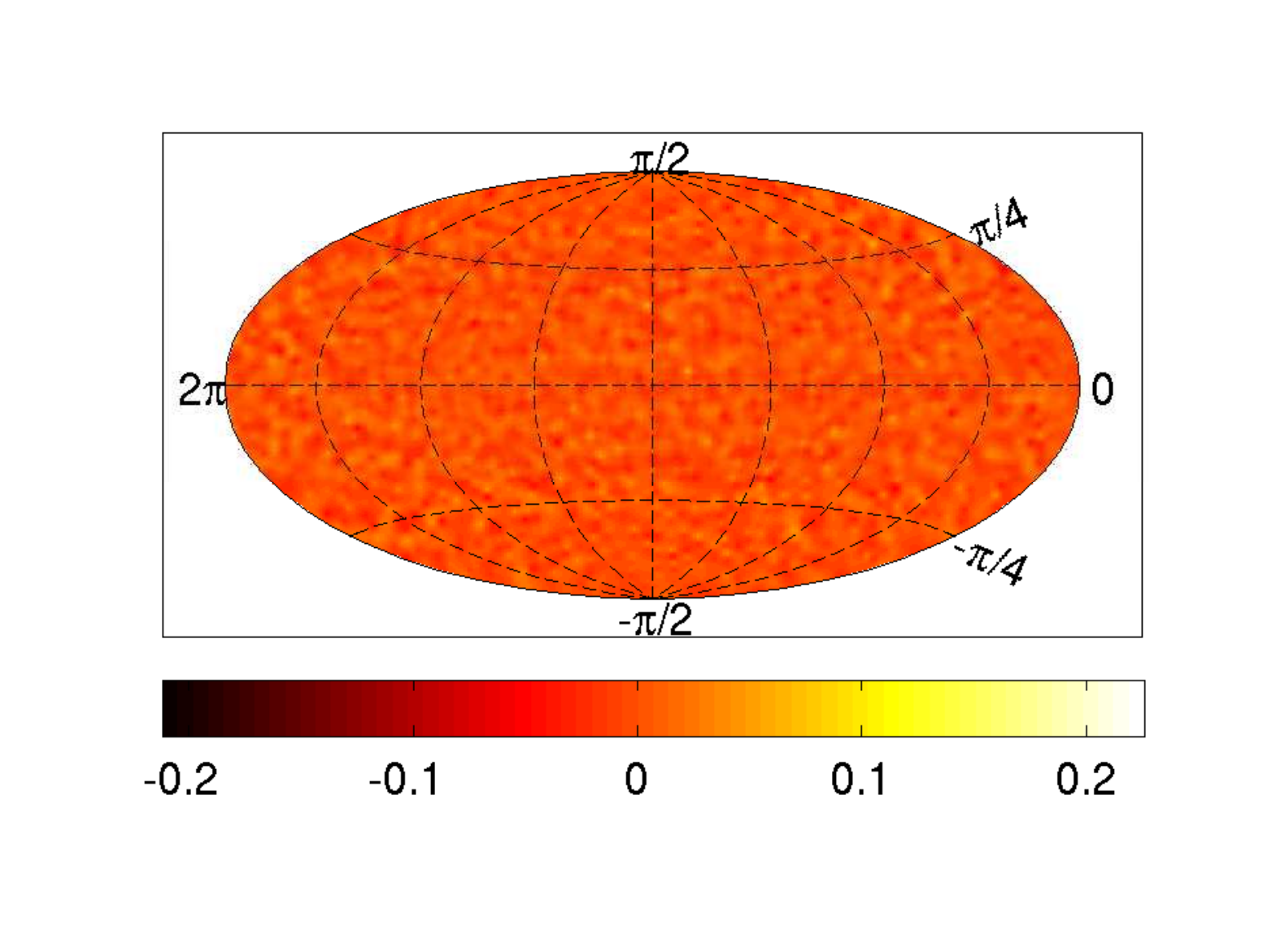}}
\caption{\label{fig:biascap}Difference between the clean maps of Fig.~\ref{fig:cleancap} and the injected map of Fig.~\ref{fig:dirtycap}(a).}
\end{center}
\end{figure}

We performed numerical comparison of map-making performance for two types of toy sky patterns -- (i)~extended, multideclination sky (Figs.~\ref{fig:dirtygal} and \ref{fig:cleangal}), mimicking the (partially masked) image of the sky constructed by the WMAP satellite~\cite{WMAP}, where essentially a modified galactic structure stands out; and (ii)~a relatively localized source peaked at the north pole (Figs.~\ref{fig:dirtycap} and \ref{fig:cleancap}).

In both cases, the dirty maps from different baselines are quite distorted compared to the injected maps. (Compare the last three plots with the first one in Fig. ~\ref{fig:dirtygal} and in Fig.~\ref{fig:dirtycap}). However, the deconvolution procedure yields reasonably resolved maps for all the baselines (Figs.~\ref{fig:cleangal} and \ref{fig:cleancap}), signifying that none of the beam matrices are completely degenerate. The clean maps from the network are, however, of better quality, as can be seen from the corresponding low NMSE. To demonstrate this visually, we also show the difference between the clean and injected maps for the respective cases in Figs.~\ref{fig:biasgal} and \ref{fig:biascap}. As expected, the difference maps for the network look less noisy and uniform across the sky than the individual baselines.

\subsubsection{Maximized-likelihood-ratio statistic}

We finally compute the MLR statistic introduced in Section~\ref{subsec:statistic} as a figure of merit and, also, to demonstrate how this statistic can be powerful in identifying signal in noisy maps.
The MLR statistic for both dirty and clean maps for the two types of sources considered here have been listed in Tables~\ref{table:CMB} and \ref{table:CAP}. 

\begin{table}[ht]
\caption{MLR statistic of dirty maps ($\lambda$) versus clean maps ($\lambda_{c}$) for the simulated maps in Figs.~\ref{fig:dirtygal} and~\ref{fig:cleangal}.}
\centering
\begin{tabular}{c| c| c}
\hline\hline
Baseline & $\lambda$ & $\lambda_{c}$ \\ [0.5ex]
\hline
H1L1 & 785.555 & 783.271 \\
L1V1 & 359.004 & 358.940 \\
H1V1 & 315.717 & 315.662 \\
H1L1V1 & 919.594 & 917.600 \\ [1ex]
\hline
\end{tabular}
\label{table:CMB}
\end{table}

\begin{table}[ht]
\caption{MLR statistic of dirty maps ($\lambda$) versus clean maps ($\lambda_{c}$) for the simulated maps in Figs.~\ref{fig:dirtycap} and~\ref{fig:cleancap}.}
\centering
\begin{tabular}{c| c| c}
\hline\hline
Baseline & $\lambda$ & $\lambda_{c}$ \\ [0.5ex]
\hline
H1L1 & 284.652 & 284.173 \\
L1V1 & 39.308 & 39.377 \\
H1V1 & 64.129 & 64.113 \\
H1L1V1 & 294.419 & 293.961 \\ [1ex]
\hline
\end{tabular}
\label{table:CAP}
\end{table}

It is intriguing to note that, for both dirty and clean maps, one obtains similar values. This suggests that a deconvolution effected with only a few tens of conjugate-gradient basis vectors does not cause a significant amount of information loss.

\begin{figure}[h!]
\begin{center}
\subfigure[~H1L1]
{\includegraphics[width=0.238\textwidth]{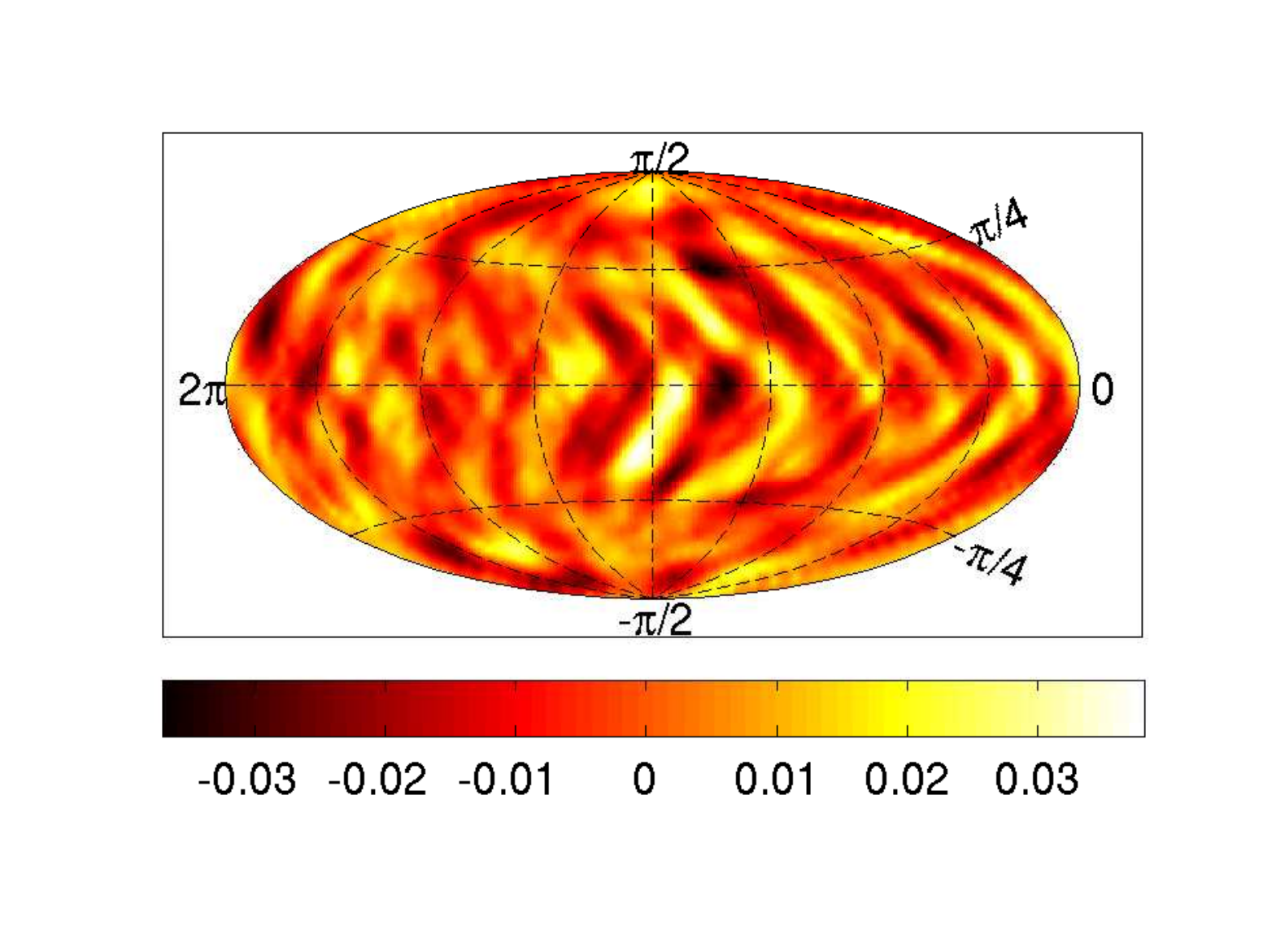}}
\subfigure[~L1V1]
{\includegraphics[width=0.238\textwidth]{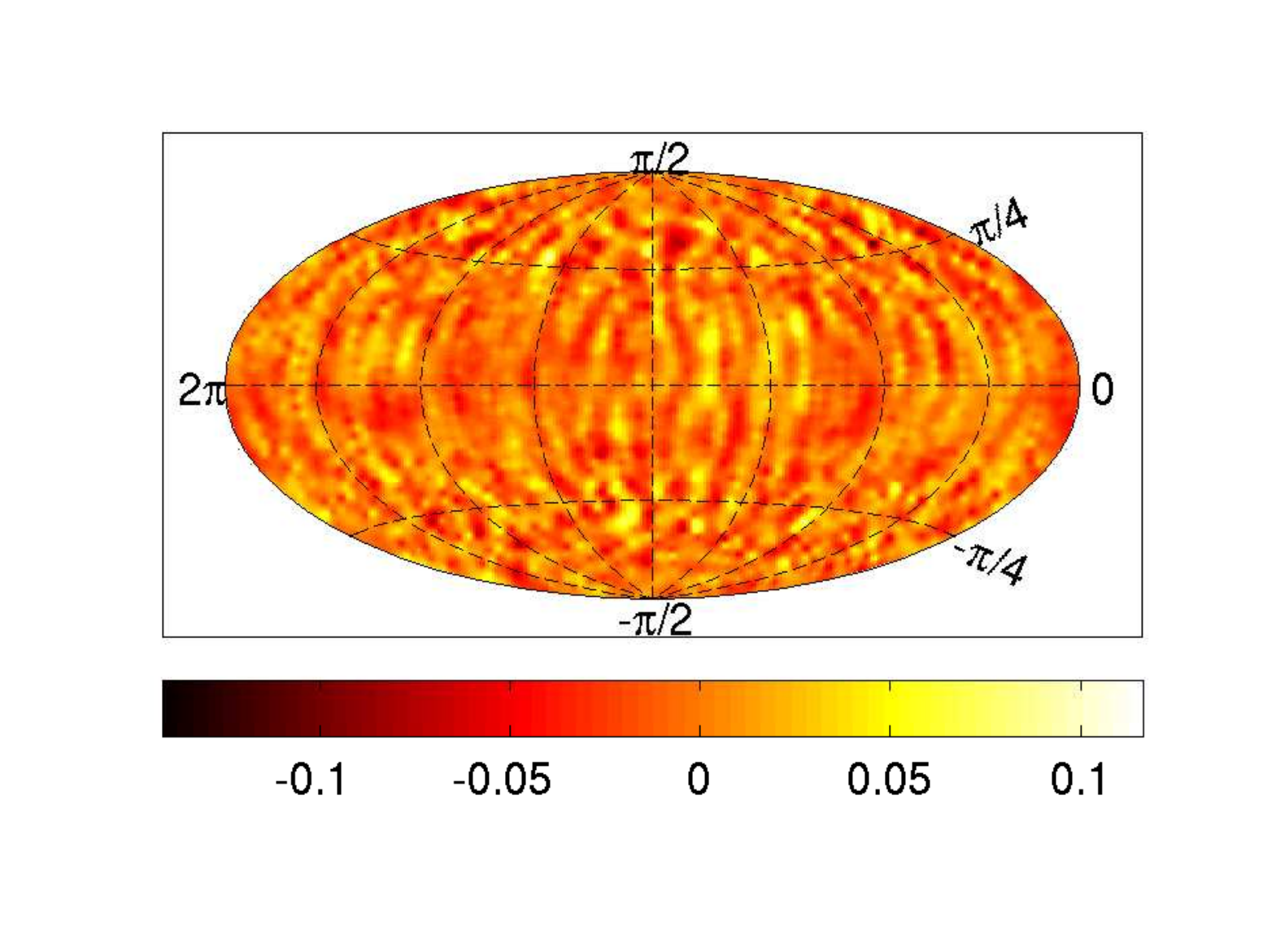}}
\subfigure[~H1V1]
{\includegraphics[width=0.238\textwidth]{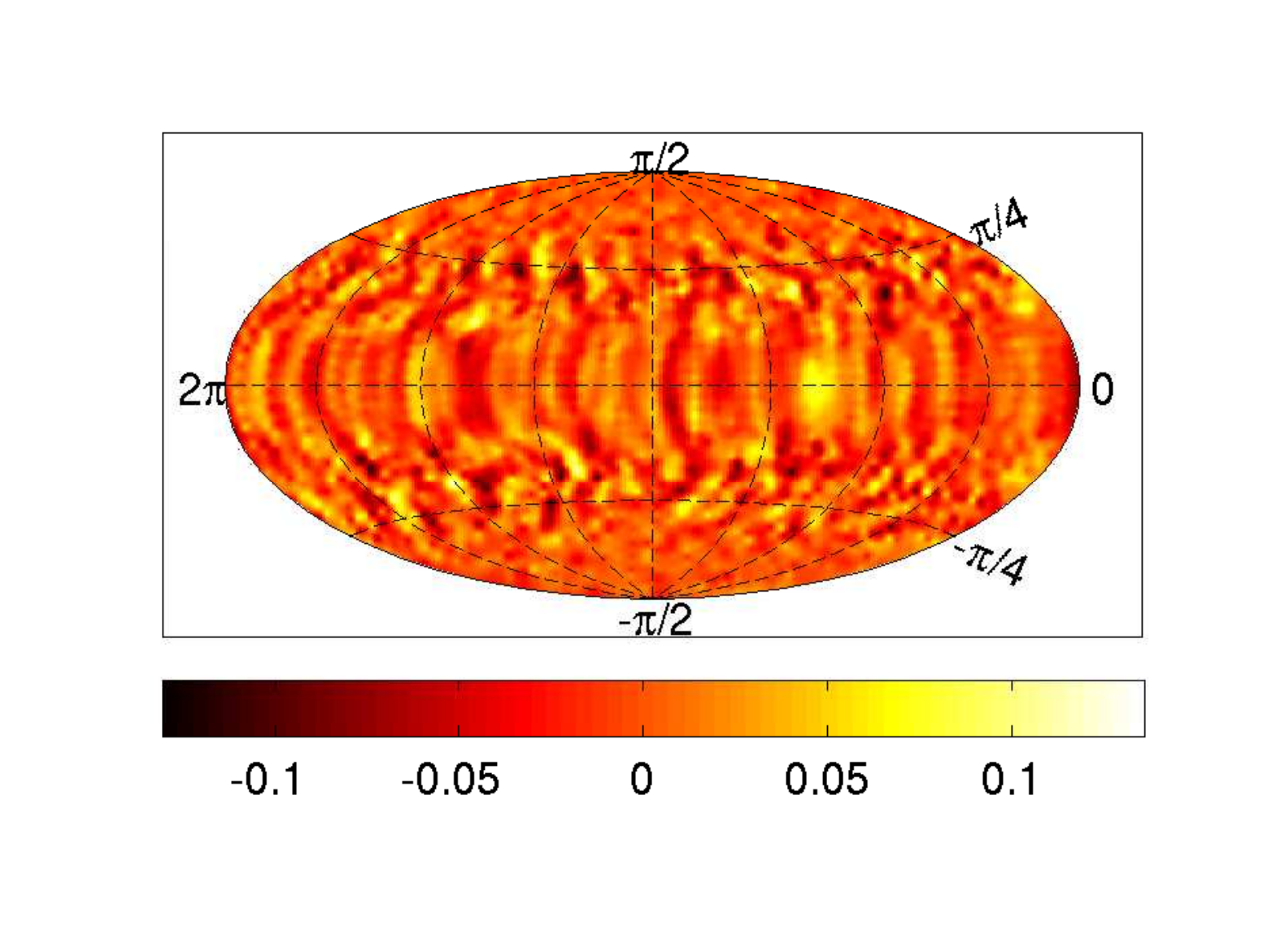}}
\caption{\label{fig:dirtynoise}Dirty maps made for simulated noise, without any injected signals.}
\end{center}
\end{figure}

\begin{figure}[h!]
\begin{center}
\subfigure[~H1L1]
{\includegraphics[width=0.238\textwidth]{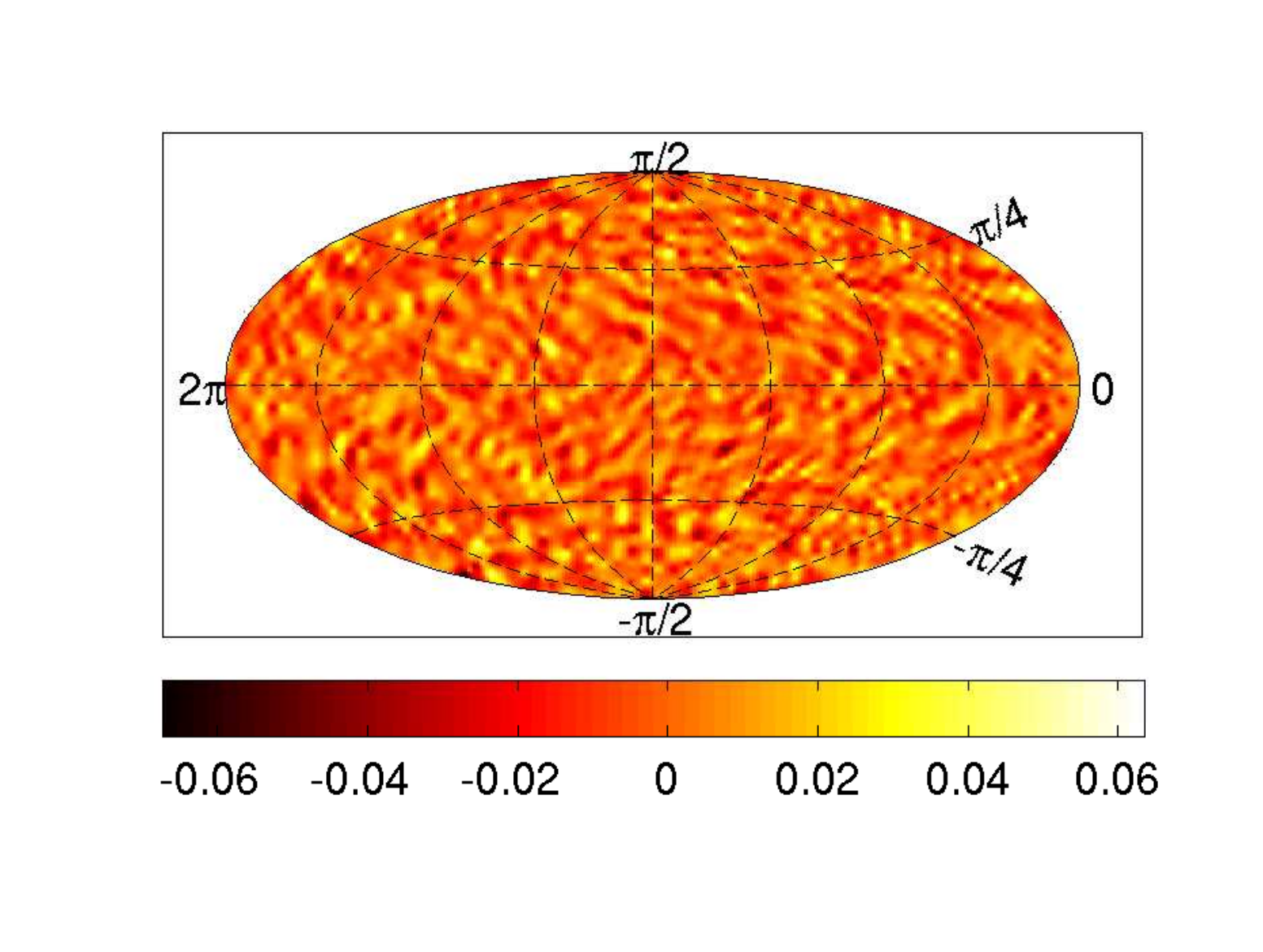}}
\subfigure[~L1V1]
{\includegraphics[width=0.238\textwidth]{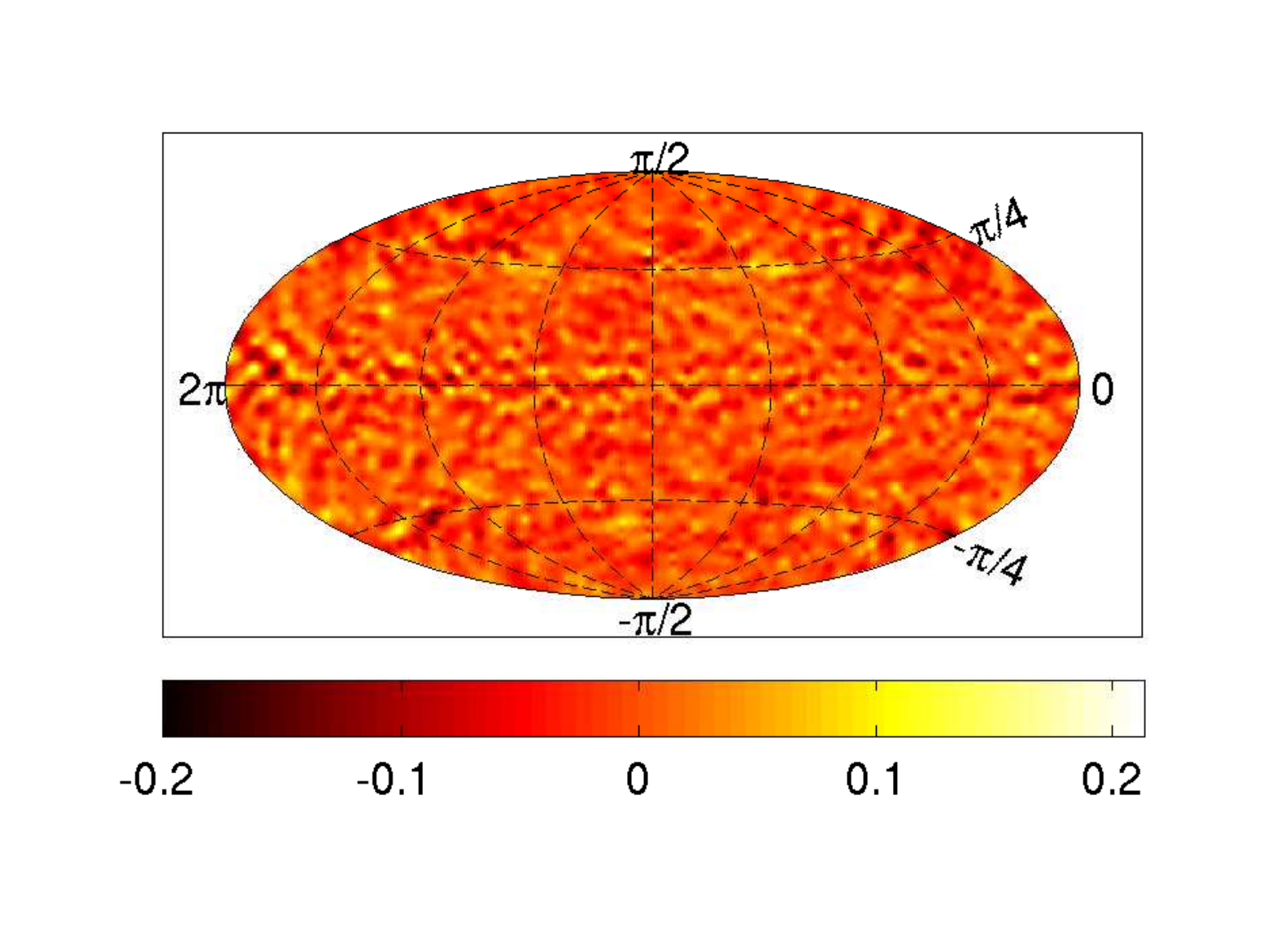}}
\subfigure[~H1V1]
{\includegraphics[width=0.238\textwidth]{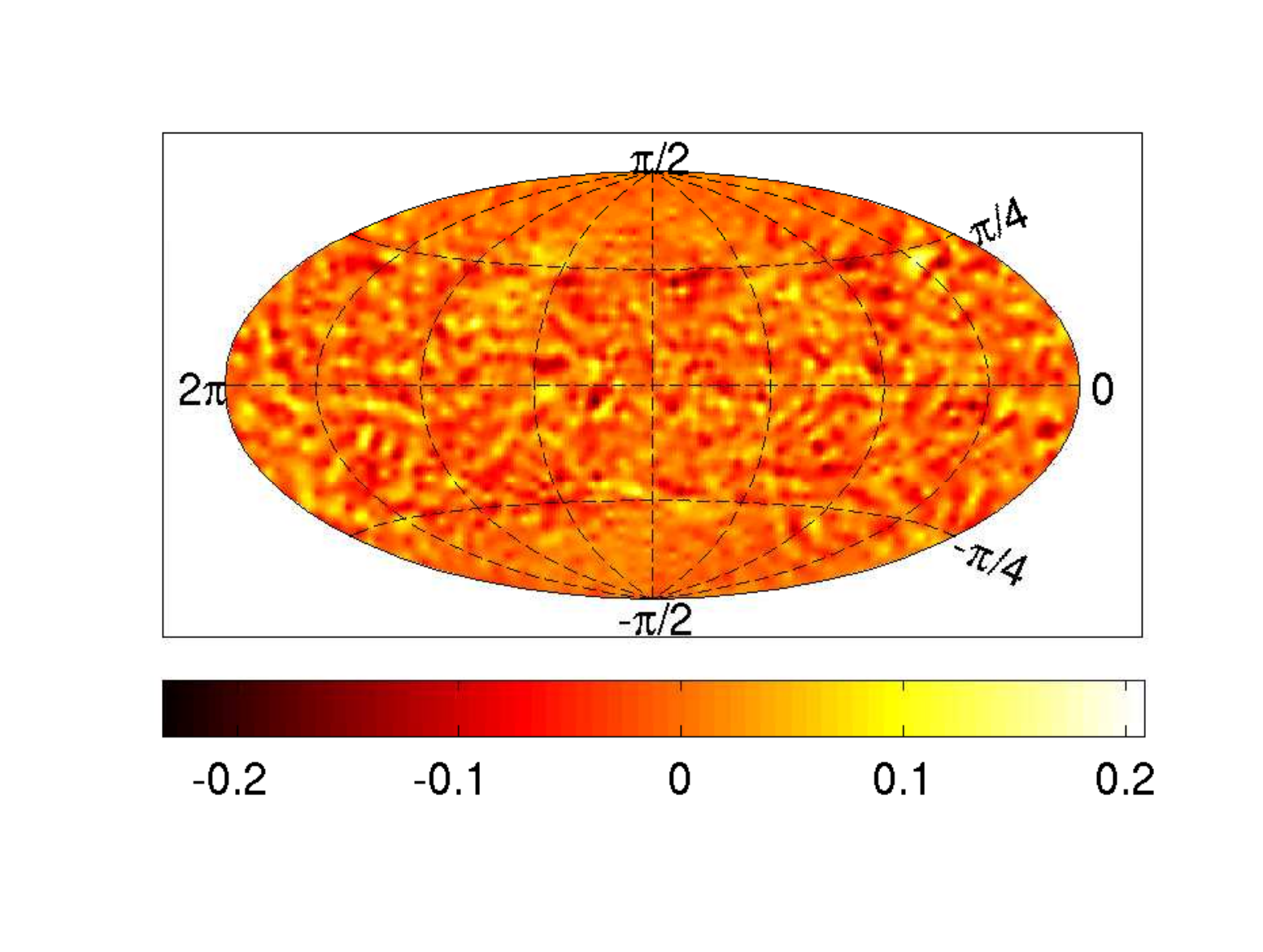}}
\subfigure[~H1L1V1]
{\includegraphics[width=0.238\textwidth]{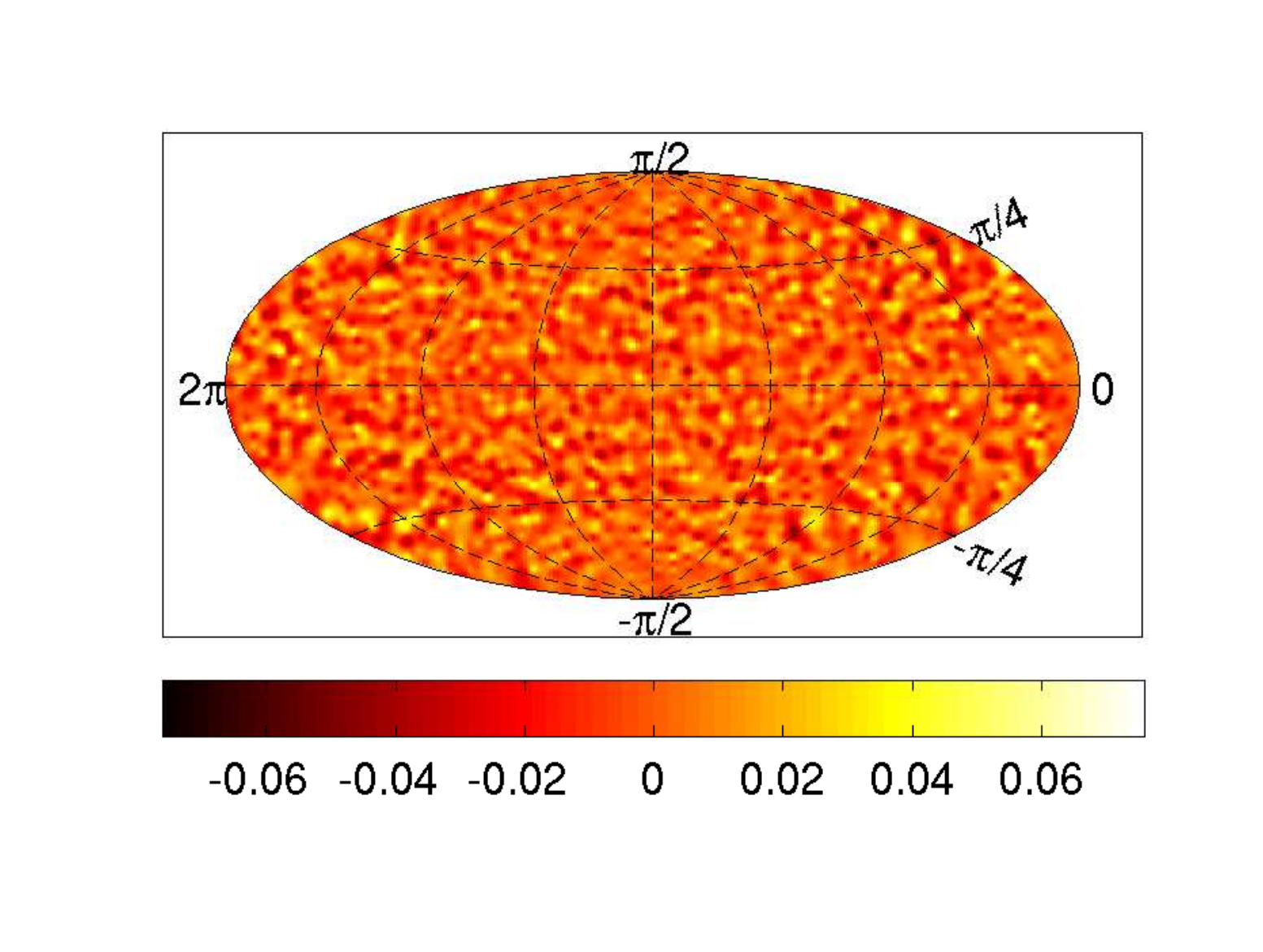}}
\caption{\label{fig:cleannoise}Clean maps obtained by the deconvolution of the dirty maps of Fig.~\ref{fig:dirtynoise}, using 20 CG iterations.}
\end{center}
\end{figure}

To understand the significance of the MLR statistic in the present context, we perform two more exercises. First, we study the no-injection case; that is, we make dirty maps of simulated noise (Fig.~\ref{fig:dirtynoise}), deconvolve it (Fig.~\ref{fig:cleannoise}), and obtain its MLR (Table~\ref{table:Noise}). The similarity of the values of this statistic for the dirty and clean maps proves the unitarity of our deconvolution method.
\begin{table}[ht]
\caption{MLR statistic of dirty maps ($\lambda$) versus clean maps ($\lambda_{c}$) for simulated noise (which actually has an extremely weak signal added) in Figs.~\ref {fig:dirtynoise} and~\ref{fig:cleannoise}.}
\centering
\begin{tabular}{c| c| c}
\hline\hline
Baseline & $\lambda$ & $\lambda_{c}$ \\ [0.5ex]
\hline
H1L1 & 0.512 & 0.433 \\
L1V1 & -1.549 & -1.542 \\
H1V1 & 1.105 & 1.120 \\
H1L1V1 & 0.208 & 0.149 \\ [1ex]
\hline
\end{tabular}
\label{table:Noise}
\end{table}
One can see that the MLR statistic (Table~\ref{table:Noise}) is small ($\approx 1$) in all these cases. We then introduce a small signal -- the same as the extended (galaxylike) source considered before, but at a much reduced strength.
\begin{figure}[h!]
\begin{center}
\subfigure[~Injected map]
{\includegraphics[width=0.238\textwidth]{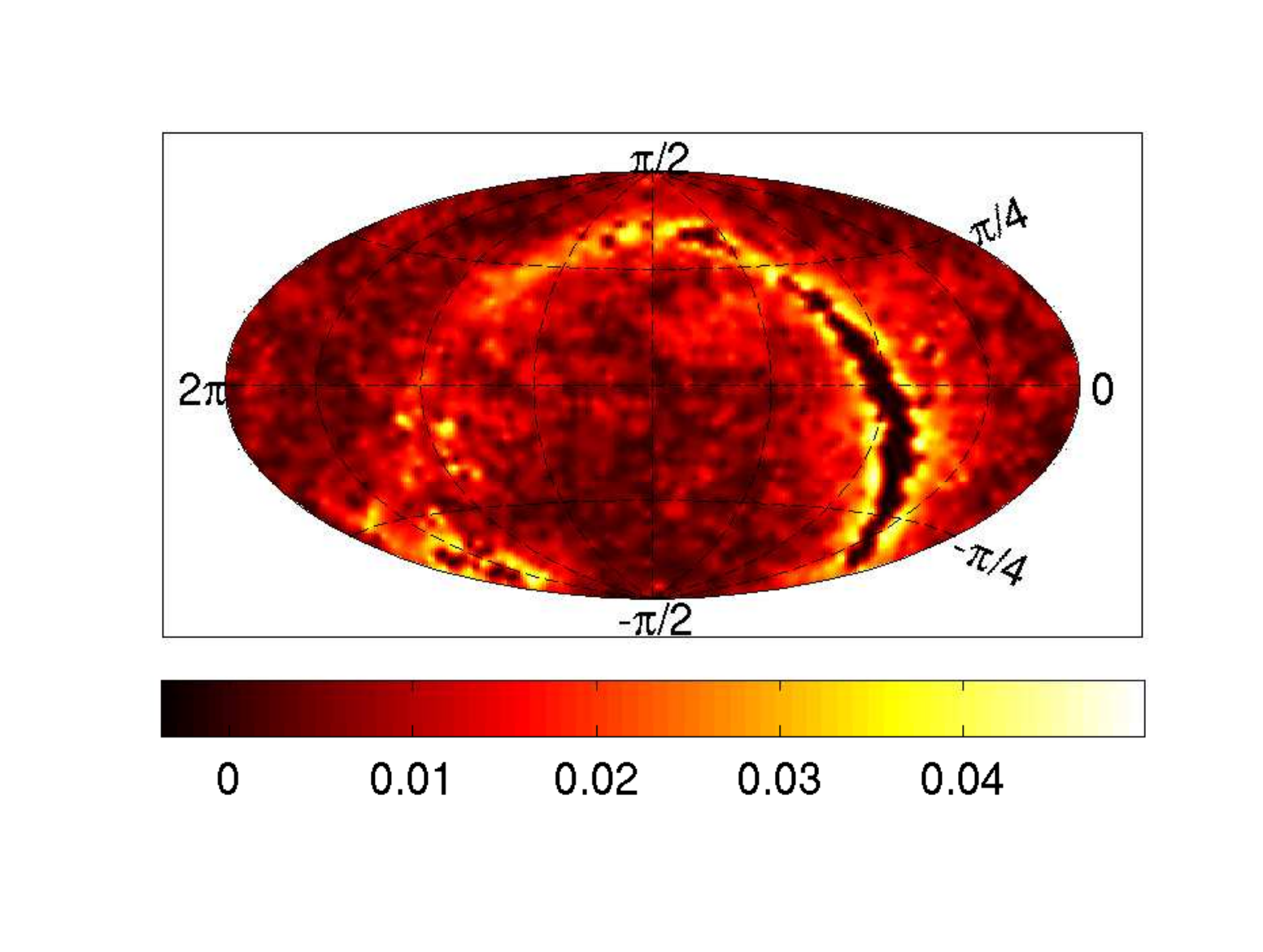}}
\subfigure[~H1L1]
{\includegraphics[width=0.238\textwidth]{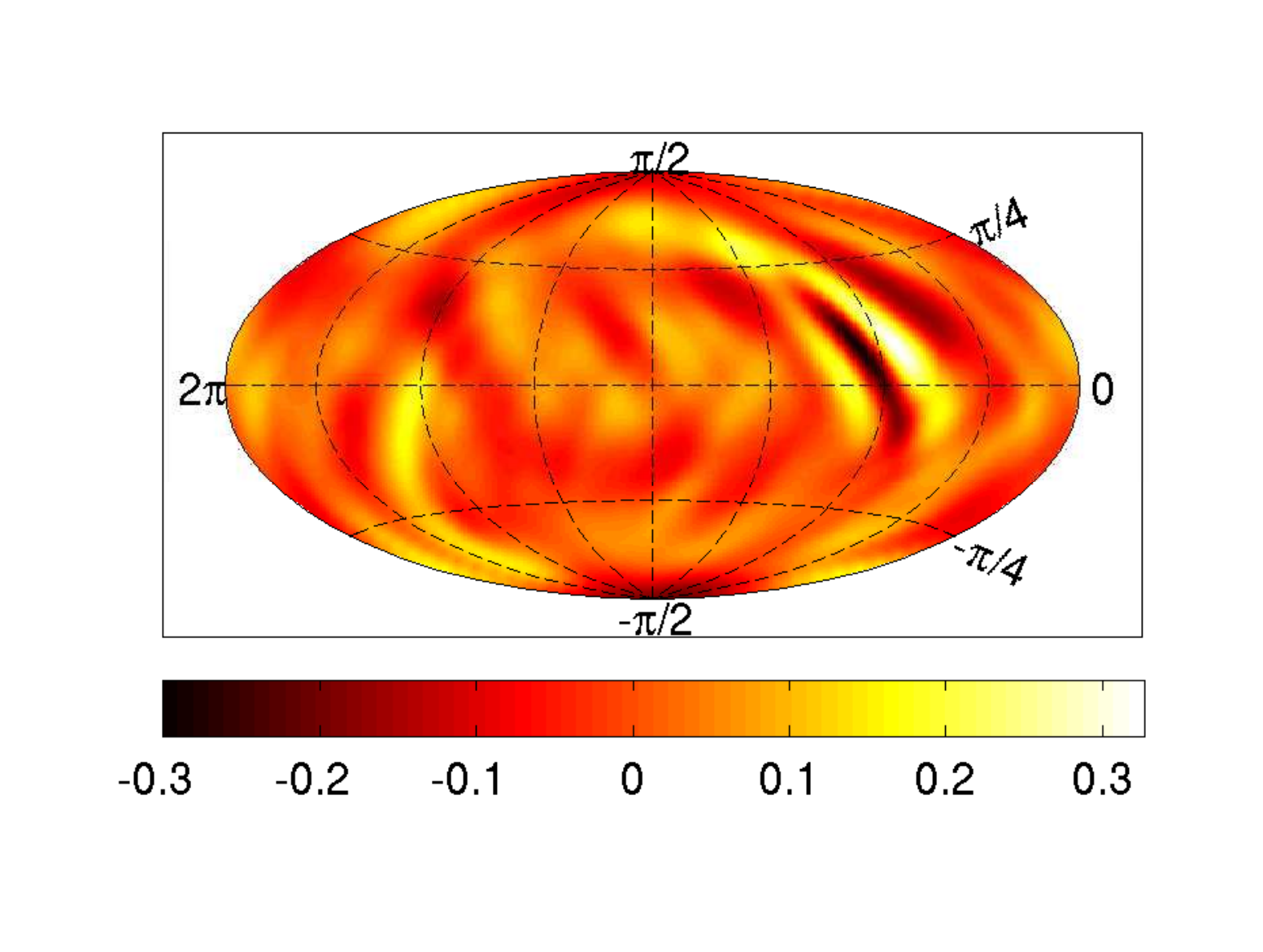}}
\subfigure[~L1V1]
{\includegraphics[width=0.238\textwidth]{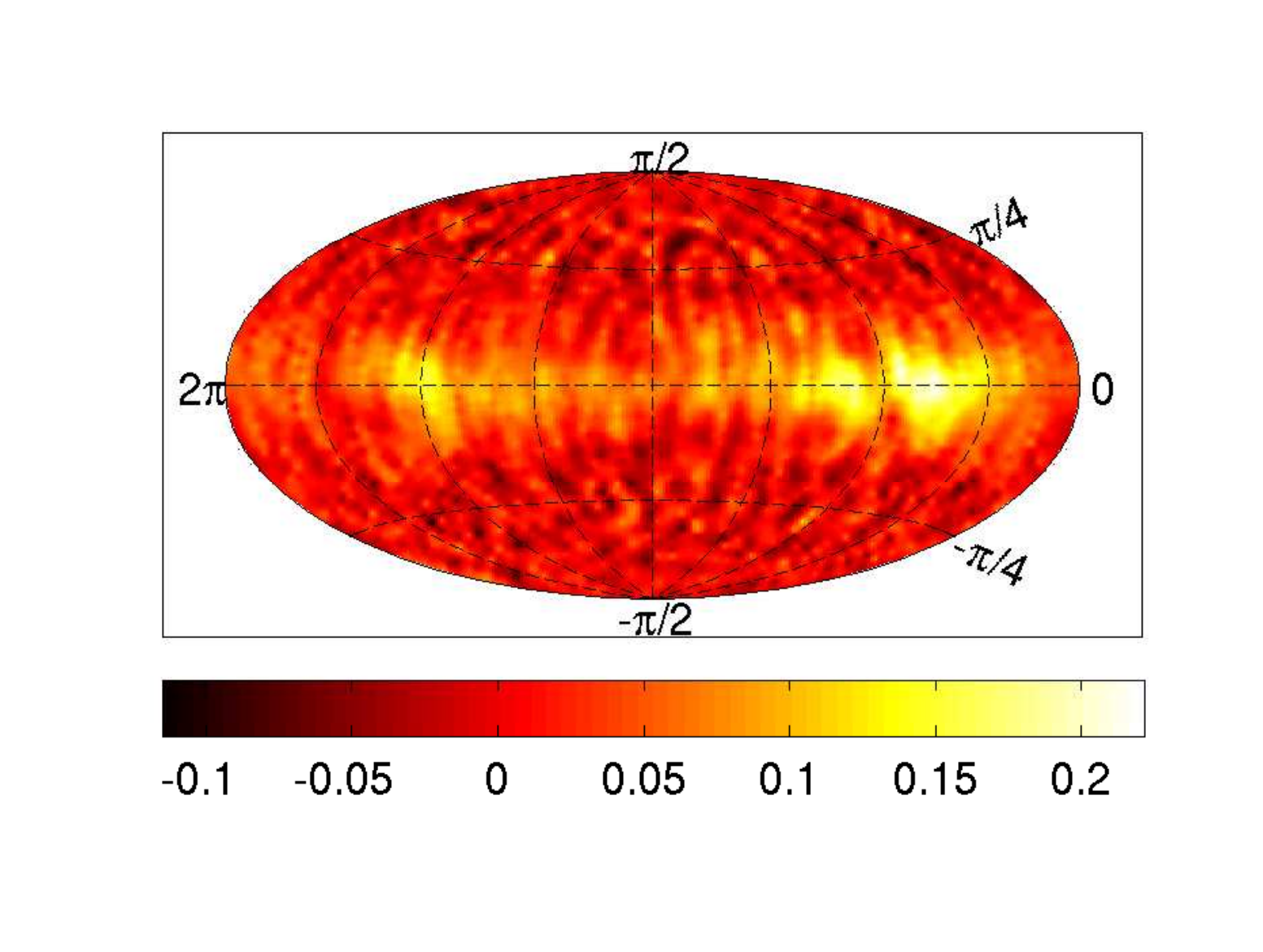}}
\subfigure[~H1V1]
{\includegraphics[width=0.238\textwidth]{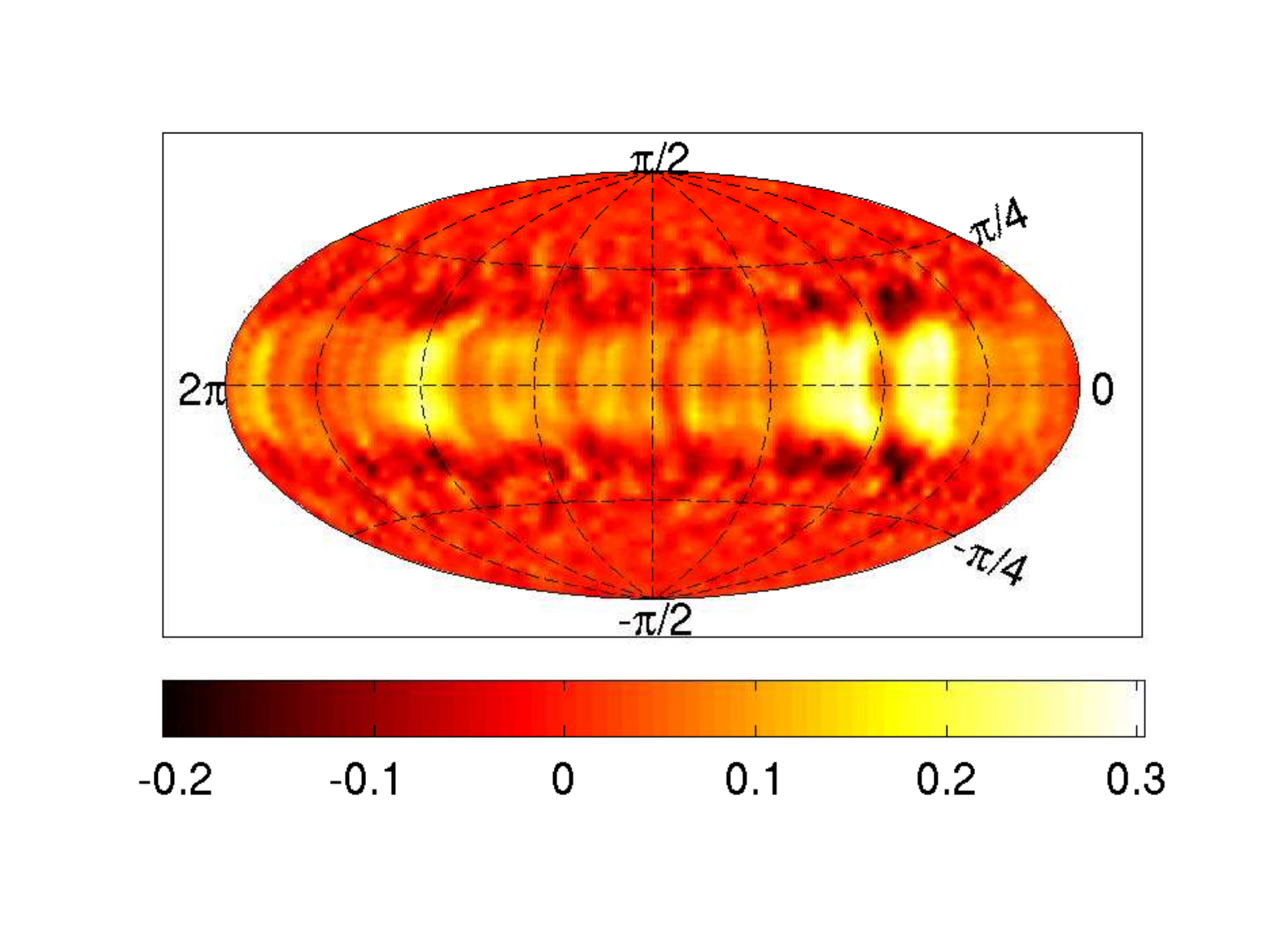}}
\caption{\label{fig:dirtyweak}The toy model of a very weak extended source is shown in (a). Dirty maps made from simulated data from three LIGO-Virgo baselines are shown in the last three panels.}
\end{center}
\end{figure}
\begin{figure}[h!]
\begin{center}
\subfigure[~H1L1 (NMSE=1.0694)]
{\includegraphics[width=0.238\textwidth]{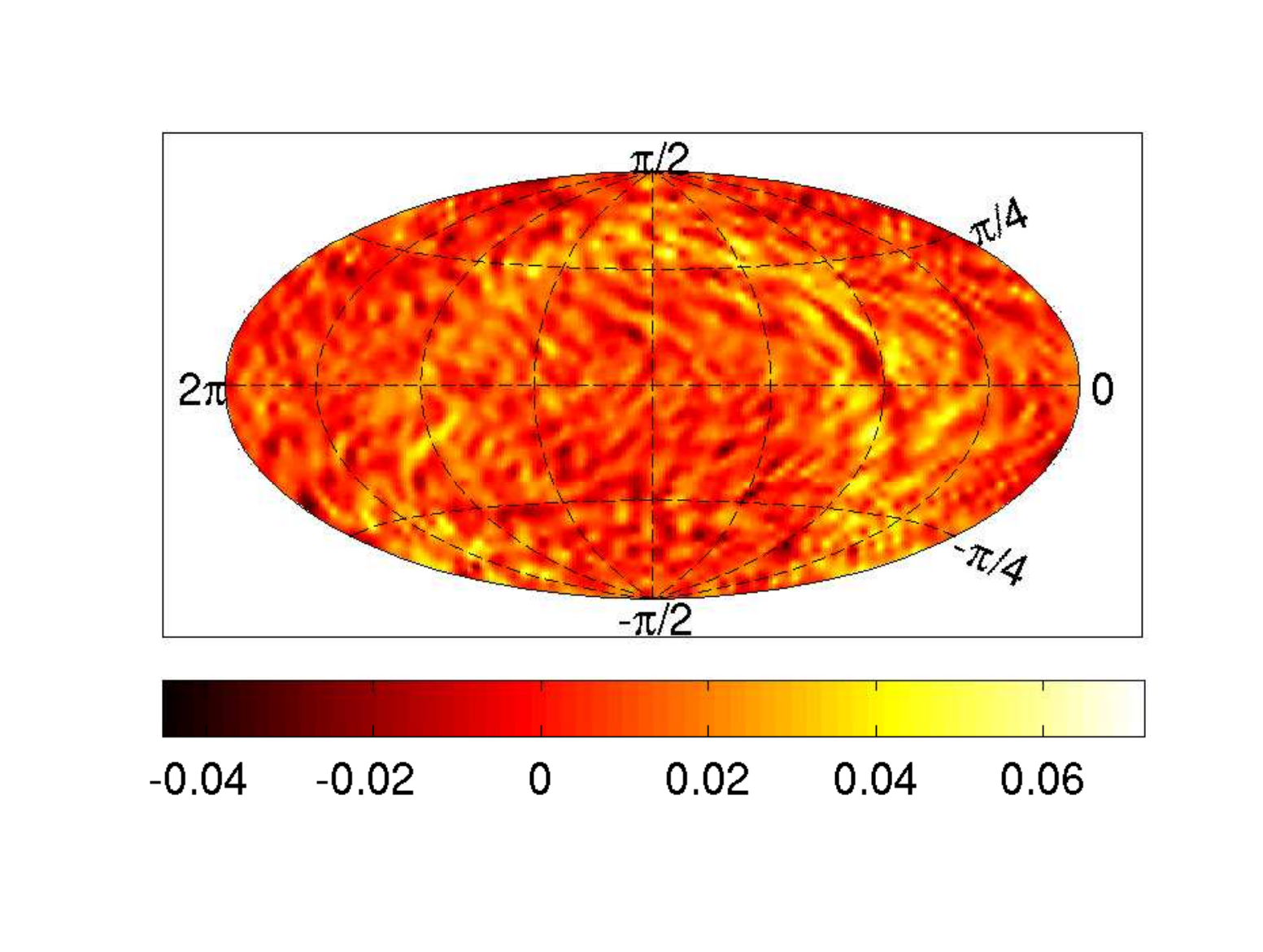}}
\subfigure[~L1V1 (NMSE=3.3557)]
{\includegraphics[width=0.238\textwidth]{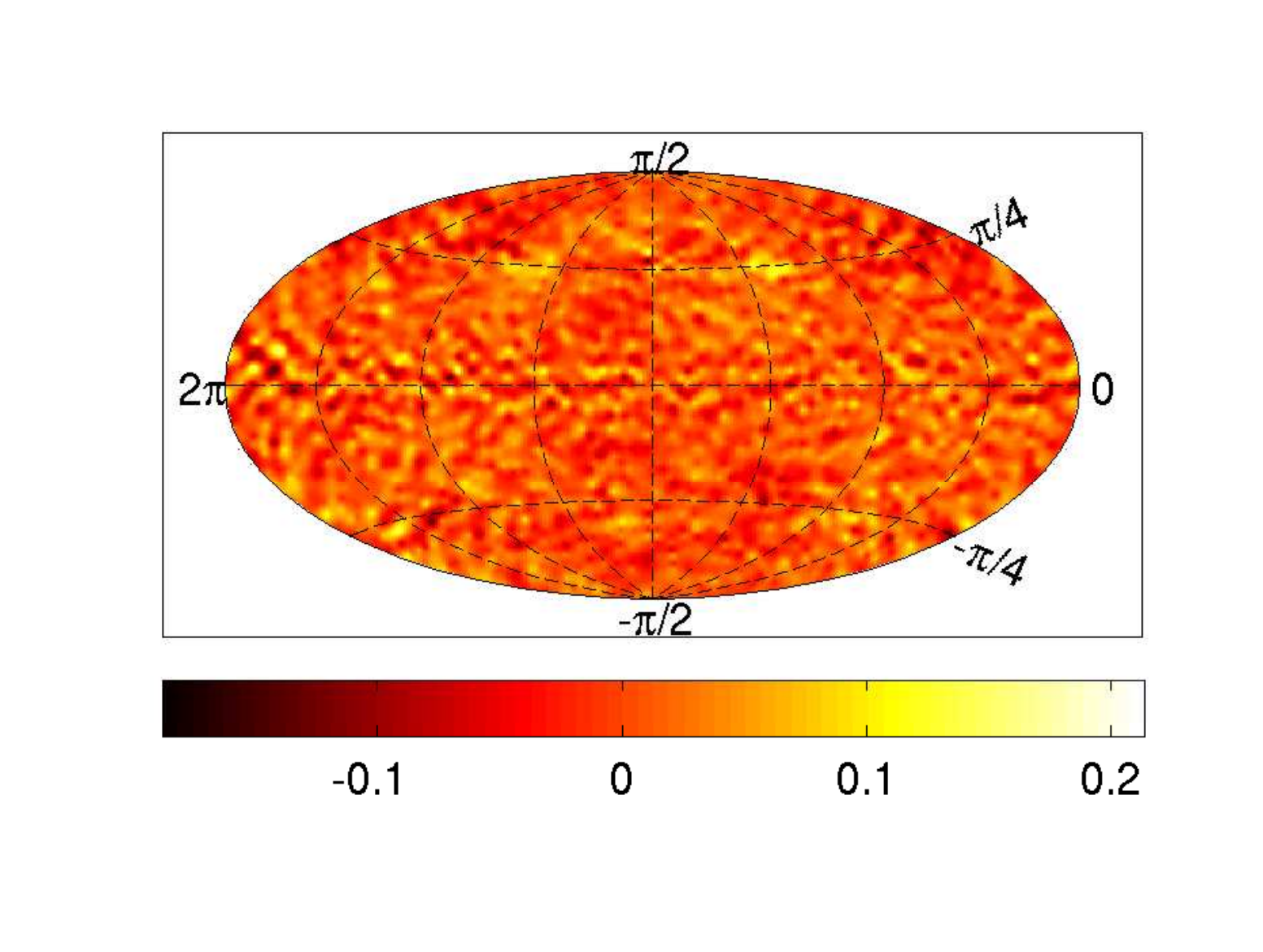}}
\subfigure[~H1V1 (NMSE=3.8406)]
{\includegraphics[width=0.238\textwidth]{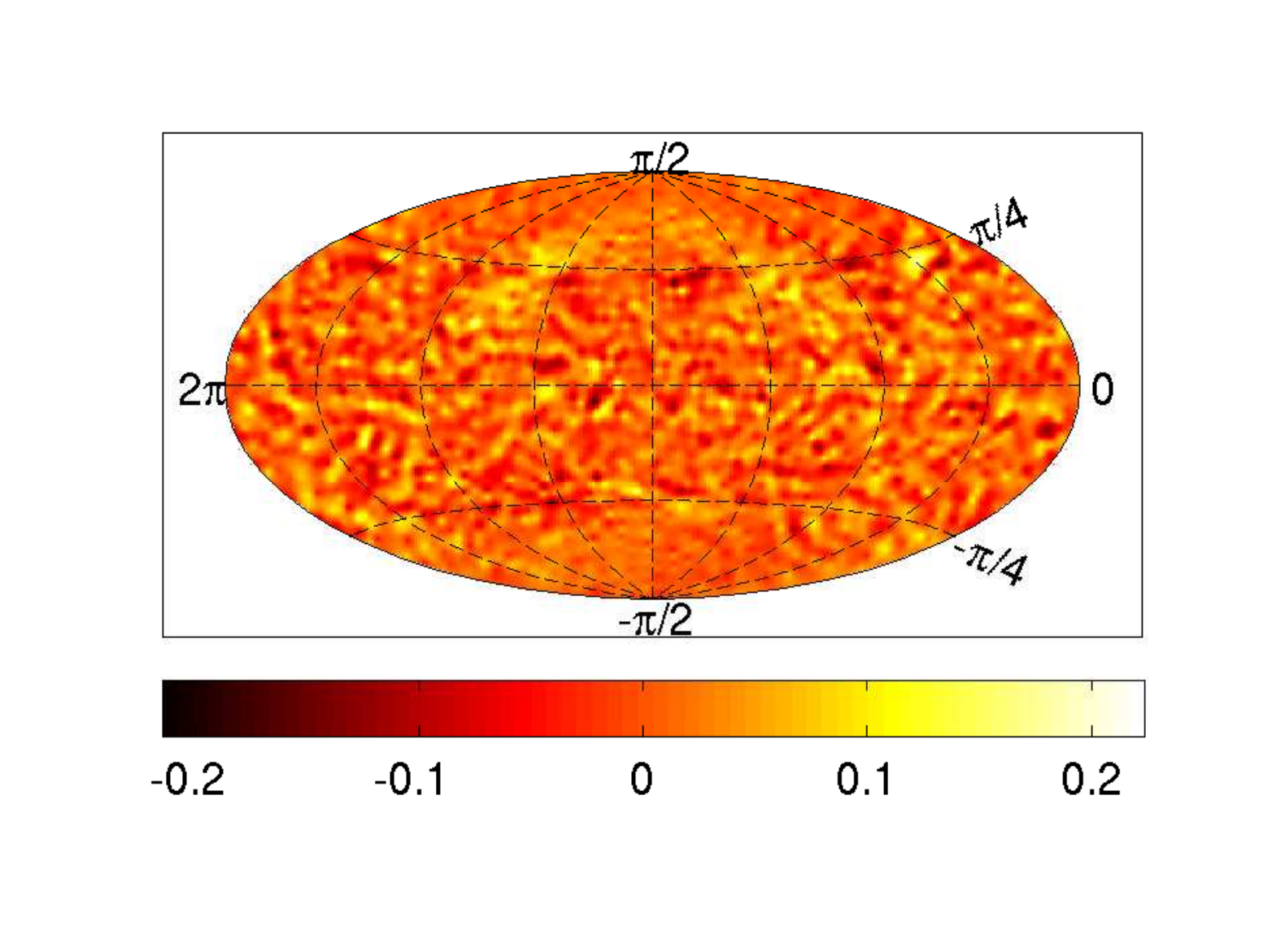}}
\subfigure[~H1L1V1 (NMSE=1.3503)]
{\includegraphics[width=0.238\textwidth]{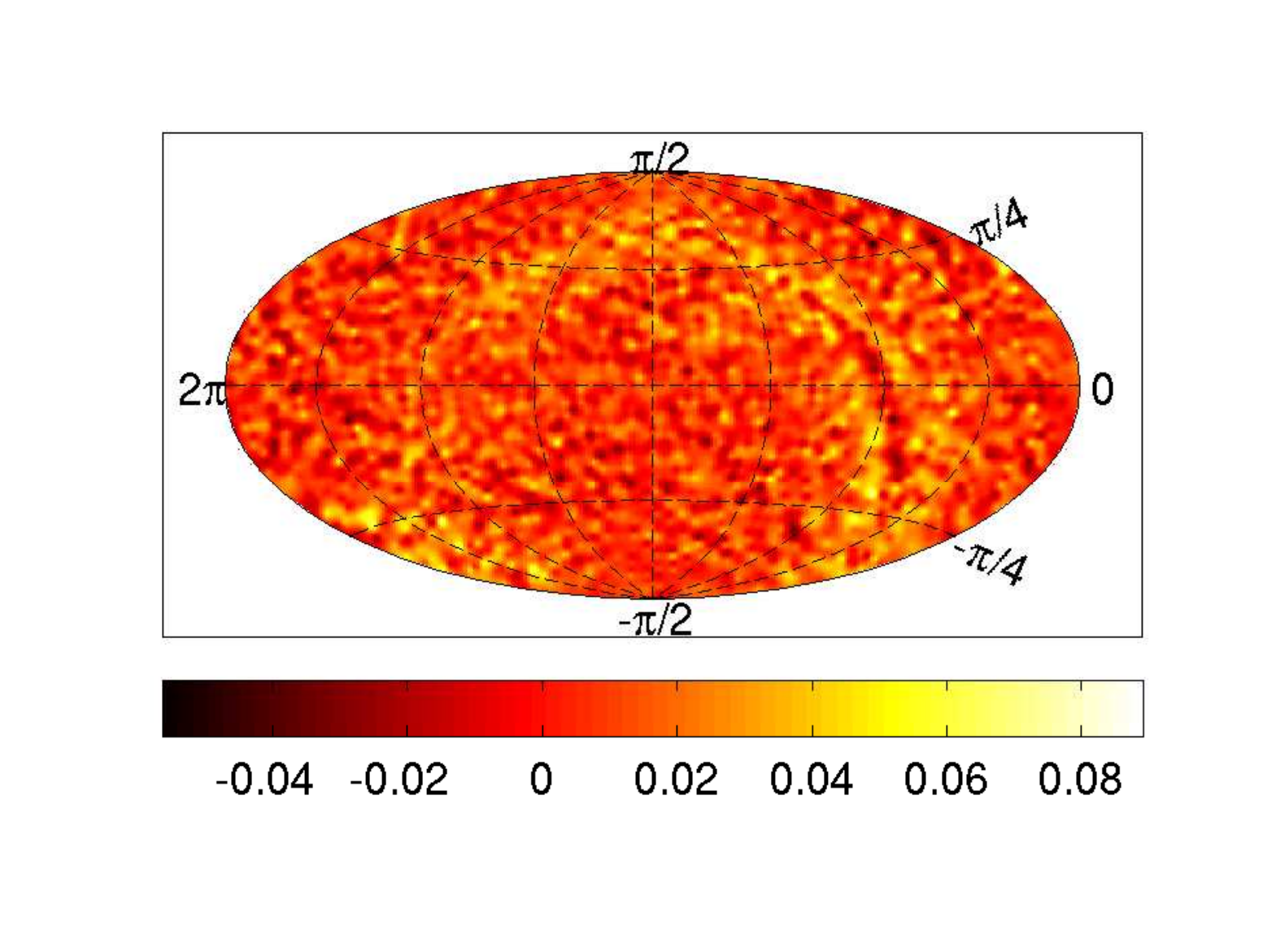}}
\caption{\label{fig:cleanweak}Clean maps obtained by the deconvolution of the dirty maps of Fig.~\ref{fig:dirtyweak}, using 20 CG iterations.}
\end{center}
\end{figure}
\begin{table}[ht]
\caption{MLR statistic of dirty maps ($\lambda$) versus clean maps ($\lambda_{c}$) for the simulated maps in Figs.~\ref{fig:dirtyweak} and~\ref{fig:cleanweak}.}
\centering
\begin{tabular}{c| c| c}
\hline\hline
Baseline & $\lambda$ & $\lambda_{c}$ \\ [0.5ex]
\hline
H1L1 & 98.643 & 98.670 \\
L1V1 & 43.520 & 43.536 \\
H1V1 & 40.432 & 40.475 \\
H1L1V1 & 115.132 & 115.176 \\ [1ex]
\hline
\end{tabular}
\label{table:Weak_CMB}
\end{table}

Visually, the dirty maps are now weaker (Fig.~\ref{fig:dirtyweak}), and clean maps almost do not show the obvious presence of any source (Fig.~\ref{fig:cleanweak}), but the MLR statistic (Table~\ref{table:Weak_CMB}) provides a clear and reliable indication of the presence of a signal, thus proving its usefulness in the search for signal in a noisy map.


\section{Discussion}
\label{sec:conclusion}

The search for an anisotropic stochastic gravitational wave background plays an important role in present GW research. In addition to setting interesting upper limits on astrophysical and cosmological backgrounds, the simplicity of the concomitant analysis reveals invaluable knowledge about the coherent performance of the GW detector network. 

So far, detailed analysis strategies have been developed to search for anisotropic background in pixel and spherical harmonic spaces, and a general maximum-likelihood-based framework has been established to search in any convenient basis. 
The spherical harmonic search has been demonstrated using a network of detectors~\cite{ThraneEtal}. In this paper, for the first time, we numerically implement the directed radiometer search, including deconvolution, for a network of detectors. These methods, in the past, were focused primarily on showing that the observed map is consistent with Gaussian noise or in estimating sky maps. The latter required the inversion of the convolution equation, which itself assumed the network of detectors to be nondegenerate. Neither of these methods may work in the presence of excessive noise and weak signal. Most importantly, a statistically meaningful, all-sky combined statistic, in the form of an optimal ``detection statistic,'' was needed in order to make precise statements about the presence or absence of a given background model in a map. Here, we proposed a MLR statistic, which yields a single number when computed on the dirty or the clean map and can be used as a detection statistic. By computing the MLR statistic for a couple of toy models of the background, we observe that the detection statistic is much larger than the noise-only case, even in the presence of weak signals that are barely visible in dirty or clean maps. We corroborated these statements with results obtained from extensive Monte Carlo simulations of a diffuse background of known shape in an ensemble of noise realizations. However, a more detailed study using signals from a variety of background models is surely worth pursuing in order to determine how accurate the templates need to be in order to extract meaningful information from weak backgrounds. 

We also compared the performance of individual baselines and the whole network for the directed radiometer search using different figures of merit. 
Evaluating the performance of a network of GW detectors in SGWB searches is relatively straightforward compared to other GW signal searches~\cite{Bose00,Pai01,WenChen04,Hayama07,Dhurandhar08, Gursel89}. This exercise was useful in drawing insights about the characteristics of a network that are particularly helpful in boosting its performance.
Our overall observation, not surprisingly, is that the network improves performance in mainly three ways, namely, (1)~by increasing the sensitivity by observing each direction a greater number of times, (2)~by observing the sky more uniformly, and (3)~by probing each direction on the sky with additional detectors on the globe. The latter two enhancements lead to better localization of pointlike sources. This can be understood via the behavior of the Fisher information matrix: More detectors reduce its degeneracy and improve the well-posedness of the inverse problem. This, in turn, leads to a more accurate production of clean maps. 

Another question worth addressing in the future is about how closely spaced must the templates be on the parameter space to maximize the chances of detection with available computational resources. Indeed, the proposal for templated searches for SGWB signals is not new to this paper. For example, it has been addressed earlier in the context of isotropic searches (see Ref. \cite{Bose:2005fm} and the references therein). Reference \cite{Bose:2005fm} also introduced a metric on the parameter space of those signals so as to enable an experimenter to infer what the principle axes are on that space and how fine a template bank one can afford based on the computational resources available. A similar study can be carried out for finding a more optimal spacing of templates for directed searches than the one used here.

Whereas results presented here were derived for Gaussian noise, the codes used can be applied to real data as well. Indeed, the performance of the proposed statistic in real data sets from the LIGO and Virgo detectors can be determined through hardware injections that were done in the recent science runs, such as the ones described in Ref. \cite{Bose:2003nb}, and supplementing them with multiple software injections to improve the statistics. The expected improvement of network sensitivity over individual baselines, as demonstrated here, merits the investment required for extending the current single-baseline analysis efforts \cite{Mitra,Ballmer} to a multibaseline one. This conclusion is strengthened by the fact that adding a detector to a baseline can potentially mitigate the contribution of cross correlated environmental noise that affects only one of the three resulting baselines. Including V1, which is on a different continental plate than the H1L1 baseline, can serve this purpose. Employing a null-stream statistic~\cite{Chatterji06,Gursel89} to complement the detection statistic might also help in discriminating against such noise.

\begin{acknowledgments}
We thank Joe Romano, Stefan Ballmer, and Warren Anderson for discussions, careful reading of the manuscript, and helpful comments. We would also like to thank Bruce Allen, Sanjeev Dhurandhar, Albert Lazzarini, Vuk Mandic, Himan Mukhopadhyay, Alan Weinstein, and  Holger Pletsch for helpful discussions. SM would like to acknowledge the Centre National d'\'Etudes Spatiales (France) for supporting part of the research. This work was supported in part by NSF Grant No. PHY-0855679. Part of the research described in this paper was carried out at the Jet Propulsion Laboratory, California Institute of Technology, under a contract with the National Aeronautics and Space Administration. LIGO was constructed by the California Institute of Technology and Massachusetts Institute of Technology with funding from the National Science Foundation and operates under cooperative agreement PHY-0757058.
\end{acknowledgments}

\appendix
\section{Parameter accuracy}
The {\textit{match}} can be rewritten as
\begin{equation}\label{eqA.1}
M = 1-g_{\alpha\beta}\;\Delta\Theta_{(k)}^{\alpha}\Delta\Theta_{(k)}^{\beta} \, , 
\end{equation}
where
\begin{eqnarray}\label{eqA2,3}
\Gamma_{(k)\alpha\beta} &=& (\text{SNR}_{(k)})^2\, g_{\alpha\beta}({\bf{\Theta}}_{(k)})\, ,\\
g_{\alpha\beta} & = & -\frac{1}{2}\left(\frac{\partial^{2}M}{\partial\Theta_{(k^{\prime})}^{\alpha}\partial\Theta_{(k^{\prime})}^{\beta}}\right)\Bigg|_{{\bf{\Theta}}_{(k^{\prime})}={\bf{\Theta}}_{(k)}}=g_{\alpha\beta}({\bf{\Theta}}_{(k)})\nonumber \, ,\\
&:=& \begin{pmatrix} g_{\mu_{(k)}\mu_{(k)}} & g_{\mu_{(k)}\phi_{(k)}} \\ g_{\phi_{(k)}\mu_{(k)}} & g_{\phi_{(k)}\phi_{(k)}}\\ \end{pmatrix} \, .
\end{eqnarray}
The components of the above $g_{\alpha\beta}$ matrix are obtained from the derivatives 
of the beam matrix:
\begin{eqnarray}\label{eqA4}
&&g_{\mu_{(k)}\mu_{(k)}} =\nonumber\\
&& \frac{1}{2(\mathcal{B}_{(k)(k)})^2}\Bigg[({\mathcal B}_{(k)(k)})\left(\frac{\partial^{2}}{\partial\mu_{k^{\prime}}\partial\mu_{k}}\mathcal{B}_{k^{\prime}k}\Big|_{k^{\prime}=k}\right) \nonumber\\
&&-\left(\frac{\partial}{\partial\mu_{k^{\prime}}}\mathcal{B}_{k^{\prime}k}\Big|_{k^{\prime}=k}\right)^2\Bigg]\, ,
\end{eqnarray}
\begin{eqnarray}\label{eqA5}
&&g_{\mu_{(k)}\phi_{(k)}} = g_{\phi_{(k)}\mu_{(k)}} =\nonumber \\
&& \frac{1}{2(\mathcal{B}_{(k)(k)})^2}\Bigg[({\mathcal B}_{(k)(k)})\left(\frac{\partial^{2}}{\partial\mu_{k^{\prime}}\partial\phi_{k}}\mathcal{B}_{k^{\prime}k}\Big|_{k^{\prime}=k}\right) \nonumber\\
&&-\left(\frac{\partial}{\partial\mu_{k^{\prime}}}\mathcal{B}_{k^{\prime}k}\Big|_{k^{\prime}=k}\right)\left(\frac{\partial}{\partial\phi_{k^{\prime}}}\mathcal{B}_{k^{\prime}k}\Big|_{k^{\prime}=k}\right)\Bigg]\, ,
\end{eqnarray}
\begin{eqnarray}\label{eqA6}
&&g_{\phi_{(k)}\phi_{(k)}} =\nonumber\\
&& \frac{1}{2(\mathcal{B}_{(k)(k)})^2}\Bigg[({\mathcal B}_{(k)(k)})\left(\frac{\partial^{2}}{\partial\phi_{k^{\prime}}\partial\phi_{k}}\mathcal{B}_{k^{\prime}k}\Big|_{k^{\prime}=k}\right) \nonumber\\
&&-\left(\frac{\partial}{\partial\phi_{k^{\prime}}}\mathcal{B}_{k^{\prime}k}\Big|_{k^{\prime}=k}\right)^2\Bigg]\, .
\end{eqnarray}
The estimation error \eqref{eq5.5} is obtained from Eqs.~\eqref{eqA2,3} and ~\eqref{eq5.4} by utilizing the fact that the inverse of the determinant of a matrix is the same as the determinant of the inverse of that matrix.

\bibliography{refs}

\end{document}